\documentclass[12pt, a4paper]{article}

\pdfoutput=1

\usepackage{amsmath}
\allowdisplaybreaks
\usepackage{amsfonts}
\usepackage{amssymb}
\usepackage{bbm}
\usepackage{verbatim}
\usepackage{dsfont}
\usepackage{booktabs}
\usepackage{slashed}
\usepackage{bm}

\usepackage{epsfig}
\usepackage{color}
\usepackage[table]{xcolor}
\usepackage{graphicx}
\usepackage[]{caption}
\usepackage[listofformat=empty,subrefformat=empty]{subfig}

\usepackage{float}
\usepackage{overpic}

\usepackage{enumerate}
\usepackage{hhline}
\usepackage{multirow}

\usepackage{cite}
\usepackage{xspace}
\usepackage{setspace}

\usepackage[pdftitle={Quantum corrections in D-brane models with broken supersymmetry},
  pdfauthor={},
  pdfsubject={},
  bookmarksopen, bookmarksnumbered, bookmarksopenlevel=2, colorlinks=false, linkcolor=blue, citecolor=blue, urlcolor=blue]{hyperref}

\DeclareMathOperator{\tr}{\text{tr}}

\newcommand{\ot}{\ensuremath{\overline{\theta}}}
\newcommand{\nt}{\ensuremath{\theta}}
\newcommand{\bz}{\overline{z}}

\newcommand{\A}{\alpha}

\newcommand{\B}{\beta}
\newcommand{\Bb}{\bar{\beta}}
\newcommand{\G}{\gamma}

\newcommand{\D}{\delta}

\newcommand{\oN}{\overline{N}}
\newcommand{\opsi}{\bar{\psi}}
\newcommand{\ophi}{\bar{\phi}}
\newcommand{\ochi}{\bar{\chi}}
\newcommand{\olambda}{\overline{\lambda}}
\newcommand{\E}{\varepsilon}
\newcommand{\e}{\epsilon}
\newcommand{\R}{\rho}
\newcommand{\pd}{\partial}
\newcommand{\opd}{\overline{\partial}}
\newcommand{\bxi}{\overline{\xi}}
\newcommand{\bM}{\overline{M}}
\newcommand{\kk}{\eta}
\newcommand{\bk}{\overline{\eta}}

\begin{document}

\thispagestyle{empty}

\begin{flushright}
CPHT-RR046.072019\\
DESY 19-125\\
\end{flushright}
\vskip .8 cm
\begin{center}
{\Large {\bf Quantum corrections for D-brane models 
\\with broken supersymmetry}}\\[12pt]

\bigskip
\bigskip 
{
{\bf{Wilfried Buchmuller$^\dagger$}\footnote{E-mail: wilfried.buchmueller@desy.de}},
{\bf{Emilian Dudas$^\ast$}\footnote{E-mail: emilian.dudas@cpht.polytechnique.fr}},
{\bf{Yoshiyuki Tatsuta$^\dagger$}\footnote{E-mail: yoshiyuki.tatsuta@desy.de}}
\bigskip}\\[0pt]
\vspace{0.23cm}
{\it $^\dagger$ Deutsches Elektronen-Synchrotron DESY, 22607 Hamburg, Germany \\ \vspace{0.2cm}
$^\ast$ CPHT, CNRS, Institute Polytechnique de Paris,  France}\\[20pt] 
\bigskip
\end{center}

\begin{abstract}
\noindent
Intersecting D-brane models and their T-dual magnetic
compactifications yield attractive models of particle physics where
magnetic flux plays a twofold role, being the source of fermion
chirality as well as supersymmetry breaking. A potential problem of
these models is the appearance of tachyons which can only be avoided in
certain regions of moduli space and in the presence of Wilson lines. We study the effective
four-dimensional field theory for an orientifold compactification of
type IIA string theory and the corresponding toroidal
compactification of type I string theory. After determining the
Kaluza-Klein and Landau-level towers of massive states in different sectors of the
model, we evaluate their contributions to the one-loop effective potential, summing over all
massive states, and we relate the result to the corresponding string
partition functions. We find that the Wilson-line effective potential
has only saddle points, and the theory is therefore driven to the tachyonic
regime. There tachyon condensation takes place and chiral fermions
acquire a mass of the order of the compactification scale.
We also find evidence for a tachyonic behaviour of the volume
moduli. More work on tachyon condensation is needed to clarify
the connection between supersymmetry breaking,
a chiral fermion spectrum and vacuum stability.
\end{abstract}

\newpage 
\setcounter{page}{2}
\setcounter{footnote}{0}

{\renewcommand{\baselinestretch}{1.5}

\section{Introduction}
\label{sec:Introduction}

Intersecting D-brane models and their T-dual magnetic
compactifications provide attractive and intuitive string theory
compactifications to four dimensions with chiral fermion spectra
\cite{Angelantonj:2002ct,Blumenhagen:2006ci}. The main emphasis in
model building has been on the construction of vacua with unbroken
$\mathcal{N}=1$ supersymmetry (for a review and references, see
\cite{Blumenhagen:2005mu,Ibanez:2012zz}), but in 
absence of any hint for supersymmetry at the Large Hadron Collider,
models where supersymmetry is broken at a high scale, in the spirit of
`split supersymmetry' \cite{ArkaniHamed:2004fb,Giudice:2004tc} or 
`split symmetries' \cite{Buchmuller:2015jna,Buchmuller:2017vho}, 
are also of current interest.

An intriguing aspect of magnetic compactifications is the connection
between fermion chirality and supersymmetry breaking
\cite{Bachas:1995ik}, which occurs in
compactifications of type I strings on tori and orbifolds
\cite{Abouelsaood:1986gd,Blumenhagen:2000wh,Angelantonj:2000hi}
and in the related intersecting D-brane models
\cite{Berkooz:1996km,Blumenhagen:2000wh,Aldazabal:2000dg}. 
This setup allows to
construct models which come very close to the Standard Model of
particle physics \cite{Aldazabal:2000cn,Ibanez:2001nd,Blumenhagen:2001te,Cvetic:2001nr}. 
Generically, magnetic compactifications have tachyonic instabilities of
Nielsen-Olesen type \cite{Nielsen:1978rm}. Originally, one could hope to relate such
an instability to electroweak symmetry breaking
\cite{Bachas:1995ik,Aldazabal:2000cn,Ibanez:2001nd}
in case of a low string scale and large extra dimensions. This
is no longer viable but the structure of the setup is rich enough to
incorporate in principle also split supersymmetry 
\cite{Antoniadis:2004dt,Antoniadis:2006eb}.

The goal of this paper is the computation of quantum corrections for
string compactifications with magnetic background flux. This is partly
motivated by the recent observation that in quantum corrections to
Wilson-line scalars large cancellations occur \cite{Buchmuller:2016gib,
Ghilencea:2017jmh,Buchmuller:2018eog,Hirose:2019ywp} due to the
presence of magnetic flux. This suggests that in appropriate compactifications
similar cancellations  may occur in quantum corrections to Higgs
masses, which would be important in view of the hierarchy problem. In
order to address these questions we extend the previous calculations for
six-dimensional field theory models to a full string compactification
on magnetized tori. Notice, that another motivation of our effective field theory approach is that, whenever supersymmetry is broken
by magnetic fluxes, in string theory  NSNS tadpoles are generated that make any quantum computation very hard, both conceptually and technically
(see, for example, \cite{Dudas:2004nd}). 

Our starting point is the construction of an intersecting brane model
with broken supersymmetry in a matter sector without tachyons and with
chiral fermions which can acquire mass via the Higgs mechanism. For
simplicity, and to facilitate the computation of quantum corrections,
we choose as unbroken gauge group $U(N)\times U(1)\times U(1)$ rather
than the Standard Model gauge group. The model has a Higgs sector and
antisymmetric tensor fields with fermions in vector-like
representations. Some scalar masses in these sectors depend on the distance
between branes that are parallel in some tori.  These moduli
correspond to Wilson-line scalars in the T-dual picture. They become
tachyonic if the branes come close to each other. At tree level the
Wilson-line potential is flat. However, as we shall see, one-loop
quantum corrections make it concave, implying that the system is driven into the
tachyonic regime of moduli space.

After determining intersection numbers and scalar masses for the
D-brane model, we turn to the T-dual magnetic compactification which
is better suited to evaluate the four-dimensional (4d) effective field theory.
Starting from the 10d $SO(32)$ Super-Yang-Mills Lagrangian expressed in terms of 
$\mathcal{N}=1$ vector and chiral superfields
\cite{Marcus:1983wb,ArkaniHamed:2001tb}, we compute the 4d effective
action for a toroidal compactification with three $U(1)$
magnetic background fluxes that break $SO(32)$ to $U(N)\times
U(1)\times U(1)$. For each sector of the model we determine the 
Kaluza-Klein (KK) and Landau-level (LL) towers
of mass eigenstates of vectors, fermions and scalars. The calculations
are based on the harmonic oscillator algebra of covariant derivatives
in a flux background \cite{Bachas:1995ik,Cremades:2004wa,Alfaro:2006is,
Hamada:2012wj,Buchmuller:2018eog}. The mass spectra
are compared with the string formula of Bachas, also in view of
supersymmetries that remain unbroken
for particular choices of magnetic fluxes in some sectors.

In the Higgs sector branes are parallel in some tori and, knowing the
spectrum of massive KK and LL states, we compute the effective potential as
function of magnetic flux and Wilson lines. The effective potential is
also obtained in the field theory limit of the corresponding string
partition function, and the two results agree. As function of the Wilson line the potential is
concave and there are no local minima. Hence, the tree level vacua
with non-vanishing Wilson lines are unstable. This is a new result
of our paper. For vanishing Wilson lines tachyon condensation takes
place and all chiral fermions acquire masses of the order of the
compactification scale.

The contributions to the effective potential from the various sectors are most easily
obtained from the corresponding string partition functions. In sectors
without Wilson lines we also calculate the effective potential as function
of the volume moduli of the three tori. We find evidence that also in
this case the system is driven to the tachyonic regime of moduli space,
which would imply that the only vacuum state corresponds to the
decompactification limit.  A further, well-known problem is the NSNS
tadpole (see, for example, \cite{Dudas:2004nd}) in case of broken supersymmetry.

The paper is organized as follows. The intersecting D-brane model 
and its T-dual magnetic compactification are discussed in Sections~\ref{sec:Dbranes} and 
\ref{sec:Tdual}, respectively. Mass eigenstates and mass spectra are
derived in Sections~\ref{sec:matter} and \ref{sec:higgs}, and the
effective one-loop potential is computed in
Section~\ref{sec:potential}. Section~\ref{sec:tachyon} deals with
tachyon condensation. The appendices \ref{app:N2N} and
\ref{app:commutators} give details concerning the embedding of the
various sectors of the model in the adjoint representation of
$SO(32)$, and in the appendices \ref{app:susy} and \ref{app:jacobi}
some formulae are collected for superfield components and Jacobi
functions, respectively.

\section{Intersecting D-brane model}
\label{sec:Dbranes}

We are interested in a D-brane model with broken supersymmetry, which
contains a `matter sector' with chiral fermions and a `Higgs sector'
with vector-like fermions such that vacuum expectation values of Higgs
fields can give mass to the chiral fermions. As a simple example, we choose the gauge group
\begin{align}\label{gaugegroup}
G = U(N) \times U(1) \times U(1)\ ,
\end{align}
corresponding to a stack of $N$ branes, $a$, and two single branes, $b$ and $c$.
The fermions are supposed to be chiral with respect to $U(1) \times U(1)$
and vector-like with respect to the `colour group' $U(N)$. Following
\cite{Blumenhagen:2000wh,Ibanez:2001nd}}, we start from 
type IIA string theory compactified on a rectangular factorized torus 
$T^6=T^2_1\times T^2_2\times T^2_3$ with real coordinates $x_4, \ldots ,
x_9$ and complex coordinates $z_i = (x_{2+2i} + i x_{3+2i})/2$,
$i=1,2,3$, with the identifications $z_i \sim z_i +L_i/2$, 
$z_i \sim z_i +iL'_i/2$. An orientifold is obtained by dividing
out the discrete symmetry $\Omega\mathcal{R}(-1)^{F_L}$,
where $\Omega$ is worldsheet parity, $F_L$ is left-moving
fermion number, and $\mathcal{R}$ is a reflection symmetry of $T^6$,
\begin{align}\label{R}
\mathcal{R}: (z_1,z_2,z_3) \rightarrow
(\bar{z}_1,\bar{z}_2,\bar{z}_3)\ .
\end{align}
The orientifold has eight $O6$-planes along Minkowski space and the
directions $x_{3+2i}$ that are invariant under $\mathcal{R}$. The
orientifold planes are
localized at the fixed points $(\hat{z}_1,\hat{z}_2,\hat{z}_3)$, $\hat{z}_i = (0,
iL'/4)$.  Each orientifold plane has RR charge $Q_{O6} = -2$ in units of a
D6-brane charge.
\begin{table}[t]
\begin{center}
\begin{tabular}{l|ccc} 
\hline
\hline
Branes, gauge groups & $(n^1,m^1)$ & $(n^2,m^2)$ & $(n^3,m^3)$ \\
\hline
$a\ ,\ U(N)$ & $(1,0)$ & $(1,2)$ & $(1,1)$ \\
$b\ ,\ U(1)$  & $(1,1)$ & $(1,l)$ & $(1,-2)$ \\
$c\ ,\ U(1)$  & $(1,1)$ & $(1,-l)$ & $(1,2)$\\
\hline
\end{tabular}
\end{center}
\caption{Intersecting D-brane model. Wrapping numbers of a stack of $N$
  branes, $a$, and two single branes, $b$ and $c$.}
\label{tab:wrapping}
\end{table}
Cancellation of the total RR charge requires 16 D6-branes together
with 16 mirror D6 branes to satisfy the reflection symmetry $\mathcal{R}$ of the
compact space. A brane $e$ is wrapped around the  1-cycles $[a_i]$
and $[b_i]$ of the 2-tori $T^2_i$ with wrapping numbers $n_e^i$ and
$m_e^i$, yielding for the wrapped 3-cycle of the brane the homology
class\footnote{We mostly follow the conventions of \cite{Ibanez:2012zz}.}
\begin{align}
[\Pi_e] = \otimes_i \left(n^i_e [a_i] + m^i_e[b_i]\right)\ .
\end{align}
The homology class $[\Pi_{e'}]$ of the mirror brane is obtained from
$[\Pi_e]$ by replacing $m^i_e$ by $-m^i_e$.
In case of stacks of $N_e$ branes, leading to gauge symmetries
$U(N_e)$, the RR tadpole cancellation condition can now be written as
\begin{align}\label{RRcancellation}  
\sum_e N_e [\Pi_e] - 2 [\Pi_{O6}] = 0\ ,
\end{align}
where $[\Pi_{O6}] = 8 \otimes_i [a_i]$ is the homology class of the
orientifold plane. 

We are interested in the gauge group $U(N)\times U(1)\times U(1)$,
corresponding to one stack of $N$ branes, $a$, with gauge group
$U(N)$, and two further single $U(1)$ branes, $b$ and
$c$. 
Table~\ref{tab:wrapping} shows a set of wrapping numbers
which can be consistent with the wanted gauge group $U(N)\times
U(1)\times U(1)$.
We have chosen all wrapping number in the $x_{3+2i}$ directions equal,
$n^i = 1$, and one wrapping number in the first torus as zero,
$m^1_a = 0$. In this case, the tadpole conditions
\eqref{RRcancellation}  read explicitly,
\begin{equation}
\begin{split}
N+2 &= 16\ ,\\
N m^2_a m^3_a + m_b^2 m^3_b + m_c^2 m^3_c &= 0\ ,\\
m_b^1 m^3_b + m_c^1 m^3_c &= 0\ ,\\
m_b^1 m^2_b + m_c^1 m^2_c &= 0\ .
\end{split}
\end{equation}
\begin{figure}[t]
\begin{center} 
\includegraphics[width = 0.7\textwidth]{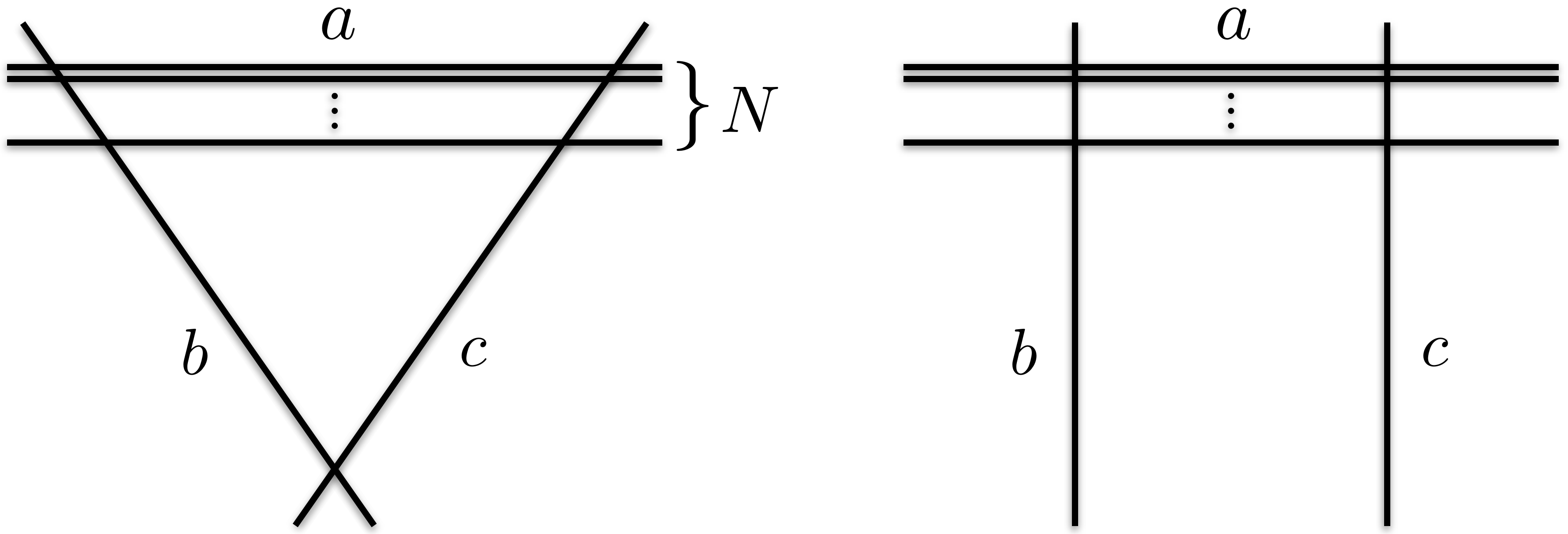}
\end{center}
\vspace{0.3cm}	
\caption{Left: intersections of brane stack $a$ with branes $b$ and
  $c$,  and brane $b$ with $c$ in the second torus $T^2_2$; right:
  intersections of  brane stack $a$ with branes $b$ and $c$ in torus
$T^2_1$ where branes $b$ and $c$ are parallel.}
\label{fig:intersections}
\end{figure}
One easily verifies that these equations are solved by the ansatz in Table~\ref{tab:wrapping},
with $N=14$, $l=7$.
The chosen wrapping numbers imply that not all branes intersect in all
tori: $a$ and $a'$, and $b$ and $c$  are parallel in the first torus,
whereas $b$ and $c'$ are parallel in the second and in the third torus.
This situation is illustrated in Figure~\ref{fig:intersections}.

On each brane an $\mathcal{N}=4$ supermultiplet of zero-modes in the
adjoint representation of the gauge group is localized. The branes
intersect at angles determined by the wrapping numbers. At these
intersections fermions and scalars in bi-fundamental representations
$(N_e,\bar{N}_f)$ are localized. For non-zero intersection numbers 
\begin{align}\label{Inumber}
I_{ef} = \otimes_i \left(n^i_em^i_f - m^i_en^i_f\right)
\end{align}
the fermion spectrum is chiral. The fermions are left-handed for
$I_{ef} > 0$ and right-handed for $I_{ef} < 0$, corresponding to
left-handed fermions in the complex conjugate representation $(\bar{N}_e,N_f)$.
%
\begin{table}[t]
\begin{center}
\begin{tabular}{c|ccc} 
\hline
\hline
Brane sector & Intersection number $I$ & 4d fermions (L)   \\
\hline
$ab + ba$ & $-3(l-2)$ & $\bar{N}_{1,0}$  \\
$ac + ca$ & $-(l+2)$ & $\bar{N}_{0,1}$  \\
$ab' + b'a$ & $l+2$ & ${N}_{1,0}$  \\
$ac' + c'a$ & $3(l-2)$ & ${N}_{0,1}$  \\
$aa'$ & $0$ & $N(N-1)/2$, $\bar{N}(\bar{N}-1)/2$\\
$bc + cb$ & $0$ & $1_{1,-1}$, $1_{-1,1}$  \\
$bc' + c'b$ & $0$ & $1_{1,1}$, $1_{-1,-1}$  \\
\hline
\end{tabular}
\end{center}
\caption{Chiral and vector-like representations of left-handed fermions at
  various brane intersections.}
\label{tab:fermions}
\end{table}
At the intersections of the brane system defined in
Table~\ref{tab:wrapping} one obtains the left-handed fermions listed
in Table~\ref{tab:fermions}. There are matter fields that carry `colour',
transforming as $N$ or $\bar{N}$ under $SU(N)$. They form a chiral representation of the
full gauge group, whereas colour singlet `Higgs fields' form 
vector-like representations. The quantum numbers of the chiral
fermions allow Yukawa couplings that are most conveniently expressed
in terms of the associated chiral superfields,
\begin{align}\label{yuk1}
\mathcal{L}_Y \supset \sum_{r,s}^{3(l-2)} y^{(1)}_{rs}
\bm{\bar{N}}^r_{1,0} \bm{N}^s_{0,1} \bm{1}_{-1,-1} +
 \sum_{r,s}^{l+2} y^{(2)}_{rs} \bm{\bar{N}}^r_{0,1} \bm{N}^s_{1,0} \bm{1}_{-1,-1}\ .
\end{align}
These couplings lead to fermion mass terms after a vacuum
expectation value $\langle 1_{-1,-1} \rangle \neq 0$ breaks the chiral
group $U(1)\times U(1)$ to the diagonal $U(1)$ subgroup. The complete
list of Yukawa couplings will be given in the subsequent section.

In the brane sector $aa'$, $bb'$ and $cc'$ chiral fermions
in symmetric and antisymmetric representations of the gauge group
occur with multiplicities 
\begin{equation}
\begin{split}
n_{\text{sym},e} &= \frac{1}{2}\left(I_{ee'}-I_{e,O6}\right) = - 4
m^1_em^2_em^3_e\left(n^1_en^2_en^3_e-1\right)\ , \\
n_{\text{asym},e} &= \frac{1}{2}\left(I_{ee'}+I_{e,O6}\right) = - 4
m^1_em^2_em^3_e\left(n^1_en^2_en^3_e+1\right) ,\; e=a,b,c\ .
\end{split}
\end{equation}
Since in our model $m^1_a=0$ and $n^1_e = n^2_e = n^3_e = 1$ for
$e=a,b,c$, there are no chiral fermions in symmetric or antisymmetric
representations. As we shall see in the following section, a
vector-like pair of fermions in the antisymmetric representation of
$SU(N)$ occurs in the $aa'$-sector. The $bb'$- and the $cc'$-sector
correspond to $U(1)$ symmetries where such representations are absent.

The masses of bi-fundamental scalars depend on the angles at which the
branes intersect. We restrict ourselves to small angles with respect
to the orientifold planes,
\begin{align}\label{thetarho}
\tan{\theta^i_e} = m^i_e \rho_i \simeq \theta^i_e \ ,\quad
\rho_i = \frac{L'_i}{L_i}\ ,\ i=1,2,3\ ,\ e=a,b,c\ ,
\end{align}
where $L_i/(2\pi)$ and $L'_i/(2\pi)$ are the two radii of the torus
$T^2_i$, respectively.
In the T-dual picture small angles correspond to large areas of the
dual tori so
that we shall be able to use a field theory approximation to string
partition functions.

At the intersection of two stacks of branes, $e$ and $f$, one then has
three light bi-fundamental scalars with masses \cite{Aldazabal:2000dg}
\begin{equation}\label{scalarmasses}
\begin{split}
4\pi\alpha' M^2_1|_{ef} &= -|\theta^1_{ef}| + |\theta^2_{ef}| +
|\theta^3_{ef}| \\
4\pi\alpha' M^2_2|_{ef} &= |\theta^1_{ef}| - |\theta^2_{ef}| +
|\theta^3_{ef}| \\
4\pi\alpha' M^2_3|_{ef} &= |\theta^1_{ef}| + |\theta^2_{ef}| - |\theta^3_{ef}| \ ,
\end{split}
\end{equation}
\newpage
\noindent
where $\theta^i_{ef} = \theta^i_e - \theta^i_f$, with $-\pi/2 \leq
\theta^i_{ef} \leq \pi/2$. For the model
defined in Table~\ref{tab:wrapping} one obtains
\begin{align}
\theta^1_a &= 0\ , & \theta^2_a &= 2\rho_2\ , & \theta^3_a &=
\rho_3\ , \nonumber\\
\theta^1_b &= \rho_1\ ,  & \theta^2_b &= l\rho_2\ , &
\theta^3_b &= -2\rho_3\ , \nonumber\\ 
\theta^1_c &= \rho_1\ ,  & \theta^2_c &= -l\rho_2\ , &
\theta^3_c &= 2\rho_3\ .
\end{align}
Using Eq.~\eqref{scalarmasses} these angles yield the scalar mass
spectrum at the various brane intersections which is listed in Table~\ref{tab:scalarmasses}.
\begin{table}[b]
\begin{center}
\begin{tabular}{c|ccc} 
\hline
\hline
Brane sectors & $4\pi\alpha'M_1^2|_{ef}$ & $4\pi\alpha'M_2^2|_{ef}$ & $4\pi\alpha'M_3^2|_{ef}$\\
\hline
$ab,ac'$ & $-\rho_1 + (l-2)\rho_2 + 3\rho_3$ & $\rho_1 - (l-2)\rho_2 + 3\rho_3$ & $\rho_1 + (l-2)\rho_2 - 3\rho_3$ \\
$ab',ac$ & $-\rho_1 + (l+2)\rho_2 + \rho_3$ & $\rho_1 - (l+2)\rho_2 + \rho_3$ & $\rho_1 + (l+2)\rho_2 - \rho_3$ \\
$aa'$ & $4\rho_2 + 2\rho_3$ & $-4\rho_2 + 2\rho_3$ & $4\rho_2 - 2\rho_3$ \\
$bc$ & $2l\rho_2 + 4\rho_3$ & $-2l\rho_2 + 4\rho_3$ & $2l\rho_2 -4\rho_3$ \\
$bc'$ & $-2\rho_1 $ & $2\rho_1$ & $2\rho_1$\\
\hline
\end{tabular}
\end{center}
\caption{Masses of scalars at various brane intersections.}
\label{tab:scalarmasses}
\end{table}

In any sector of states stretched between any two stacks of branes
$e$, $f$, some supersymmetry is preserved provided that angles fulfill
the following conditions \cite{Berkooz:1996km},
\begin{align}
&\theta_{ef}^1 \pm  \theta_{ef}^2 \pm  \theta_{ef}^3 \not = 0 \ :  \quad\hspace{0.7cm}  {\cal N} = 4 \to {\cal N} = 0 \ , \nonumber \\
&\theta_{ef}^1 \pm  \theta_{ef}^2 \pm  \theta_{ef}^3 = 0 \ :  \quad\hspace{0.7cm}  {\cal N} = 4 \to {\cal N} = 1 \ , \nonumber \\
&\theta_{ef}^1 \pm  \theta_{ef}^2 = 0  \ , \; \theta_{ef}^3  = 0 \ :  \quad  {\cal N} = 4 \to {\cal N} = 2  \ . \label{susyangles}
\end{align}
 
In the considered model, a tachyon occurs in the $bc'$-sector,
\begin{align}
m^2_1|_{bc'} \propto -2\rho_1 < 0\ .
\end{align} 
A further tachyon appears either in the $aa'$-sector or in the $bc$-sector. 
In both sectors the flux in the first torus is zero, i.e.~$\theta^1 =
0$. Choosing $\theta^2 = \theta^3$ yields two massless scalars,
avoiding tachyons. For the $aa'$-sector this means
\begin{align}\label{rho32}
\rho_3 = 2\rho_2\ , 
\end{align}
which avoids coloured tachyons
but implies a second tachyon in the $bc$-sector,
\begin{align}
m^2_1|_{bc} \propto (-2l + 4)\rho_2 < 0\ .
\end{align} 
Together with the two
fermionic zero-modes the $aa'$-sector then forms a subsystem with
$\mathcal{N}=2$ supersymmetry. For the choice $l\rho_2=2\rho_3$
the roles the $aa'$-sector and the $bc$-sector are reversed.

\begin{figure}[t]
\begin{center} 
\includegraphics[width = 0.4\textwidth]{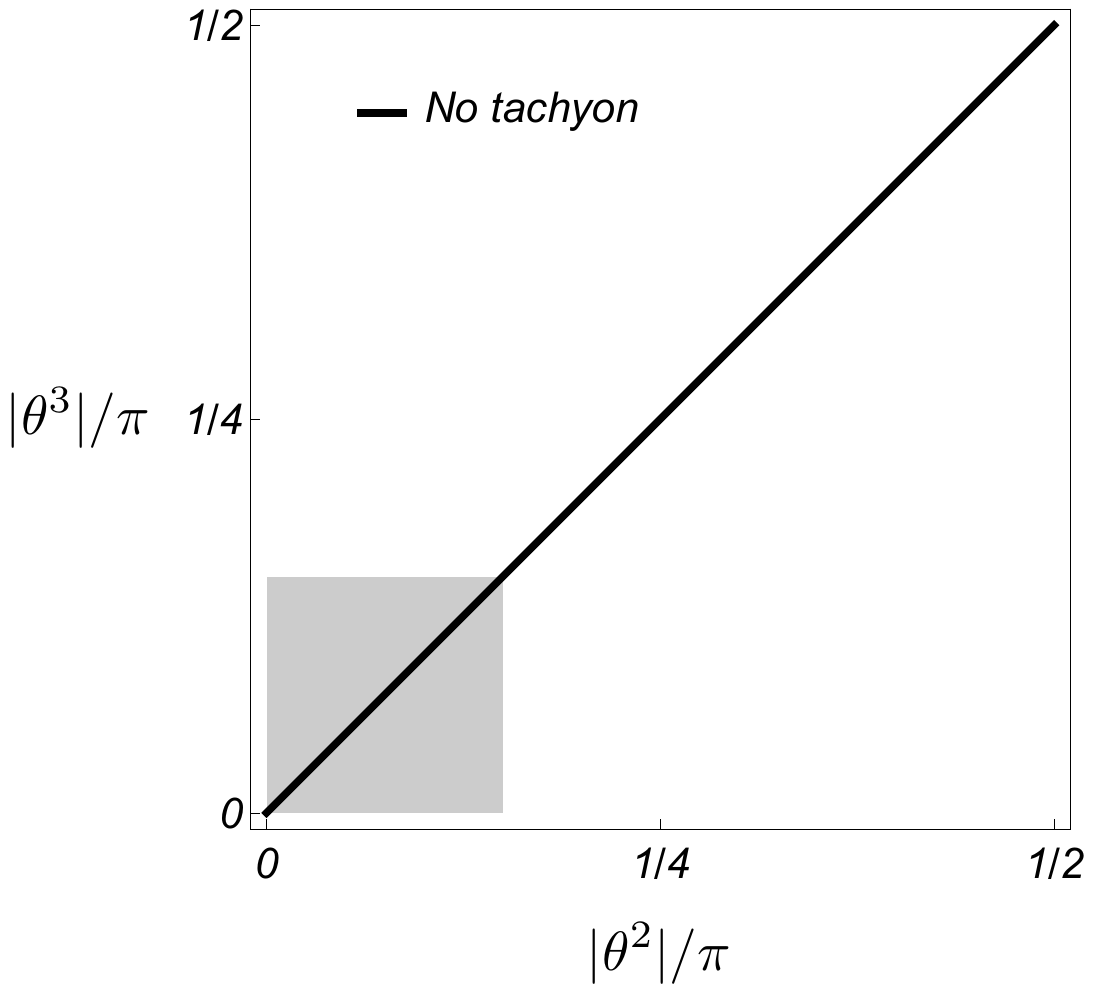}
\hspace{1.2cm}\includegraphics[width = 0.45\textwidth]{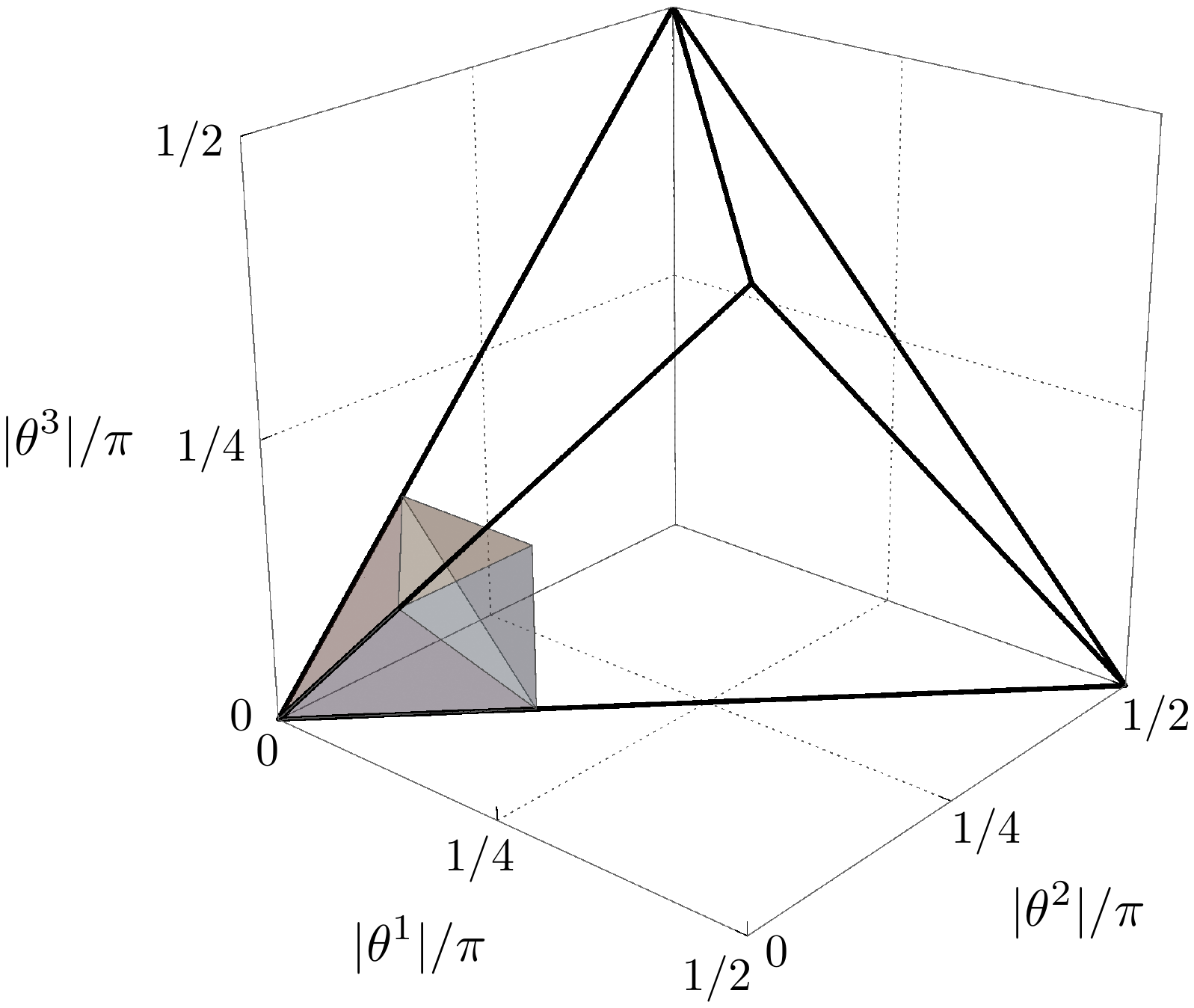}
\end{center}
\caption{Domain of angles for which no tachyons appear. Left: line in
  $\theta^2-\theta^3$-plane in case of no flux in first torus, i.e.~$\theta^1=0$.
Right: tetrahedron  in case of fluxes in all tori. The gray areas
indicate small-angle domains with $|\theta^i|/\pi < 0.15$.}
\label{fig:modulispace}
\end{figure}

The condition for absence of tachyons in the $ab$- and
$ac'$-sector reads
\begin{align}\label{ab}
(8-l)\rho_2 < \rho_1 < (l+4) \rho_2\ ,
\end{align}
and for the $ab'$- and $ac$-sector one obtains
\begin{align}\label{ac}
l\rho_2 < \rho_1 < (l+4) \rho_2\ .
\end{align}
With $l=7$, the last condition \eqref{ac} is the stronger one and implies
condition \eqref{ab}. Hence, once the conditions
\eqref{rho32} and \eqref{ac} are satisfied, all scalars in the sectors
$ab$, $ac'$, $ab'$ and $ac$ are massive and
supersymmetry is completely broken. The angles satisfying these
conditions form a tetrahedron \cite{Ibanez:2001nd}. It is illustrated
in Figure~\ref{fig:modulispace}, together with a domain of small
angles.

The appearance of tachyons is a generic feature of non-supersymmetric
intersecting D-brane models. However, it is argued that such
tachyons can be removed by couplings to moduli fields that parametrize
the distance between branes in tori where they are parallel. In the
T-dual picture discussed in the following section these moduli correspond
to Wilson-lines $\xi$, $\xi'$ that acquire
vacuum expectation values (see, for example,
\cite{Ibanez:2001nd,Antoniadis:2006eb}). 
In the present model the
corresponding superpotential terms would have the form
(in superfield notation, see Table~\ref{tab:fermions}),
\begin{align}
W_{\xi,\xi'} = \lambda_1~\xi~\mathbf{1}_{1,-1} \mathbf{1}_{-1,1} + \lambda_2~\xi'~
\mathbf{1}_{1,1} \mathbf{1}_{-1,-1}\ .
\end{align}
Clearly, existence and stability of a ground state require an appropriate potential for
$\xi$, $\xi'$. At tree-level the potential is flat.
To compute the one-loop quantum correction to the potential is an essential goal of this
paper. To achieve this we first construct the T-dual type I string
compactification on a magnetized torus, which allows a straightforward
computation of the full mass spectrum of the model as well as Yukawa couplings.

\section{T-dual toroidal flux compactification}
\label{sec:Tdual}

The intersecting D-brane model constructed in the previous section is
T-dual to a type I compactification on a magnetized dual rectangular torus
$T^2_1\times T^2_2\times T^2_3$ with the identifications 
\begin{align}
z_i \sim z_i + L_i/2\ , \quad z_i \sim z_i + 2\pi^2\alpha'/L'_i \ ,
\end{align}
where the angles $\theta^i_{e}$ between brane $e$ and the orientifold plane 
are related to magnetic flux densities in the 2-tori $T^2_i$ \cite{Berkooz:1996km},
\begin{align}
\tan \theta^i_{e} = 2\pi\alpha' g f^i_I\ . 
\end{align}
Here $g$ is the gauge coupling, brane $e$ ($e=a,b,c$) has a $U(1)$ group with
Cartan generator $H_I$ ($I=0,1,2$), and
$f^i_I$ is the corresponding flux density in the torus $T^2_i$.
Using Eq.~\eqref{thetarho},
$\tan{\theta^i_e}=m^i_I\rho_i$,  this
implies the Dirac quantization condition for the flux densities $f^i_I$,
\begin{align}\label{Dquantization}
g\int_{T^2_i} f^i_I = \frac{4\pi^2\alpha'}{\rho_i}gf^i_I = 2\pi m^i_I \ .
\end{align} 
For small angles, corresponding to small flux densities, one
has\footnote{In the following we shall use the notations $f^i_e$,
  $m^i_e$ and $f^i_I$, $m^i_I$ in parallel, according to convenience.}
\begin{align}
\theta^i_e \simeq m^i_I \rho_i = 2\pi\alpha' gf^i_I\ , \quad \rho_i = L'_i/L_i \ .
\end{align} 

The considered D-brane model has three stacks of branes and therefore
three $U(1)$ factors, $U(1)_a$, $U(1)_b$ and
$U(1)_c$. Correspondingly, each torus $T^2_i$ can have 
three flux densities $f^i_I$, which allow to break $SO(32)$ to the
gauge group of the D-brane model,
\begin{align}
SO(32) \supset U(16) \supset U(14) \times U(1) \times U(1)\ .
\end{align}
The corresponding decomposition of the adjoint representation reads 
(see appendices \ref{app:N2N} and \ref{app:commutators}, $N=14$),
\begin{align}\label{adj}
SO(32) \sim \left(
\begin{array}{c|c|c|c|c|c} 
U(N) & N_{-1,0} & N_{0,-1} &  A &  N_{1,0} & N_{0,1} \\
aa & ab & ac & aa' & ab' & ac' \\
\hline
\oN_{1,0} & U(1) & 1_{1,-1} & N_{1,0} & 0 & 1_{1,1}\\
ba & bb & bc & ba' & bb' & bc'\\
\hline
\oN_{0,1} & 1_{-1,1}  & U(1) & N_{0,1} & 1_{1,1} & 0 \\
ca & cb & cc & ca' & cb' & cc' \\ 
\hline
A^*  & \oN_{-1,0} & \oN_{0,-1} & U(N)^* & \oN_{1,0} & \oN_{0,1}\\
a'a & a'b & a'c & a'a' & a'b' & a'c'\\
\hline
\oN_{-1,0} & 0 & 1_{-1,-1} & N_{-1,0} & U(1)^* & 1_{-1,1}\\
b'a & b'b & b'c & b'a' & b'b' & b'c' \\
\hline
\oN_{0,-1} & 1_{-1,-1} & 0 & N_{0,-1} & 1_{1,-1} & U(1)^* \\
c'a & c'b & c'c & c'a' & c'b' & c'c' 
\end{array}
\right)\ .
\end{align}
Each block is labeled by the related brane intersection. The
upper left and the lower right quadrant correspond to the adjoint
representation of $U(16)$, whereas the upper right and the lower left
quadrant represent the antisymmetric representation of $U(16)$, 
decomposed with respect to $U(14)\times U(1)\times U(1)$. 

The representation in the block $ef$ feels the magnetic flux
$f^i_{ef}=f^i_e-f^i_f$ in torus $T^2_i$. According to the index theorem the
multiplicities of chiral zero-modes are given by
\begin{align}
I_{ef} = \left(\frac{g}{2\pi}\right)^3 \prod_i \int_{T^2_i}
f^i_{ef}\ =\ \prod_i (m^i_e-m^i_f) \ .
\end{align} 
Because of Eq.~\eqref{Dquantization} these multiplicities agree with the
intersection numbers of the D-brane model given in Table~\ref{tab:fermions}.

The starting point for the computation of the 4d effective action is
the 10d Super-Yang-Mills action with $\mathcal{N}=4$ supersymmetry and gauge
group $SO(32)$, which is conveniently expressed in term of 4d vector
superfields $V$ and chiral superfields $\phi$ \cite{Marcus:1983wb,
ArkaniHamed:2001tb},
\begin{align}
S_{10} =\int d^{10} x \bigg\{ &\frac{1}{k} \int d^2 \theta 
\tr \Big[ \frac{1}{4} W W + 
\frac{1}{2}\E_{ijk} \phi^i\Big(\pd_j\phi^k
    +\frac{g}{3\sqrt{2}}\left[\phi^j,\phi^k\right]  \Big) \Big] 
+ \text{h.c.}  \\
+ &\frac{1}{k} \int d^4 \theta \frac{1}{g^2}\tr \Big[ 
\big(-\sqrt{2} \opd_i +
      g \ophi^i \big) e^{gV} \big(\sqrt{2} \pd_i + g \phi^i
    \big) e^{-gV} +\opd_i e^{gV} \pd_i e^{-gV} \Big]
\bigg\} \nonumber \,.
\end{align}
Here $W$ is the field strength of the vector field\footnote{We
  use the conventions of \cite{Wess:1992cp}, and we have dropped the
  WZW term that vanishes in WZ gauge, $V^3=0$.}, $i,j,k=1,2,3$ label the three 2-tori,
and our trace convention is $\tr
\left( T_a T_b \right) = k\delta_{ab}$. Expanding the exponentials,
integrating some of the terms by part, and using the WZ gauge $V^3 =
0$, one obtains
\begin{align}\label{10daction}
S_{10} = \int d^{10} x \bigg\{ \frac{1}{k} \int d^2 \theta 
  \tr &\Big[ \frac{1}{4} W W + 
\frac{1}{2}\E_{ijk} \phi^i\Big(\pd_j\phi^k
    +\frac{g}{3\sqrt{2}}[\phi^j,\phi^k]  \Big) \Big]
+ \text{h.c.}  \nonumber \\
 + \frac{1}{k} \int d^4 \theta \tr &\Big[\ophi^i \phi^i 
+ \sqrt{2}(\pd_i \ophi^i + \opd_i \phi^i) V  
 -  g [\ophi^i, \phi^i] V \nonumber\\ 
  &+ \Big(\opd_i V - \frac{g}{\sqrt{2}} [\ophi^i,V]\Big) 
\Big(\pd_i V +\frac{g}{\sqrt{2}} [\phi^i,V] \Big) \Big] \bigg\}\ .
\end{align}
Note, that in this action the invariance with respect to 4
supersymmetry transformations is manifest whereas the invariance
with respect to 12 further supersymmetry transformations is hidden.
This will be important in our discussion of supersymmetry breaking by
magnetic fluxes in the following sections.

Vector and chiral superfields are conveniently decomposed into the
different sectors indicated in Eq.~\eqref{adj}. The unbroken group is
$H = U(N)\times U(1)\times U(1) \subset U(N+2)$ with the 
$U(1)$ and $SU(N)$ generators\footnote{A sum over repeated indices is understood.} 
\begin{equation}
H_0 = \frac{1}{\sqrt{N}}  T_{\alpha\alpha}\ , \quad
H_1 = T_{N+1,N+1}\ , \quad
H_2 = T_{N+2,N+2}\ , \quad \tilde{T}_{\A\B} \ .
\end{equation}
In terms of the generators of $H$ and $SO(32)/H$, vector 
superfields can be expressed as (see Appendix~\ref{app:commutators})
\begin{equation}\label{expV}
\begin{split}
V = &\ V_{\A\B}\tilde{T}_{\A\B} + V_I H_I + V^{-0}_\A T^{-0}_\A + V^{+0}_\A T^{+0}_\A +
V^{0-}_\A T^{0-}_\A + V^{0+}_\A T^{0+}_\A \\
&+ V^{+-} T^{+-} + V^{-+} T^{-+} 
+ \frac{1}{2} V^{+}_{\G\D} X^{+}_{\G\D} + \frac{1}{2} V^{-}_{\G\D} X^{-}_{\G\D} 
+ \tilde{V}^{+0}_\A X^{+0}_\A \\
&+ \tilde{V}^{-0}_\A X^{-0}_\A +\tilde{V}^{0+}_\A X^{0+}_\A
+\tilde{V}^{0-}_\A X^{0-}_\A + V^{++} X^{+-} + V^{--} X^{--} \ .
\end{split}
\end{equation}
The charges with respect to $H_1$ and $H_2$ are indicated
explicitly. The fields $V^{-0}_\A$,  $V^{0-}_\A$, $\tilde{V}^{+0}_\A$ and
$\tilde{V}^{0+}$ transform in the fundamental, and the fields
$V^{+0}_\A$,  $V^{0+}_\A$, $\tilde{V}^{-0}_\A$ and $\tilde{V}^{0-}$
in the anti-fundamental representation of $SU(N)$, respectively. 
$V^+_{\G\D}$ is an antisymmetric tensor of $SU(N)$ and $V^-_{\G\D}$
is the complex conjugate representation. $V^\pm_{\G\D}$ are neutral
with respect to $H_1$ and $H_2$. Here, the superscript denotes the
charge with respect to $H_0$. Analogously, the
decomposition of the chiral and antichiral superfields is given
by\footnote{Note that $\ophi^{-0}_\alpha$ stands for
$\overline{\phi^{-0}_\alpha}$.}
\begin{equation}\label{expfi}
\begin{split}
\phi =&\ \phi_{\A\B}\tilde{T}_{\A\B} + \chi_I H_I + \phi^{-0}_\A T^{-0}_\A + \phi^{+0}_\A T^{+0}_\A +
\phi^{0-}_\A T^{0-}_\A + \phi^{0+}_\A T^{0+}_\A \\
&+ \phi^{+-} T^{+-} + \phi^{-+} T^{-+} 
+ \frac{1}{2} \phi^{+}_{\G\D} X^{+}_{\G\D} + \frac{1}{2} \phi^{-}_{\G\D} X^{-}_{\G\D} 
+ \tilde{\phi}^{+0}_\A X^{+0}_\A \\
&+ \tilde{\phi}^{-0}_\A X^{-0}_\A +\tilde{\phi}^{0+}_\A X^{0+}_\A
+\tilde{\phi}^{0-}_\A X^{0-}_\A +\phi^{++} X^{++} +\phi^{--} X^{--}\ ,
\end{split}
\end{equation}
\begin{equation}\label{expfib}
\begin{split}
\ophi =&\ \ophi_{\A\B}\tilde{T}_{\B\A} + \ochi_I H_I + \ophi^{-0}_\A T^{+0}_\A + \ophi^{+0}_\A T^{-0}_\A +
\ophi^{0-}_\A T^{0+}_\A + \ophi^{0+}_\A T^{0-}_\A \\
&+ \ophi^{+-} T^{-+} + \ophi^{-+} T^{+-} 
+ \frac{1}{2} \ophi^{+}_{\G\D} X^{-}_{\G\D} + \frac{1}{2} \ophi^{-}_{\G\D} X^{+}_{\G\D} 
+ \bar{\tilde{\phi}}^{+0}_\A X^{-0}_\A \\
&+ \bar{\tilde{\phi}}^{-0}_\A X^{+0}_\A +\bar{\tilde{\phi}}^{0+}_\A X^{0-}_\A
+\bar{\tilde{\phi}}^{0-}_\A X^{0+}_\A + \ophi^{++} X^{--} + \ophi^{--}
X^{++} \ .
\end{split}
\end{equation}

In order to compute the mass spectrum caused by the magnetic fluxes
and also for a discussion of tachyon condensation one has to know the
Yukawa couplings of the model. They are obtained from the cubic gauge
coupling in the action \eqref{10daction} and the commutators listed in
Appendix~\ref{app:commutators}. A straightforward calculation yields
the result
\begin{equation}\label{10dY}
\mathcal{L}_\text{Y} = \frac{1}{k}\int d^2 \theta 
\frac{g}{6\sqrt{2}}\E_{ijk} \text{tr}\big[\phi^i
    [\phi^j,\phi^k] \big] 
=\frac{g}{\sqrt{2}}\E_{ijk}\int d^2 \theta
\big(W^1_{ijk} + W^2_{ijk}\big) \ ,
\end{equation}
where $W^1$ and $W^2$ describe couplings without and with $SU(N)$
fields, respectively,
\begin{align}
W^1_{ijk} = &\ \frac{1}{\sqrt{N}}\chi^i_0\big(\phi^{j-0}_\A \phi^{k+0}_\A 
+ \phi^{j0-}_\A \phi^{k0+}_\A - \tilde{\phi}^{j+0}_\A\tilde{\phi}^{k-0}_\A 
- \tilde{\phi}^{j0+}_\A\tilde{\phi}^{k0-}_\A
+\phi^{j+}_{\A\B}\phi^{k-}_{\A\B}\big) \nonumber\\
&+ \chi^i_1\big(-\phi^{j-0}_\A\phi^{k+0}_\A - \tilde{\phi}^{j+0}_\A \tilde{\phi}^{k-0}_\A 
+\phi^{j+-}\phi^{k-+} + \tilde{\phi}^{j++}\tilde{\phi}^{k--}\big) \nonumber\\
&+ \chi^i_2\big(-\phi^{j0-}_\A\phi^{k0+}_\A 
- \tilde{\phi}^{j0+}_\A \tilde{\phi}^{k0-}_\A -\phi^{j+-}\phi^{k-+} 
+ \tilde{\phi}^{j++}\tilde{\phi}^{k--}\big)\nonumber\\
&-\phi^{i+-}\big(\phi^{j-0}_\A \phi^{k0+}_\A +\tilde{\phi}^{j0+}_\A\tilde{\phi}^{k-0}_\A\big)
-\phi^{i-+}\big(\phi^{j0-}_\A \phi^{k+0}_\A +\tilde{\phi}^{j+0}_\A\tilde{\phi}^{k0-}_\A\big)
\nonumber\\
&+\phi^{i++}(-\phi^{j-0}_\A\tilde{\phi}^{k0-}_\A
+\phi^{j0-}_\A\tilde{\phi}^{k-0}_\A)
+ \phi^{i--}(\phi^{j+0}_\A\tilde{\phi}^{k0+}_\A -
\phi^{j0+}_\A\tilde{\phi}^{k+0}_\A )\ \nonumber\\
&-\phi^{i+}_{\A\B} (\phi^{j+0}_\B \tilde{\phi}^{k-0}_{\A}  + \phi^{j0+}_\B \tilde{\phi}^{k0-}_{\A})
-\phi^{i-}_{\A\B} (\phi^{j-0}_\B \tilde{\phi}^{k+0}_{\A}  + \phi^{j0-}_\B
\tilde{\phi}^{k0+}_{\A}) \ ,\label{Wyuk1}\\
W^2_{ijk} = &\ \phi^i_{\A\B} \big(\phi^{j-0}_\B \phi^{k+0}_\A 
+ \phi^{j0-}_\B \phi^{k0+}_\A - \tilde{\phi}^{j+0}_\B\tilde{\phi}^{k-0}_\A 
- \tilde{\phi}^{j0+}_{\B}\tilde{\phi}^{k0-}_\A \nonumber\\
&\quad+\phi^{j+}_{\G\B}\phi^{k-}_{\G\A} - \phi^{j-}_{\G\A}\phi^{k+}_{\G\B}
+\phi^j_{\B\G}\phi^k_{\G\A} - \phi^j_{\G\A}\phi^k_{\B\G}\big) \ .\label{Wyuk2}
\end{align}
Note, that these couplings involve 10d fields. The 4d effective
Lagrangian is obtained by performing a mode expansion for all fields
and by
evaluating the overlap integrals of products of mode functions.

The gauge group $SO(32)$ is broken to the subgroup $U(N)\times
U(1)\times U(1)$ by a background of the $U(1)$ gauge fields in the compact
dimensions,
\begin{align}\label{background}
\langle \chi^i_I \rangle = \frac{1}{\sqrt{2}} f^i_I \bar{z}_i + \xi^i_I\  ,
\end{align}
corresponding to Wilson lines and magnetic fluxes in the three 2-tori
($\chi^i_I|_{\theta=\bar{\theta}=0} = (A_{I,3+2i} + i A_{I,2+2i})/\sqrt{2}$),
\begin{align}
\langle F_{2+2i,3+2i}\rangle \D_{ij} = \opd_i \langle\chi^j_I \rangle
= f^i_I \D_{ij} \ .
\end{align}
The mass spectrum of the charged fields is obtained by calculating the
quadratic part of the effective action in this gauge field
background. 

Each pair of fields in Eq.~\eqref{adj}, such as $(A,A^*)$,
$(N_{-1,0},\overline{N}_{1,0})$ etc, feels magnetic
fluxes $f^i_I$ in the three tori. The mass spectrum of each sector
$ef$ is then characterized by Landau levels $(n_1,n_2,n_3)$ and internal
helicities $(\sigma_1,\sigma_2,\sigma_3)$ in the three tori, with $n_i
\in \mathbb{N}$ and $\sigma_i = 0, \pm 1/2, \pm 1$. Hence, for each triple of
Landau levels one obtains two 4d complex vector states, eight 4d Weyl
fermions and six complex 4d scalars. Their masses have been obtained 
in a type I string compactification on a magnetized torus ($f^i_{ef} =
f^i_e - f^i_f$) \cite{Bachas:1995ik},
\begin{align}\label{master}
M^2_{ef}(n;\sigma) = g\sum_i((2n_i+1)|f_{ef}^i| + 2f_{ef}^i\sigma_i)\ . 
\end{align}
Here $\sigma = (\sigma_1,\sigma_2,\sigma_3)$ takes the values
$(0,0,0)$, $(\pm 1/2,\pm 1/2,\pm 1/2)$ and $(\pm,0,0)$, $(0,\pm,0)$,
$(0,0,\pm)$ for vectors, fermions and scalars, respectively.
Contrary to what one might expect, these masses are not
associated with a single set $(n_1,n_2,n_3)$ of Landau levels in a
mode expansion of the 10d fields in Eq.~\eqref{adj}. As we shall see
in the following section, the magnetic fluxes mix neighboring levels in the 
Kaluza-Klein towers, and mass eigenstates are linear combinations of
different Landau levels.

In this T-dual internal magnetic field description that we will mainly
focus on, supersymmetry breaking for generic magnetic fields is captured by the (internal helicity) spin-magnetic field
coupling in the mass formula  (\ref{master}). The special values of magnetic fields for which some supersymmetry is preserved can be understood in various ways. One of them is by checking
the boson and fermion mass formulae and the flux value parameters for
which there is boson-fermion degeneracy.  Equivalently, the scalar
potential that we will compute in Section 6 will turn out to 
vanish precisely for  these flux values. Another way to understand supersymmetry breaking and preservation is by writing the gaugino variation for the Super-Yang-Mills theory directly
in ten dimension, before compactification, which reads
\begin{equation}
\delta \lambda^a = - \frac{1}{4} \Gamma^{PQ} F_{PQ}^a \epsilon \ , \label{susymag1}
\end{equation}
where $\Gamma^{PQ}  = \frac{1}{2} (\Gamma^P \Gamma^Q - \Gamma^Q \Gamma^P )$, $F_{PQ}^a $ is the 10d Yang-Mills field strength and $\epsilon$ are the 10d supersymmetry parameters.  The number of preserved supercharges is given by the 
number of independent spinors $\epsilon$ annihilated by the operator
\begin{equation}
\Gamma (f) = \Gamma^{PQ} \langle F_{PQ}^I \rangle \sim \Gamma^{45} f_1 + \Gamma^{67} f_2 + \Gamma^{89} f_3 \ ,
\end{equation}
where we defined the fluxes $f_k = \langle F_{2+2k,3+2k} \rangle $, $k=1,2,3$, and $I$ is a Cartan subalgebra generator supporting the magnetic flux.  
A well known and convenient Fock space basis for fermions is obtained
by introducing the creation and annihilation operators (see, for example, \cite{Polchinski:1998rr})   
\begin{equation}
b_k^{\dagger} = \frac{1}{2} (\Gamma^{2+2k} - i \Gamma^{3+2k}) \quad , \quad b_k = \frac{1}{2} (\Gamma^{2+2k} + i \Gamma^{3+2k}) \ , \label{susymag2}
\end{equation}
with $k=1,2,3$ denoting the complex internal space degrees of freedom. Using 
\begin{equation}
b_k^{\dagger}  b_k - b_k b_k^{\dagger} = i  \Gamma^{2+2k} \Gamma^{3+2k} = - 2 J_{2+2k,3+2k}  \ , \label{susymag3}
\end{equation} 
where $J_{2+2k,3+2k}$ are rotations generators in the internal space, one can rewrite the operator $\Gamma$ as
\begin{equation}
\Gamma (f) \sim \sum_{k=1}^3 f_k (b_k^{\dagger}  b_k - b_k b_k^{\dagger})  \ . \label{susymag4}
\end{equation}
Then, by explicit construction, one can show that $\Gamma (f) \epsilon = 0$ for
\begin{equation}
\begin{split}
&\epsilon_0 = | 0 \rangle \ \hspace{0.5cm}  {\rm and}  \ \ b_1^{\dagger}
b_2^{\dagger} b_3^{\dagger}  | 0 \rangle \ ,\quad {\rm if} \quad f_1 + f_2 + f_3 = 0 \ , \\ 
& \epsilon_1 =b_1^{\dagger}  | 0 \rangle \ \  {\rm and} \hspace{0.6cm}
b_2^{\dagger} b_3^{\dagger}  | 0 \rangle \ , \quad {\rm if} \quad - f_1 + f_2 + f_3 = 0 \ ,  \\
& \epsilon_2 =b_2^{\dagger}  | 0 \rangle  \ \ {\rm and} \hspace{0.6cm}
b_3^{\dagger} b_1^{\dagger}  | 0 \rangle\ , \quad {\rm if} \quad  f_1 - f_2 + f_3 = 0 \ ,  \\
& \epsilon_3 =b_3^{\dagger}  | 0 \rangle  \ \ {\rm and}  \hspace{0.6cm}
b_1^{\dagger} b_2^{\dagger}  | 0 \rangle\ , \quad {\rm if} \quad  f_1 + f_2 - f_3 = 0 \  . \label{susymag5}
\end{split}
\end{equation}
These relations match the field theory limit of the intersecting brane supersymmetry conditions (\ref{susyangles}).
As we will see explicitly in the following sections, the effective
theory does not easily capture the supersymmetry restoration points in moduli space. The reason
is that the supercharge corresponding to $\epsilon_0$ is aligned with
the superspace expansion, whereas the other preserved supercharges corresponding to $\epsilon_{1,2,3}$ are 
not, and the corresponding supersymmetries are hidden in an effective Lagrangian that at first sight looks non-supersymmetric.

In later sections we will discuss tachyon condensation, which requires to add the fluctuations around the magnetic background  (\ref{susymag1}). In this case, the operator  $\Gamma$
is changed according to 
\begin{align}
\Gamma  \equiv \Gamma^{PQ} F_{PQ}^I = &2 \sum_{k=1}^3 (c_{AB}^I \phi^k_A  {\bar \phi}^k_B -i f^k_I) (b_k^{\dagger}  b_k - b_k b_k^{\dagger}) \  \nonumber \\
& + 4  c_{AB}^I \sum_{i<j}  \left[ b_i^\dagger b_j^\dagger  \phi^i_A \phi^j_B  +b_i b_j  {\bar \phi}^i_A {\bar \phi}^j_B +b_i^\dagger b_j \phi^i_A {\bar \phi}^j_B 
+ b_i b_j^\dagger  {\bar \phi}^i_A \phi^j_B   \right] \ , \label{susymag6}
\end{align}
where $c_{AB}^I$ are the structure constants of the 10d Yang-Mills
gauge group and $A,B$ are indices of the adjoint representation.  
Acting with the operator $\Gamma$ on the spinors $Q \equiv (\epsilon_0, \epsilon_1, \epsilon_2 , \epsilon_3)^T$  one defines a $4 \times 4$ matrix ${\cal M}$ according to
\begin{equation}
\Gamma  Q = {\cal M} Q \ . \label{susymag7}
\end{equation}
Notice that the spinors $\epsilon_0,\epsilon_i$ do not carry flux charge, since they transform only under the $SU(4)$ R-symmetry group, which commutes with the gauge group generators. They should
be understood as the constant zero modes of the  KK reduction from 10d to 4d. 
 Labeling the four rows and columns by $0,1,2,3$, the matrix elements are computed to be
\begin{align}\label{susymag8}
{\cal M}_{00} &= - 2 i (f^1_I+f^2_I+f^3_I) - 2 c_{AB}^I {\bar \phi}^k_A \phi^k_B \ , \ {\cal M}_{0i} = - 4 c_{AB}^I \epsilon_{ijk} {\bar \phi}^j_A {\bar \phi}^k_B \ , \\
 {\cal M}_{i0} &=  4 c_{AB}^I \epsilon_{ijk} {\phi}^j_A { \phi}^k_B \ , \
 {\cal M}_{ij} =  2 i (f^1_I+f^2_I+f^3_I - 2 f^i_I ) \delta_{ij} +  2 c_{AB}^I ( {\bar \phi}^k_A \phi^k_B  \delta_{ij} - 2 {\bar \phi}^i_A \phi^j_B  )  \ . \nonumber
\end{align}
After compactification to four dimensions, quantities like $\phi^j_A {\bar \phi}^k_B$ should be understood as integrated over the internal space, leading to a sum over Landau levels
$\sum_{n,n'} \phi^j_{A,nn'} {\bar \phi}^k_{B,nn'}$. 
 As before, the number of zero eigenvalues of the matrix ${\cal M}$ is the number of unbroken supersymmetries in four dimensions. Let us study some simple examples: 
 \begin{itemize}
\item One flux, say $f^1_I=f^2_I= 0$, $f^3_I\not=0$.
 In the absence
 of vev's for $\phi$'s, there is no zero eigenvalue according to   (\ref{susymag5}) and all supersymmetries are broken. However, by giving a vacuum expectation value 
 $ \phi_3 \not =0$, one can set to zero all matrix elements and restore full ${\cal N}=4$ supersymmetry by choosing $c_{AB}^I  \phi^3_A   {\bar \phi}^3_B = i f^3_I$.   Notice that the vev's
  concern fields charged under the (Cartan) generator $H_I$.     
 
 \item Two fluxes, say $f^1_I=0, f^2_I, f^3_I \not=0$.
  
  In this case, in the absence of vev's for $\phi$'s, all supersymmetries are generically broken, except for   $f^2_I = \pm f^3_I $, which preserves ${\cal N}=2$ supersymmetry. 
  For $f^2_I \not=  f^3_I $, one can easily find vev's restoring ${\cal N}=2$ supersymmetry:
\begin{equation}
c_{AB}^I    {\bar \phi}^2_A \phi^3_B = 0 \ , \ i (f^2_I - f^3_I) = 2 c_{AB}^I  (  {\bar \phi}^2_A \phi^2_B -  {\bar \phi}^3_A \phi^3_B)  \ , \label{susymag9}
\end{equation}
which can be satisfied for example for $\phi^2_B=0$ and   $i (f^2_I - f^3_I) = - 2 c_{AB}^I  {\bar \phi}^3_A \phi^3_B$.
 One can also search the existence of an ${\cal N}=4$ vacuum. It seems natural to assume $\phi^1_A=0$, both
since this field is not tachyonic for such fluxes and since in this case the matrix ${\cal M}$ has a simpler block-diagonal form of two $2 \times 2$ matrices. The conditions for the
existence of an ${\cal N}=4$ vacuum are
\begin{equation}
\begin{split}
\tr [  {\phi}^2, {\bar \phi}^2 ] H_I &= -  f^2_I \ , \  \tr [  {\phi}^3, {\bar \phi}^3 ] H_I =  - f^3_I  \ , \\ 
\tr [  {\phi}^2, {\bar \phi}^3 ] H_I &=0 \ , \  \tr [  {\phi}^2, {\phi}^3 ] H_I =0 \ . \label{susymag10}  
\end{split}
\end{equation}
 
 \item Three fluxes, say $f^1_I, f^2_I, f^3_I \not=0$.
  
  In this case, in the absence of vev's for $\phi$'s, all supersymmetries are generically broken, except for   $f^1_I \pm f^2_I  \pm f^3_I =0$, which preserves ${\cal N}=1$ supersymmetry. 
  For   $f^1_I \pm f^2_I  \pm f^3_I \not=0$ , one can easily find
  vev's restoring ${\cal N}=1$ supersymmetry by switching on only one vev. For example, one can choose 
 $ \phi^2=\phi^3=0$ and 
  \begin{equation}
 i (f^1_I \pm f^2_I \pm f^3_I) =  c_{AB}^I \phi^1_A {\bar \phi}^1_B    \ , \label{susymag12}
\end{equation}
for any (single) choice of signs. 
 The case of vev's restoring more supersymmetries seems similar to the
 previous example with two fluxes. In order to obtain ${\cal N}=4$ 
 supersymmetry, one would need vev's for the three $\phi^i$'s 	and satisfy 
  \begin{equation}
 \tr [  {\phi}^i, {\bar \phi}^i ] H_I = - f^i_I \ , \  \epsilon_{ijk} \tr [  {\phi}^i, {\phi}^j ] H_I =  0 \ , \  \epsilon_{ijk} \tr [  {\phi}^i, {\bar \phi}^j ] H_I =  0 \ .   \label{susymag13}
 \end{equation} 
 This seems always possible. 
  \end{itemize}
   
 In all cases, one should also impose the D-term conditions for the
 charged generators. They are more complicated than the ones for the Cartan generators written above. The reason is that in addition to bilinear
 terms  similar to the ones for Cartan generators (for example, $\epsilon_{ijk}  \tr [  {\phi}^i, {\bar \phi}^j ] E_{\alpha}$), there are also terms linear in the charged fields coming from the covariant derivative acting on
 charged fields, which have a non-constant profile in the internal space. These terms, of the type  $\sqrt{f_I^i} a^i_{\alpha} \tr (\phi^j E_\alpha) $ or $\sqrt{f_I^i} a^{i\dagger}_{\alpha} \tr (\phi^j E_\alpha) $, 
 depending on the sign of the flux, can be computed in explicit cases
 and will be displayed explicitly in Section 7. However, a general
 expression for these terms, and a general analysis of the charged D-term conditions
 is beyond the scope of this paper. Therefore, at this point
we leave open the question whether or not there are vacua with full ${\cal N}=4$ supersymmetry in the case of arbitrary fluxes.

\section{Matter sector}
\label{sec:matter}

In this section we consider potentially tachyon-free sectors of the model,
i.e., the antisymmetric tensor with vector-like massless fermions,
and the fields in fundamental and anti-fundamental representations with
chiral fermions.

\subsection{Antisymmetric tensor ($aa'$-sector)}

Let us start with the antisymmetric tensor fields, $V^\pm_{\G\D}$,
$\phi^\pm_{\G\D}$ and $\ophi^\pm_{\G\D}$. These fields have charge
$\pm 2/\sqrt{N}$ with respect to $H_0$ and charge zero with respect to 
$H_{1}$ and $H_2$. For
simplicity, we choose $\xi^i_0 = 0$. According to
Table~\ref{tab:wrapping}, the flux in the first torus vanishes
and the fluxes in the second and third torus satisfy the
quantization conditions ($\rho_i=L'_i/L_i$),
\begin{align}\label{fq1}
\frac{g}{2\pi\sqrt{N}} \int_{T^2_2} f^2_0  
= \frac{g}{\sqrt{N}} \frac{2\pi\alpha'}{\rho_2} f^2_0 =2\ , \quad
\frac{g}{2\pi\sqrt{N}} \int_{T^2_3} f^3_0  
= \frac{g}{\sqrt{N}} \frac{2\pi\alpha'}{\rho_3} f^3_0 =1 \ ,
\end{align}
which yields the flux densities
\begin{align}\label{fluxaa'}
gf^2_0 = 2\sqrt{N} \frac{\rho_2}{2\pi\alpha'} \equiv g\sqrt{N}f_2\ , \quad 
gf^3_0 = \sqrt{N}\frac{\rho_3}{2\pi\alpha'} \equiv g\sqrt{N}f_3\ .
\end{align}
For the special choice $\rho_3 = 2\rho_2$ in Eq.~\eqref{rho32}, the
flux density is the same in both tori, i.e., $f^2_0 = f^3_0$ or $f_2
= f_3$. 

Using the relevant commutators in Eq.~\eqref{comaa'}, 
\begin{equation}
\begin{split}
[H_0,X^{\pm}_{\A\B}] &= \pm \frac{2}{\sqrt{N}} X^{\pm}_{\A\B}\ ,\\
[X^{+}_{\A\B}, X^{-}_{\G\D}] &= \frac{2}{\sqrt{N}}(\D_{\A\G} \D_{\B\D} - \D_{\B\G} \D_{\A\D}) + \ldots ,
\end{split}
\end{equation}
it is
straightforward to derive the quadratic part and the cubic couplings involving
the neutral fields $\chi^i_0\equiv \chi^i$ and $\ochi^i_0 \equiv \ochi^i$,
\begin{align}
\label{anti1} 
S_{10} \supset \int d^{10} x \bigg\{ \int d^2 \theta 
  &\Big( \frac{1}{4} W_0 W_0 + \frac{1}{4} W^+_{\G\D}W^-_{\G\D}  
+ \frac{1}{2}\E_{ijk} 
 \phi^{i+}_{\G\D}\Big(\pd_j - \frac{\sqrt{2}g}{\sqrt{N}} \chi^j\Big)\phi^{k-}_{\G\D} \Big)
+ \text{h.c.}  \nonumber \\
 + \int d^4 \theta &\Big(\ochi^i\chi^i + \frac{1}{2}\ophi^{i+}_{\G\D} \phi^{i+}_{\G\D} 
+ \frac{1}{2}\ophi^{i-}_{\G\D} \phi^{i-}_{\G\D} 
+ \sqrt{2}(\pd_i \ochi^i + \opd_i \chi^i) V_0  \nonumber\\
&+\frac{g}{\sqrt{N}}\big(\ophi^{i+}_{\G\D} \phi^{i+}_{\G\D} 
- \ophi^{i-}_{\G\D} \phi^{i-}_{\G\D} \big)V_0 \nonumber\\
&+\frac{1}{\sqrt{2}}\Big(\Big(\pd_i - \frac{\sqrt{2}g}{\sqrt{N}}\chi^i\Big)\ophi^{i+}_{\G\D} 
+ \Big(\opd_i + \frac{\sqrt{2}g}{\sqrt{N}}\ochi^i\Big)\phi^{i-}_{\G\D} \Big)V^+_{\G\D} 
\nonumber\\
&+\frac{1}{\sqrt{2}}\Big(\Big(\pd_i + \frac{\sqrt{2}g}{\sqrt{N}}\chi^i\Big)\ophi^{i-}_{\G\D} 
+ \Big(\opd_i - \frac{\sqrt{2}g}{\sqrt{N}}\ochi^i\Big)\phi^{i+}_{\G\D} \Big)V^-_{\G\D} 
\nonumber\\
&+\frac{1}{2}\Big(\opd_i + \frac{\sqrt{2}g}{\sqrt{N}}\ochi^i\Big)V^-_{\G\D} 
\Big(\pd_i + \frac{\sqrt{2}g}{\sqrt{N}}\chi^i\Big)V^+_{\G\D} \nonumber\\
&+\frac{1}{2}\Big(\opd_i - \frac{\sqrt{2}g}{\sqrt{N}}\ochi^i\Big)V^+_{\G\D} 
\Big(\pd_i - \frac{\sqrt{2}g}{\sqrt{N}}\chi^i\Big)V^-_{\G\D} \Big)\bigg\}\ .
\end{align}
There is no $H_0$ flux in the first torus. To obtain the lowest mass
eigenstates, we can therefore neglect the dependence of the fields on
$z_1$. Inserting the background flux in the second and third torus 
yields covariant derivatives $\pd_{2} \pm 2gf_2\bz_{2}$, $\opd_{2} \pm 2gf_2 z_{2}$,
and $\pd_{3} \pm g f_3 \bz_{3}$,   $\opd_{3} \pm g f_3 z_{3}$,
which form a harmonic
oscillator algebra.  The fields can therefore be expanded in the
corresponding set of orthonormal eigenfunctions.

For a flux density $gf = 2\pi M$, $M\in \mathbb{N}$,  one
defines two pairs of annihilation and creation operators \cite{Buchmuller:2018eog},
\begin{equation}\label{anncrea}
\begin{split}
\quad &a_+ =\frac{i}{\sqrt{2gf}} \left(\pd + gf\bar{z}\right)\,, \quad
a_+^{\dagger} =\frac{i}{\sqrt{2gf}} \left(\opd - gfz\right)\,,\\
\quad &a_- =\frac{i}{\sqrt{2gf}} \left(\opd + gfz\right)\,, \quad
a_-^{\dagger} =\frac{i}{\sqrt{2gf}} \left(\pd - gf\bar{z}\right)\,, 
\end{split}
\end{equation}
which satisfy the commutation relations
\begin{align}
[a_\pm,a^\dagger_\pm]=1\ , \quad [a_\pm,a_\mp]=0\ , \quad
[a_\pm,a^\dagger_\mp]=0\ . 
\end{align}
The ground state wave functions are determined by
\begin{align}\label{actionancr}
a_+ \xi_{0,j} = 0\,, \quad a_- \overline{\xi}_{0,j}=0\,,
\end{align}
where $j = 0,\ldots |M|-1$ labels the degeneracy. An orthonormal set
of higher mode functions is given by
\begin{align} 
\xi_{n,j} = \frac{i^n}{\sqrt{n!}} \left(a_+^\dagger\right)^n \xi_{0,j}\,,\quad
\overline{\xi}_{n,j} = \frac{i^n}{\sqrt{n!}}
\left(a_-^\dagger\right)^n \overline{\xi}_{0,j}\,.
\end{align}
Annihilation and creation operators act on these mode functions as
\begin{equation}\label{axi}
\begin{split}
a_+ \xi_{n,j} &= i \sqrt{n}\ \xi_{n-1,j}\,, \quad a_+^\dagger \xi_{n,j}
= -i\sqrt{n+1}\ \xi_{n+1,j}\,,\\
a_- \overline{\xi}_{n,j} &= i \sqrt{n}\ \overline{\xi}_{n-1,j}\,, \quad a_-^\dagger \overline{\xi}_{n,j}
= -i\sqrt{n+1}\ \overline{\xi}_{n+1,j}\,.
\end{split}
\end{equation}
The mode expansions of the fields with positive and negative
charge read
\begin{equation}\label{KK1}
\begin{split}
\phi^+ &= \sum_{nj} \phi^+_{nj}\xi_{nj}\ ,\quad \phi^- = \sum_{n,j}
\phi^-_{n,j}\overline{\xi}_{n,j}\ ,  \quad \ophi^+ = \sum_{n,j}
\ophi^+_{n,j}\overline{\xi}_{n,j} \ , \\
\ophi^- &= \sum_{n,j} \ophi^-_{n,j}\xi_{n,j}\ , \quad V^+ = \sum_{n,j}
V^+_{n,j}\xi_{n,j}\ , \quad V^- = \sum_{n,j}
V^-_{n,j}\overline{\xi}_{n,j} \,.
\end{split}
\end{equation}

The antisymmetric tensor fields feel flux in the second and third
torus. Hence, there are two sets of annihilation and creation
operators, $a^2_\pm$, $a^{2\dagger}_\pm$ and $a^3_\pm$, $a^{3\dagger}_\pm$.
Suppressing tensor indices, i.e. $W^+_{\G\D}W^-_{\G\D}/2 \equiv
W^+W^-$ etc., and using  $f^2_0 \equiv \sqrt{N}f_2$ and $f^3_0 \equiv \sqrt{N}f_3$,
one obtains from Eq.~\eqref{anti1} 
\begin{align}
S_{10} \supset \int d^{10} x \bigg\{ \int d^2 \theta 
  &\Big( \frac{1}{4} W_0 W_0 + \frac{1}{2} W^+W^-
-i\sqrt{2gf_2}\big(\phi^{1+} a^{2\dagger}_-\phi^{3-} - \phi^{3+} a^{2\dagger}_-\phi^{1-}\big)
\nonumber\\
&-i\sqrt{2gf_3}\big(\phi^{2+} a^{3\dagger}_-\phi^{1-} - \phi^{1+} a^{3\dagger}_-\phi^{2-}\big)\Big)
+ \text{h.c.}  \nonumber \\
 + \int d^4 \theta &\Big(\ophi^{i+} \phi^{i+} + \ophi^{i-} \phi^{i-}
+ \big(2\sqrt{N}(f_2 + f_3) +\frac{2g}{\sqrt{N}}(\ophi^{i+}\phi^{i+}
- \ophi^{i-} \phi^{i-})\big)V_0  \nonumber\\
&-2i\sqrt{g}\big(\sqrt{f_2}\big(a^{2\dagger}_-\ophi^{2+} + a^{2}_-\phi^{2-}\big)
+ \sqrt{f_3}\big(a^{3\dagger}_-\ophi^{3+} + a^{3}_-\phi^{3-}\big)\big)V^+ 
\nonumber\\
&-2i\sqrt{g}\big(\sqrt{f_2}\big(a^{2\dagger}_+\phi^{2+} + a^{2}_+\ophi^{2-}\big)
+ \sqrt{f_3}\big(a^{3\dagger}_+\phi^{3+} + a^{3}_+\ophi^{3-}\big)\big)V^- \nonumber\\
&- 2gf_2 \big(a^{2}_- V^- a^{2}_+V^+ + a^{2\dagger}_+ V^+a^{2\dagger}_-V^-\big)\nonumber\\
&- 2gf_3\big(a^{3}_- V^-a^{3}_+V^+ + a^{3\dagger}_- V^-a^{3\dagger}_+V^+\big) \Big)\bigg\}\ . 
\end{align}
The fields have a double expansion in two sets of
mode functions\footnote{More precisely, the
  fields depend on $z_i$ and $\bar{z}_i$.}
\begin{align}\label{KK2}
\phi^+(x;z_1,z_2,z_3) &= \sum_{nj,n'j'}
\phi^+_{nj,n'j'}(x)\xi_{n,j}(z_2)\xi_{n',j'}(z_3)\ , \nonumber\\
\phi^-(x;z_1,z_2,z_3) &= \sum_{nj,n'j'}
\phi^-_{nj,n'j'}(x)\overline{\xi}_{n,j}(z_2)\overline{\xi}_{n',j'}(z_3)\
, \quad \text{\rm etc.}\ ,
\end{align}
where, for simplicity, we only consider the lowest KK mode in the
first torus. 
According to the quantization condition \eqref{fq1} the
multiplicity in the second and third torus is two and one,
respectively, giving a total multiplicity of two for all fields.

Inserting the mode expansion of the fields in the 10d action, 
using Eq.~\eqref{actionancr} and the orthonormality of the mode
functions, and dropping the indices $j,j'$ that label the degeneracy,
one arrives at the 4d effective Lagrangian
\begin{align}
\mathcal{L}_{4} \supset  &\int d^2 \theta 
  \Big( \frac{1}{4} W_0W_0 +  \sum_{nn'} \Big(\frac{1}{2} W^+_{n,n'}W^-_{n,n'}\nonumber\\
&\hspace{1cm}-\sqrt{2gf_2(n+1)}\big(\phi^{1+}_{n+1,n'}\phi^{3-}_{n,n'} - \phi^{3+}_{n+1,n'}\phi^{1-}_{n,n'}\big)
\nonumber\\
&\hspace{1cm}-\sqrt{2gf_3(n'+1)}\big(\phi^{2+}_{n,n'+1} \phi^{1-}_{n,n'} - \phi^{1+}_{n,n'+1} \phi^{2-}_{n,n'}\big)
\Big)\Big) + \text{h.c.} \nonumber \\
 + &\int d^4 \theta \Big(2\sqrt{N}(f_2 + f_3)V_0 + \sum_{nn'}
\Big(\ophi^{i+}_{n,n'} \phi^{i+}_{n,n'} + \ophi^{i-}_{n,n'} \phi^{i-}_{n,n'}
\nonumber\\
&\hspace{1cm}+ \frac{2g}{\sqrt{N}}\big(\ophi^{i+}_{n,n'}\phi^{i+}_{n,n'}
- \ophi^{i-}_{n,n'} \phi^{i-}_{n,n'}\big)V_0  \nonumber\\
&\hspace{1cm}-2\big(\big(\sqrt{gf_2}\big(\sqrt{n}\ophi^{2+}_{n-1,n'} -
\sqrt{n+1}\phi^{2-}_{n+1,n'}\big) \nonumber\\
&\hspace{1cm} \qquad + \sqrt{gf_3}\big(\sqrt{n'}\ophi^{3+}_{n,n'-1} 
- \sqrt{n'+1}\phi^{3-}_{n,n'+1}\big)\big)V^+_{n,n'} + \text{h.c.} \big)
\nonumber\\
&\hspace{1cm} + 2M^2_{n,n'}V^+_{n,n'}V^-_{n,n'}\Big)\Big)\ , \label{Laa'basic}
\end{align}
where
\begin{align}
M_{n,n'} = (gf_2(2n+1)+gf_3(2n'+1))^{1/2}\ .
\end{align}
The magnetic flux mixes different Landau levels of the KK towers and
it is therefore convenient to introduce linear combinations of the
original chiral superfields,
\begin{align}
\phi^-_{n,n'} &= \frac{1}{\mu_{n,n'}}\big(\sqrt{2gf_2n}\ \phi^{3-}_{n-1,n'}
-\sqrt{2gf_3n'}\ \phi^{2-}_{n,n'-1}\big) ,\; (n,n')\neq 0\ ;\;
\phi^-_{0,0} = 0\ ,\label{aa1}\\
\chi^-_{n,n'} &=
\frac{1}{\mu_{n+1,n'+1}}\big(\sqrt{2gf_3(n'+1)}\ \phi^{3-}_{n,n'+1}
+\sqrt{2gf_2(n+1)}\ \phi^{2-}_{n+1,n'}\big)\ ,\label{aa2}\\
\phi^+_{n,n'} &=
\frac{1}{\mu_{n+1,n'+1}}\big(\sqrt{2gf_3(n'+1)}\ \phi^{2+}_{n,n'+1}
-\sqrt{2gf_2(n+1)}\ \phi^{3+}_{n+1,n'}\big)\ ,\label{aa3}\\
\chi^+_{n,n'} &= \frac{1}{\mu_{n,n'}}\big(\sqrt{2gf_2n}\ \phi^{2+}_{n-1,n'}
+\sqrt{2gf_3n'}\ \phi^{3+}_{n,n'-1}\big) ,\; (n,n')\neq 0\ ;\;
\chi^+_{0,0} = 0\ , \label{aa4}
\end{align}
with
\begin{align}
\mu_{n,n'} =(2gf_2n + 2gf_3n')^{1/2}\ .
\end{align}
In terms of the new fields the 4d Lagrangian reads
\begin{align}\label{L4daasuper}
\mathcal{L}_{4} \supset &\int d^2 \theta \Big( \frac{1}{4} W_0W_0
  + \sum_{nn'} \Big( \frac{1}{2} W^+_{n,n'}W^-_{n,n'}
-\mu_{n,n'} \phi^{1+}_{n,n'}\phi^{-}_{n,n'} 
\nonumber\\
&\hspace{2cm}-\mu_{n+1,n'+1} \phi^{1-}_{n,n'} \phi^{+}_{n,n'} \Big)\Big)
+ \text{h.c.}  \nonumber \\
 + &\int d^4 \theta \Big(2\sqrt{N}(f_2 + f_3)V_0 +
\sum_{nn'}\Big(|\phi^{1+}_{n,n'}|^2 + |\phi^{1-}_{n,n'}|^2 
+ |\phi^{+}_{n,n'}|^2 + |\phi^{-}_{n,n'}|^2 \nonumber\\
&\hspace{1cm}+ |\chi^{+}_{n,n'}|^2 + |\chi^{-}_{n,n'}|^2
+ \frac{2g}{\sqrt{N}}\big(|\phi^{1+}_{n,n'}|^2
+ |\phi^{+}_{n,n'}|^2 + |\chi^{+}_{n,n'}|^2 \nonumber\\
&\hspace{1cm}\quad\quad\quad - |\phi^{1-}_{n,n'}|^2 - |\phi^{-}_{n,n'}|^2 -
|\chi^{-}_{n,n'}|^2\big)V_0  \nonumber\\
&\hspace{1cm}-\sqrt{2}\big(\big(\mu_{n,n'} \ochi^+_{n,n'} 
- \mu_{n+1,n'+1} \chi^{-}_{n,n'}\big)V^+_{n,n'} 
+ \text{h.c.} \big)
\nonumber\\
&\hspace{1cm} + 2M^2_{n,n'}V^+_{n,n'}V^-_{n,n'}\Big)\Big)\ .
\end{align}

So far the diagonalization could be performed in terms of
superfields. Since the magnetic flux breaks supersymmetry, one has to
expand the superfields in components\footnote{Note, that we use the
  same symbol for the chiral superfield and its scalar component.} in the final step
(cf.~Appendix~\ref{app:susy}),
\begin{align}
\phi = (\phi,\psi,F) \ , \quad V = (A_\mu,\lambda,D) \ .
\end{align}
The mixing term between chiral and vector superfields then leads to a
charged D-term and a derivative coupling between Goldstone bosons and
vector fields,
\begin{align}
\int d^4 \theta
&\big(\mu_{n,n'} \ochi^+_{n,n'} 
- \mu_{n+1,n'+1}\chi^{-}_{n,n'}\big)V^+_{n,n'}
\nonumber\\
=&\frac{1}{2}\big(\mu_{n,n'} \ochi^+_{n,n'} 
- \mu_{n+1,n'+1} \chi^{-}_{n,n'}\big)D^+ 
-\frac{i}{\sqrt{2}}M_{n,n'} \pd_\mu \Pi^-_{n,n'} A^{+\mu}_{n,n'}\ .
\end{align}
Here the Goldstone fields $\Pi^-$ and the orthogonal complex scalars $\Sigma^-$, formed
from the complex scalars $\ochi^+$ and $\chi^-$, are given by
\begin{align}
\Pi^-_{n,n'} &= \frac{1}{\sqrt{2}M_{n,n'}}
\big(\mu_{n,n'} \ochi^+_{n,n'} + \mu_{n+1,n'+1} \chi^{-}_{n,n'}\big) \ ,\\
\Sigma^-_{n,n'} &= \frac{1}{\sqrt{2}M_{n,n'}}
\big(\mu_{n+1,n'+1} \ochi^+_{n,n'} - \mu_{n,n'} \chi^{-}_{n,n'}\big) \ .
\end{align}
The vector bosons of the tower of Landau levels acquire their mass by
the St\"uckelberg mechanism, and a shift of the vector bosons, 
\begin{align}
A_{n,n'}^{-\mu} \rightarrow A_{n,n'}^{-\mu} 
+ \frac{i}{M_{n,n'}}\ \pd_\mu\Pi^{-}_{n,n'}\ ,
\end{align}
cancels the mixings with the Goldstone bosons as well as the kinetic
terms of the Goldstone bosons. Finally, eliminating all F- and D-terms via
their equations of motion, one obtains the bosonic mass terms
\begin{align} \label{Laa'b}
\mathcal{L}^b_4 \supset - \sum_{n,n'} \Big(
&M^2_{n,n'}
\big(A^{+\mu}_{n,n' \mu}A^{-\mu}_{n,n'} + |\phi^{1+}_{n,n'}|^2 + |\phi^{1-}_{n,n'}|^2 
+ |\Sigma^-_{n,n'}|^2\big) \nonumber\\
&+ (M^2_{n,n'} - 2gf_2 - 2gf_3) |\phi^-_{n,n'}|^2 
+ (M^2_{n,n'} + 2gf_2 + 2gf_3) |\phi^+_{n,n'}|^2\Big) \ ,
\end{align}
where it is important to remember that $\phi^-_{0,0} = \Sigma^-_{0,0}
= 0$. 

Consider first the lowest lying scalars,
\begin{align}
\mathcal{L}^b_4 \supset 
- g(f_2 + f_3) (|\phi^{1+}_{0,0}|^2 + |\phi^{1-}_{0,0}|^2) 
- g(-f_2 + f_3) |\phi^-_{0,1}|^2 - g(f_2 - f_3) |\phi^-_{1,0}|^2 \ .
\end{align}
These masses are in agreement with the ones given in
Table~\ref{tab:scalarmasses} for the $aa'$-sector.
The comparison with the string formula \eqref{master} is more subtle.
The mass spectrum of $\phi^{1\pm}$ corresponds to $M^2_{aa'}(0,n,n';\pm,0,0)$.
Since $\phi^-_{0,0} = 0$, one can write
\begin{align}
\sum_{n,n'} (M^2_{n,n'} - 2gf_2 - 2gf_3) |\phi^-_{n,n'}|^2 &= \nonumber\\
\sum_{n,n'}\big((M^2_{n,n'} - 2gf_2) |\phi^-_{n,n'}|^2
&+ (M^2_{n,n'} - 2gf_3) |\phi^-_{n,n'}|^2- M^2_{n,n'} |\phi^-_{n,n'}|^2 \big)\ . 
\end{align}
Hence, the spectrum of $\phi^-$ together with one polarization state of the vector
corresponds to the spectrum $M^2_{aa'}(0,n,n';0,-,0)$ together with
$M^2_{aa'}(0,n,n';0,0,-)$. Analogously, the spectra of $\Sigma^-$ and
$\phi^+$ correspond to $M^2_{aa'}(0,n,n';0,+,0)$ together with
$M^2_{aa'}(0,n,n';0,0,+)$. Since in the string formula \eqref{master}
massive vectors are only counted with two polarization states, the
entire spectra of Eqs.~\eqref{master} and \eqref{Laa'b}
agree. However, there is no direct correspondence for individual
Landau levels.

Denoting the Weyl fermions contained in the superfields $\phi^{1\pm}$,
$\phi^\pm$, $\chi^\pm$ and $V^+=V^{-\dagger}$ by $\psi^{1\pm}$, $\psi^\pm$, 
$\omega^\pm$ and $\lambda^\pm$, respectively, one finds for
the fermionic mass terms of the 4d Lagrangian \eqref{L4daasuper} 
(cf.~Appendix \ref{app:susy}),
\begin{align}\label{Laa'f}
\mathcal{L}_{4} \supset \sum_{nn'} \Big(
 &\mu_{n,n'} \big(\psi^{1+}_{n,n'} \psi^{-}_{n,n'}
 +i\omega^+_{n,n'} \lambda^-_{n,n'}\big) \nonumber\\
&+\mu_{n+1,n'+1} \big(\psi^{1-}_{n,n'} \psi^{+}_{n,n'} 
+i \omega^{-}_{n,n'}\lambda^+_{n,n'}\big) \Big)
+ \text{h.c.} \ .
\end{align}
Note that by definition, $\psi^{-}_{0,0} = \omega^+_{0,0} =0$
(cf.~Eqs.~\eqref{aa1}, \eqref{aa4}). Clearly,
the spectrum contains two zero-modes, $\psi^{1+}_{0,0}$ and
$\lambda^-_{0,0}$.

The structure of the $bc$-sector is identical to the one in the
$aa'$-sector. Also in the $bc$-sector the flux vanishes in the first
torus, see Table~\ref{tab:wrapping}, and only the flux densities
$f^i_a$, $i=2,3$, have to be replaced by $f^i_b$ and $f^i_c$, which
corresponds to a redefinition of $f_2$ and $f_3$ in Eq.~\eqref{fluxaa'}.

\subsection{Chiral matter ($ab$-sector)}
\label{sec:ab}

We now turn to the chiral `matter sector' and consider the vector and
chiral superfields $V^{-0}_\alpha, \phi^{-0}_\alpha \sim N_{-1,0}$
and $V^{+0}_\alpha, \phi^{+0}_\alpha \sim \oN_{1,0}$. For simplicity,
we drop the superscripts $``0"$ referring to zero $H_2$ charge in
the following. The commutators of the corresponding $SO(32)$ matrices 
are given in Eq.~\eqref{comab},
\begin{equation}\label{com2ab}
[H_0,T^{\mp 0}_\A] = \pm \frac{1}{\sqrt{N}} T^{\mp 0}_\A , \;
[H_1,T^{\mp 0}_\A] = \mp T^{\mp 0}_\A ,\;
[T^{- 0}_\A, T^{+ 0}_\B] = \D_{\A\B}\Big(\frac{1}{\sqrt{N}}H_0
-H_1\Big) +\ \ldots
\end{equation}
For anti-chiral superfields signs are exchanged.
According to
Table~\ref{tab:wrapping}, the flux densities $f^i_1$ in the three tori satisfy the
quantization conditions ($\rho_i=L'_i/L_i$)
\begin{equation}\label{fq2}
\begin{split}
\frac{g}{2\pi} \int_{T^2_1} f^1_1  
&= \frac{2\pi\alpha'}{\rho_1} gf^1_1 =1\ , \quad 
\frac{g}{2\pi} \int_{T^2_2} f^2_1 
= \frac{2\pi\alpha'}{\rho_2} gf^2_1 =l\ , \\
\frac{g}{2\pi} \int_{T^2} f^3_1 
&= \frac{2\pi\alpha'}{\rho_3} gf^3_1 =-2 \ .
\end{split}
\end{equation}
Combining the $H_1$ flux densities with the $H_0$ flux densities
given in Eq.~\eqref{fq1}, one obtains for the total flux densities, i.e.~the
differences between $f^i_0/\sqrt{N}$ and $f^i_1$, in the three tori
\begin{equation}
\begin{split}
-&gf^1_1 = -\frac{\rho_1}{2\pi\alpha'} \equiv -2gf_1\ ,\quad
\frac{g}{\sqrt{N}}f^2_0 - gf^2_1 = \frac{(2-l) \rho_2}{2\pi\alpha'}
\equiv - 2gf_2\ , \\
&\frac{g}{\sqrt{N}}f^3_0- gf^3_1  = \frac{3\rho_3}{2\pi\alpha'} \equiv 2gf_3\ .
\end{split}
\end{equation}
Note that the flux parameters $f_i$ are all positive.

Using Eqs.~\eqref{expV}, \eqref{expfi} and \eqref{com2ab}, one
obtains from the action \eqref{10daction} the relevant terms for the
generation of boson and fermion masses,
\begin{align}\label{anti1} 
\mathcal{L}_{10} \supset 
\int d^2 \theta 
  &\Big( \frac{1}{4} W_0 W_0 + \frac{1}{4} W_1 W_1 + \frac{1}{2}
  W^+_{\A}W^-_{\A}  \nonumber\\
&+ \E_{ijk} 
 \phi^{i-}_{\A}\big(\pd_j - \frac{g}{\sqrt{2N}}\chi^j_0  +\frac{g}{\sqrt{2}}\chi^j_1                      \big)\phi^{k+}_{\A} \Big)
+ \text{h.c.}  \nonumber \\
 + \int d^4 \theta &\Big(\ochi^i_0\chi^i_0  + \ochi^i_1\chi^i_1 
+ \ophi^{i-}_{\A} \phi^{i-}_{\A}  + \ophi^{i+}_{\A} \phi^{i+}_{\A}  \nonumber\\
&+ \sqrt{2}\big(\pd_i \ochi^i_0 + \opd_i \chi^i_0\big) V_0  
+ \sqrt{2}\big(\pd_i \ochi^i_1 + \opd_i \chi^i_1\big) V_1 \nonumber\\
&+\frac{g}{\sqrt{N}}\big(\ophi^{i-}_{\A} \phi^{i-}_{\A} 
- \ophi^{i+}_{\A} \phi^{i+}_{\A} \big)V_0 - g\big(\ophi^{i-}_{\A} \phi^{i-}_{\A} 
- \ophi^{i+}_{\A} \phi^{i+}_{\A} \big)V_1 \nonumber\\
&+\sqrt{2}\Big(\big(\pd_i - \frac{g}{\sqrt{2N}}\chi^i_0
+ \frac{g}{\sqrt{2}}\chi^i_1\big)\ophi^{i-}_{\A} 
+ \big(\opd_i + \frac{g}{\sqrt{2N}}\ochi^i_0
- \frac{g}{\sqrt{2}}\ochi^i_1\big)\phi^{i+}_{\A} \Big)V^-_{\A} 
\nonumber\\
&+\sqrt{2}\Big(\big(\pd_i + \frac{g}{\sqrt{2N}}\chi^i_0
- \frac{g}{\sqrt{2}}\chi^i_1\big)\ophi^{i+}_{\A} 
+ \big(\opd_i - \frac{g}{\sqrt{2N}}\ochi^i_0
+ \frac{g}{\sqrt{2}}\ochi^i_1\big)\phi^{i-}_{\A} \Big)V^+_{\A} 
\nonumber\\
&+\big(\opd_i + \frac{g}{\sqrt{2N}}\ochi^i_0  -\frac{g}{\sqrt{2}}\ochi^i_1\big)V^{+}_{\A} 
\big(\pd_i +\frac{g}{\sqrt{2N}}\chi^i_0-
\frac{g}{\sqrt{2}}\chi^i_1\big)V^-_{\A} \nonumber\\
&+\big(\opd_i - \frac{g}{\sqrt{2N}}\ochi^i_0  +\frac{g}{\sqrt{2}}\ochi^i_1\big)V^{-}_{\A} 
\big(\pd_i -\frac{g}{\sqrt{2N}}\chi^i_0
+\frac{g}{\sqrt{2}}\chi^i_1\big)V^{+}_{\A}\Big) \ .
\end{align}
Replacing the scalar fields $\chi^i_0$ and $\chi^i_1$ by the flux
densities \eqref{fq1} and \eqref{fq2}, respectively, one obtains
covariant derivatives. Using Eqs.~\eqref{anncrea} they can be replaced
by annihilation and creation operators that now act on the coordinates
of all three tori,
\begin{align}\label{matter2}
\mathcal{L}_{10} \supset \int d^2 \theta 
  &\Big( \frac{1}{4} W_0 W_0 + \frac{1}{4} W_1 W_1 +\frac{1}{2} W^+W^-
-i\sqrt{2gf_1}(\phi^{3-} a^{1}_+\phi^{2+} - \phi^{2-} a^{1}_+\phi^{3+})\nonumber\\
-i&\sqrt{2gf_2}(\phi^{1-} a^{2}_+\phi^{3+} - \phi^{3-} a^{2}_+\phi^{1+}) 
-i\sqrt{2gf_3}(\phi^{2-} a^{3\dagger}_-\phi^{1+} - \phi^{1-} a^{3\dagger}_-\phi^{2+})\Big)
+ \text{h.c.}  \nonumber \\
 + \int d^4 \theta &\Big(\ochi^i_0\chi^i_0  + \ochi^i_1\chi^i_1 + \ophi^{i+} \phi^{i+}
+ \ophi^{i-} \phi^{i-}
+ 2\big(f^2_0 + f^3_0 -\frac{g}{\sqrt{N}}(\ophi^{i+}\phi^{i+}
- \ophi^{i-} \phi^{i-})\big)V_0  \nonumber\\
&+2\big(f^1_1 + f^2_1 +f^3_1 + g(\ophi^{i+}\phi^{i+}
- \ophi^{i-} \phi^{i-})\big)V_1  \nonumber\\
&-2i\big(\big(\sqrt{gf_1}(a^1_+\ophi^{1-} + a^{1\dagger}_+\phi^{1+})
+\sqrt{gf_2}(a^2_+\ophi^{2-} + a^{2\dagger}_+\phi^{2+}) \nonumber\\
&+\sqrt{gf_3}(a^{3\dagger}_-\ophi^{3-} +
a^{3}_-\phi^{3+})\big)V^-
+ \text{h.c.}\big) \nonumber\\
&- 2gf_1 \big(a^{1\dagger}_+ V^+ a^{1\dagger}_-V^- + a^{1}_- V^- a^{1}_+V^+\big) 
- 2gf_2 \big(a^{2\dagger}_+ V^+ a^{2\dagger}_-V^- \nonumber\\
&+a^{2}_- V^- a^{2}_+V^+\big) - 2gf_3 \big(a^{3\dagger}_+ V^+a^{3\dagger}_-V^- 
+ a^{3}_- V^- a^{3}_+V^+\big) \Big)\ .
\end{align}
The fields have a triple expansion in three sets of
mode functions
\begin{align}\label{KK2}
\phi^{i-}(x;z_1,z_2,z_3) &= \sum_{n_1j_1,n_2j_2,n_3j_3}
\phi^{i-}_{n_1j_1,n_2j_2,n_3j_3}(x) \xi_{n_1j_1}(z_1)\xi_{n_2j_2}(z_2)\xi_{n_3j_3}(z_3)\ , \nonumber\\
\phi^{i+}(x;z_1,z_2,z_3) &= \sum_{n_1j_1,n_2j_2,n_3j_3}
\phi^{i+}_{n_1j_1,n_2j_2,n_3j_3}(x) \overline{\xi}_{n_1j_1}(z_1)\overline{\xi}_{n_2j_2}(z_2)\overline{\xi}_{n_3j_3}(z_3)\
, \quad \text{\rm etc.}\ .
\end{align}

As in the discussion of antisymmetric tensor fields, one can now form
linear combinations of the six chiral superfields $\phi^{i+}$ and
$\phi^{i-}$ such that two new fields, $\Xi^+$ and $\Xi^-$, mix with
the vectorfield and the other four, $\phi^\pm$ and $\Phi^\pm$, form
pairwise superpotential mass terms. It is straightforward to verify
that this is achieved in a two-step process,
\begin{align}
\phi^-_{n_1,n_2,n_3} &= \frac{1}{\mu_{n_1,n_2}}\big(-\sqrt{2gf_1n_1}\ \phi^{2-}_{n_1-1,n_2,n_3}
+\sqrt{2gf_2n_2}\ \phi^{1-}_{n_1,n_2-1,n_3}\big)\ ,\nonumber\\
& \qquad  (n_1,n_2)\neq (0,0)\ ;\quad \phi^-_{0,0,n_3} = 0\ ,\label{ab1one}\\
\phi^+_{n_1,n_2,n_3} &=
\frac{1}{\mu_{n_1+1,n_2+1}}\big(\sqrt{2gf_1(n_1+1)}\ \phi^{2+}_{n_1+1,n_2,n_3}
-\sqrt{2gf_2(n_2+1)}\ \phi^{1+}_{n_1,n_2+1,n_3}\big)\ ,\\
\chi^+_{n_1,n_2,n_3} &= \frac{1}{\mu_{n_1,n_2}}\big(\sqrt{2gf_1n_1}\ \phi^{1+}_{n_1-1,n_2,n_3}
+\sqrt{2gf_2n_2}\ \phi^{2+}_{n_1,n_2-1,n_3}\big)\ ,\nonumber\\
& \qquad (n_1,n_2)\neq (0,0)\ ;\quad \chi^+_{0,0,n_3} = 0\ ,\label{ab2}\\
\chi^-_{n_1,n_2,n_3} &=
\frac{1}{\mu_{n_1+1,n_2+1}}\big(\sqrt{2gf_1(n_1+1)}\ \phi^{1-}_{n_1+1,n_2,n_3}
+\sqrt{2gf_2(n_2+1)}\ \phi^{2-}_{n_1,n_2+1,n_3}\big)\ ,
\end{align}
with $\mu_{n_1,n_2} = (2gf_1n_1+2gf_2n_2)^{1/2}$, and, as the second step,
\begin{align}
\Phi^+_{n_1,n_2,n_3} &=
\frac{1}{\mu_{n_1,n_2,n_3}}\big(\sqrt{2gf_3n_3}\ \chi^{+}_{n_1,n_2,n_3-1}
+\mu_{n_1,n_2}\ \phi^{3+}_{n_1,n_2,n_3}\big)\ ,\nonumber\\
& \qquad (n_1,n_2,n_3)\neq (0,0,0)\ , \quad \Phi^+_{0,0,0} = \phi^{3+}_{0,0,0}\ ,\label{ab3one}\\
\Phi^-_{n_1,n_2,n_3} &=
\frac{1}{\mu_{n_1+1,n_2+1,n_3+1}}\big(\sqrt{2gf_3(n_3+1)}\ \chi^{-}_{n_1,n_2,n_3+1}
+\mu_{n_1+1,n_2+1}\ \phi^{3-}_{n_1,n_2,n_3}\big)\ ,\\
\Xi^+_{n_1,n_2,n_3} &=
\frac{1}{\mu_{n_1,n_2,n_3+1}}\big(\mu_{n_1,n_2}\ \chi^+_{n_1,n_2,n_3}
-\sqrt{2gf_3(n_3+1)}\ \phi^{3+}_{n_1,n_2,n_3+1}\big)\ ,\\
\Xi^-_{n_1,n_2,n_3} &=
\frac{1}{\mu_{n_1+1,n_2+1,n_3}}\big(\mu_{n_1+1,n_2+1}\
\chi^-_{n_1,n_2,n_3} -\sqrt{2gf_3n_3}\ \phi^{3-}_{n_1,n_2,n_3-1}\big)\ ,
\end{align}
where
\begin{align}
\mu_{n_1,n_2,n_3} = (2gf_1n_1+2gf_2n_2+2gf_3n_3)^{1/2}\ .
\end{align}
Note, that in Eq.~\eqref{ab3one} the field $\Phi^+_{0,0,0}$ is determined
from the requirement $\sum_n (|\Phi^+_n|^2 + |\Xi^+_n|^2) = 
\sum_n (|\chi^+_n|^2 + |\phi^{3+}_n|^2)$, where $(n_1,n_2,n_3)  \equiv n$.
In terms of the new fields the 4d Lagrangian reads, 
\begin{align}\label{L4absuper}
\mathcal{L}_{4} \supset \int d^2 \theta 
  &\Big( \frac{1}{4} W_0 W_0 + \frac{1}{4} W_1 W_1+ \sum_{n}\Big(\frac{1}{2} W^+_{n}W^-_{n}
-\mu_{n_1,n_2,n_3}\ \phi^{-}_{n}\Phi^{+}_{n} 
\nonumber\\
&\hspace{4cm}-\mu_{n_1+1,n_2+1,n_3+1}\ \phi^{+}_{n} \Phi^{-}_{n} \Big)\Big)
+ \text{h.c.}  \nonumber \\
 + \int d^4 \theta &\Big(2(f^2_0 + f^3_0)V_0 + 2(f^1_1+f^2_1 +
 f^3_1)V_1 \nonumber\\
&+ \sum_n \Big(|\phi^{+}_{n}|^2 + |\phi^{-}_{n}|^2 
+ |\Phi^{+}_{n}|^2 + |\Phi^{-}_{n}|^2 + |\Xi^{+}_{n}|^2 + |\Xi^{-}_{n}|^2
\nonumber\\
&\hspace{1cm}-\frac{g}{\sqrt{N}}(|\phi^{-}_{n}|^2 + |\Phi^{-}_{n}|^2 +
|\Xi^{-}_{n}|^2 - |\phi^{+}_{n}|^2 - |\Phi^{+}_{n}|^2 -|\Xi^{+}_{n}|^2\big)V_0  \nonumber\\
&\hspace{1cm}+ g(|\phi^{-}_{n}|^2 + |\Phi^{-}_{n}|^2 + |\Xi^{-}_{n}|^2 
- |\phi^{+}_{n}|^2 - |\Phi^{+}_{n}|^2 -|\Xi^{+}_{n}|^2\big)V_1  \nonumber\\
&\hspace{1cm}+\sqrt{2}\big(\big(\mu_{n_1+1,n_2+1,n_3}\ \overline{\Xi}^-_{n} 
- \mu_{n_1,n_2,n_3+1}\ \Xi^{+}_{n}\big)V^-_{n} + \text{h.c.} \big)
\nonumber\\
&\hspace{1cm} + 2M^2_{n_1,n_2,n_3}V^+_{n}V^-_{n}\Big)\Big)\ ,
\end{align}
where
\begin{align}
M_{n_1,n_2,n_3} = (gf_1(2n_1+1)+gf_2(2n_2+1) + gf_3(2n_3+1))^{1/2}\ .
\end{align}

At this step, supersymmetry breaking by the flux induced D-terms has
to be taken into account, and vector and scalar masses have to be
calculated by eliminating all auxiliary F- and D-terms. The mixing
between $\Xi^+_n$ and $\overline{\Xi}^-_n$ yields
the Goldstone fields $\Pi^+_n$ and the orthogonal complex scalars $\Sigma^+_n$, 
\begin{align}
\Pi^+_{n} &= \frac{1}{\sqrt{2}M_{n_1,n_2,n_3}} \big(\mu_{n_1+1,n_2+1,n_3}\ \overline{\Xi}^-_{n} 
+\mu_{n_1,n_2,n_3+1}\ \Xi^{+}_{n}\big) \ ,\\
\Sigma^+_{n} &= \frac{1}{\sqrt{2}M_{n_1,n_2,n_3}} \big(- \mu_{n_1,n_2,n_3+1}\ \overline{\Xi}^-_{n} 
+ \mu_{n_1+1,n_2+1,n_3}\ \Xi^{+}_{n}\big)\ .
\end{align}
The vector bosons of the tower of Landau levels acquire their mass by
the St\"uckelberg mechanism, and a shift of the vector bosons, 
\begin{align}
A_{n}^{+\mu} \rightarrow A_{n}^{+\mu} 
- \frac{i}{M_{n_1,n_2,n_3}}\ \pd_\mu\Pi^{+}_n\ ,
\end{align}
cancels the mixings with the Goldstone bosons as well as the kinetic
terms of the Goldstone bosons.
The final result for the bosonic part of the 4d Lagrangian
\eqref{L4absuper} reads
\begin{equation} \label{Labb}
\begin{split}
\mathcal{L}^b_4 \supset - \sum_{n} \Big(
& M^2_{n_1,n_2,n_3}\big(A^{+\mu}_{n}A^{-\mu}_{n} + |\Sigma^+_{n}|^2\big) \\
&+ \big(M^2_{n_1,n_2,n_3} - 2gf_1 - 2gf_2 \big) |\phi^-_{n}|^2
+ \big(M^2_{n_1,n_2,n_3} + 2gf_1 + 2gf_2 \big) |\phi^+_{n}|^2 \\
&+ \big(M^2_{n_1,n_2,n_3} + 2gf_3\big) |\Phi^-_{n}|^2 
+ \big(M^2_{n_1,n_2,n_3} - 2gf_3\big)|\Phi^+_{n}|^2\Big) \ ,
\end{split}
\end{equation}
where, by definition, $\phi^-_{0,0,n_3} = 0$ and
$\Phi^+_{0,0,0} =\phi^{3+}_{0,0,0}$ (see~Eqs.~\eqref{ab1one},\eqref{ab3one}). 
The scalars with smallest masses are $\phi^{-}_{0,1,0}$,
$\phi^{-}_{1,0,0}$ and $\Phi^{+}_{0,0,0}$,
\begin{equation}
\begin{split}
\mathcal{L}^b_4 \supset 
- g&(-f_1 + f_2 + f_3) |\phi^{-}_{0,1,0}|^2 -g(f_1-f_2+f_3)
|\phi^{-}_{1,0,0}|^2 \\
& - g(f_1 + f_2 - f_3) |\Phi^+_{0,0,0}|^2 \ .
\end{split}
\end{equation}
These masses are in agreement with the ones given in
Table~\ref{tab:scalarmasses} for the $ab$-sector.

Denoting the Weyl fermions contained in the superfields $\phi^{\pm}$,
$\Phi^\pm$, $\Xi^\pm$ and $V^+=V^{-\dagger}$ by $\psi^{\pm}$, $\psi^{'\pm}$, 
$\omega^\pm$ and $\lambda^\pm$, respectively, one finds for
the fermionic mass terms of the 4d Lagrangian \eqref{L4absuper} 
(cf.~Appendix \ref{app:susy}),
\begin{equation}\label{Labf}
\begin{split}
\mathcal{L}_{4} \supset \sum_{n} 
 &\Big( \mu_{n_1,n_2,n_3}\ \psi^{-}_{n}\psi^{'+}_{n} 
 +i \mu_{n_1,n_2,n_3+1}\ \omega^+_{n} \lambda^-_{n} \\
&+\mu_{n_1+1,n_2+1,n_3+1}\ \psi^{+}_{n} \psi^{'-}_{n} 
+i \mu_{n_1+1,n_2+1,n_3}\ \omega^{-}_{n}\lambda^+_{n} \Big)
+ \text{h.c.} 
\end{split}
\end{equation}
Note, that by definition, $\psi^{-}_{0,0,n_3} = 0$. Hence,
the spectrum contains one zero-mode, $\psi^{'+}_{0,0,0} \subset
\Phi^+_{0,0,0}$. 

The number of flux quanta in the first, second and third torus is $1$,
$l-2$ and $3$, respectively. All fields therefore have a multiplicity
of $3(l-2)$, in agreement with the
intersection number for the $ab$-sector listed in
Table~\ref{tab:fermions}. The multiplicity of fields is labeled by the indices
$j_1,j_2,j_3$. The quadratic 
part of the 4d Lagrangian, given in Eqs.~\eqref{Labb} and
\eqref{Labf}, is diagonal and the same for all fields. However, due to the
non-trivial profile of the mode functions in the compact space,
Yukawa couplings depend on $j_1,j_2,j_3$.

\section{Higgs sector}
\label{sec:higgs}

In the $bc$- and $bc'$-sectors of the D-brane model there are no
chiral fermions, and both sectors contain a tachyon, see
Table~\ref{tab:scalarmasses}. In this section we will analyze the $bc'$-sector in
detail. According to Table~\ref{tab:wrapping}, brane $b$ and brane $c'$ are
parallel in two tori, their distances being moduli. In the T-dual flux
compactification these moduli correspond to Wilson lines. From
Table~\ref{tab:wrapping} and Eq.~\eqref{background} one obtains for
the background fields in the three tori,
\begin{equation}\label{fq3}
\begin{split}
gf^1_1 &= \frac{\rho_1}{2\pi\alpha'}\ \bz_1\ , \quad
 g\chi^2_1 = \frac{l\rho_2}{2\pi\alpha'}\ \frac{\bz_2}{\sqrt{2}} + g\xi^2_1\ , \quad
 g\chi^3_1 = -\frac{2\rho_3}{2\pi\alpha'}\ \frac{\bz_3}{\sqrt{2}} + g\xi^3_1\ , \\
gf^1_2 &= \frac{\rho_1}{2\pi\alpha'}\ \bz_1\ , \quad
 g\chi^2_2 = -\frac{l\rho_2}{2\pi\alpha'}\ \frac{\bz_2}{\sqrt{2}} + g\xi^2_2\ , \quad
 g\chi^3_2 = \frac{\rho_3}{2\pi\alpha'}\ \frac{\bz_3}{\sqrt{2}} + g\xi^3_2\ . 
\end{split}
\end{equation}
The $bc'$-sector contains the vector and chiral superfields 
$V^{++}$, $\phi^{i++}$, $V^{--}$ and $\phi^{i--}$. The charges with
respect to $H_1$ and $H_2$ are identical. For notational simplicity, 
we shall drop one of the superscripts in the following.
The commutators of the relevant $SO(32)$ matrices are given in Eq.~\eqref{combc'},
\begin{equation}
[H_1,X^{\pm\pm}] = \pm X^{\pm\pm}\ , \quad
[H_2,X^{\pm\pm}] = \pm X^{\pm\pm}\ , \quad
[X^{++}, X^{--}] = H_1 + H_2\ . \nonumber
\end{equation}
Combining the $H_1$ and $H_2$ background fields in Eq.~\eqref{fq3},
one obtains for the total flux densities and Wilson lines in the three tori
\begin{equation}\label{gb3}
\begin{split}
g(f^1_1+f^1_2) &= \frac{\rho_1}{\pi\alpha'}\bz_1 \equiv
4gf\bz_1\ , \quad
g(\chi^2_1+\chi^2_2) = g(\xi^2_1 + \xi^2_2) \equiv g\sqrt{2}\xi_2\ ,\\
g(\chi^3_1+\chi^3_2) &= g(\xi^3_1 + \xi^3_2) \equiv g\sqrt{2}\xi_3\ .
\end{split}
\end{equation}

Using Eqs.~\eqref{10daction}, \eqref{expV}, \eqref{expfi} and
\eqref{combc'}, and inserting the background fields \eqref{gb3},
one obtains for the quadratic part of the 10d Lagrangian,
\begin{align}
\mathcal{L}_{10} \supset \int d^2 \theta 
 &\Big( \frac{1}{4} W_1W_1  + \frac{1}{4} W_2W_2  + \frac{1}{2} W^+W^-  \nonumber\\
&+\phi^{3+}(\pd_1 - 2gf\bz_1)\phi^{2-} 
- \phi^{2+}(\pd_1 - 2gf\bz_1)\phi^{3-} \nonumber\\
&+ \phi^{1+}(\pd_2 - g\xi_2)\phi^{3-} 
- \phi^{3+}(\pd_2 - g\xi_2)\phi^{1-} \nonumber\\
&+ \phi^{2+}(\pd_3 - g\xi_3)\phi^{1-} 
- \phi^{1+}(\pd_3 - g\xi_3)\phi^{2-}\Big)
+ \text{h.c.}  \nonumber \\
 + \int d^4 \theta &\Big( \ophi^{i+}\phi^{i+} + \ophi^{i-}\phi^{i-} 
+ 4f \big(V_1 +V_2\big) 
+g\big(\ophi^{i+}\phi^{i+} - \ophi^{i-}\phi^{i-}\big) \big(V_1 + V_2\big) \nonumber\\
&+\sqrt{2}\Big(\big(\big(\pd_1 + 2gf\bz_1\big)\ophi^{1-} 
+\big(\pd_2 + g\xi_2\big)\ophi^{2-} +\big(\pd_3 + g\xi_3\big)\ophi^{3-} \nonumber\\
&+\big(\opd_1 - 2gfz_1\big)\phi^{1+} 
+\big(\opd_2 - g\bxi_2\big)\phi^{2+} +\big(\opd_3 - g\bxi_3\big)\phi^{3+} \big)V^- 
+ \text{h.c.}\Big)\nonumber\\
&+\big(\opd_1 - 2gfz_1\big)V^{+} \big(\pd_1 - 2gf\bz_1\big)V^{-} 
+\big(\opd_2 - g\bxi_2\big)V^{+} \big(\pd_2 - g\xi_2\big)V^{-} \nonumber\\
&+\big(\opd_3 - g\bxi_3\big)V^{+} \big(\pd_3 - g\xi_3\big)V^{-} 
+\big(\opd_1 + 2gfz_1\big)V^{-} \big(\pd_1 + 2gf\bz_1\big)V^{+} \nonumber\\ 
&+\big(\opd_2 + g\bxi_2\big)V^{-} \big(\pd_2 + g\xi_2\big)V^{+} 
+\big(\opd_3 + g\bxi_3\big)V^{+} \big(\pd_3 + g\xi_3\big)V^{-} \Big)\ .
\end{align}
The fields feel magnetic flux only in the first torus. Hence the mode
functions are harmonic oscillator wave functions in the first torus
and ordinary KK mode functions in the second and third
torus,
\begin{equation}\label{KK3}
\begin{split}
\phi^{i+}(x;z_1,z_2,z_3) &= \sum_{nj,ml,m'l'}
\phi^{i+}_{nj,ml,m'l'}(x) \xi_{nj}(z_1)\bk_{ml}(z_2)\bk_{m'l'}(z_3)\ , \\
\phi^{i-}(x;z_1,z_2,z_3) &= \sum_{nj,ml,m'l'}
\phi^{i-}_{nj,ml,m'l'}(x) \bxi_{nj}(z_1)\kk_{ml}(z_2)\kk_{m'l'}(z_3)\ , 
\quad \text{\rm etc.}\ ,
\end{split}
\end{equation}
where
\begin{equation}
\kk_{lm}(z) = e^{z\mu_{lm}-\bz\overline{\mu}_{lm}} = \bk_{-l-m}\ , \quad
\mu_{lm} =2\pi(m+il) \equiv \mu_\kk \ .
\end{equation}
Replacing covariant derivatives with flux by annihilation and creation
operators according to Eq.~\eqref{anncrea}, inserting the mode
expansion \eqref{KK3} for the second and third torus and keeping for
the two $U(1)$ factors only the lowest mode, one arrives at
\begin{align}\label{fluxkk}
\mathcal{L}_{10} \supset \int d^2 \theta 
 &\Big( \frac{1}{4} W_1W_1  + \frac{1}{4} W_2W_2 +
 \sum_{\kk\kk'}\Big(\frac{1}{2} W^+_{\bk\bk'}W^-_{\kk\kk'} \nonumber\\
& -i\sqrt{4gf}\big(\phi^{3+}_{\bk\bk'} a_-^\dagger\phi^{2-}_{\kk\kk'} 
- \phi^{2+}_{\bk\bk'}a_-^\dagger\phi^{3-}_{\kk\kk'}\big) 
+ \phi^{1+}_{\bk\bk'}M_\kk\phi^{3-}_{\kk\kk'} \nonumber\\
&- \phi^{3+}_{\bk\bk'}M_\kk\phi^{1-}_{\kk\kk'} 
+ \phi^{2+}_{\bk\bk'}M_{\kk'}\phi^{1-}_{\kk\kk'} 
- \phi^{1+}_{\bk\bk'}M_{\kk'}\phi^{2-}_{\kk\kk'}\Big)\Big)
+ \text{h.c.}  \nonumber \\
 + \int d^4 \theta &\Big( 4f\big(V_1 +V_2\big)
+ \sum_{\kk,\kk'}\Big(\ophi^{i+}_{\bk\bk'}\phi^{i+}_{\kk\kk'} 
+ \ophi^{i-}_{\bk\bk'}\phi^{i-}_{\kk\kk'} \nonumber\\
&+g \big(V_1 +V_2\big)
\big(|\phi^{i+}_{\kk\kk'}|^2-|\phi^{i-}_{\kk\kk'}|^2\big) 
+\sqrt{2}\Big(\big(-i\sqrt{4gf}\big(a_+\ophi^{1-}_{\bk\bk'} +
a_+^\dagger\phi^{1+}_{\bk\bk'}\big) \nonumber\\
&\hspace{1cm}-M_\kk\ophi^{2-}_{\bk\bk'} - M_{\kk'}\ophi^{3-}_{\bk\bk'} 
+\bM_\kk\phi^{2+}_{\bk\bk'} + \bM_{\kk'}\phi^{3+}_{\bk\bk'}\big)V^-_{\kk\kk'} 
+ \text{h.c.}\Big)\\
&-4gf\big(a_+^\dagger V^{+}_{\bk\bk'} a_-^\dagger V^-_{\kk\kk'} 
+ a_- V^-_{\bk\bk'} a_+V^+_{\kk\kk'}\big) 
+2\big(|M_\kk|^2 + |M_{\kk'}|^2\big)V^{+}_{\bk\bk'} V^{-}_{\kk\kk'}
\Big)\Big)\ ,\nonumber
\end{align}
where
\begin{equation}
M_\kk = \mu_\kk - g\xi_2\ , \quad M_{\kk'} = \mu_{\kk'} - g\xi_3\ 
\end{equation}
are mass terms that depend on the Wilson lines.

Consider first the case without flux, i.e., $f=0$. In this case
supersymmetry is unbroken and, for simplicity, we restrict ourselves
to mode functions \eqref{KK3} that are
constant in the first torus. Then one can easily diagonalize the
Lagrangian. Defining the superfields\footnote{The following discussion
  holds for $M_{\kk\kk'} \neq 0$. For  $M_{\kk\kk'} = M_{\kk} =
  M_{\kk'} = 0$, the fields $\phi^{2\pm}_{00}$ and $\phi^{3\pm}_{00}$
  do not mix.}
\begin{equation}
\begin{split}
\phi^+_{\bk\bk'} &=
\frac{1}{|M_{\kk\kk'}|}\big(M_{\kk'}\phi^{2+}_{\bk\bk'} - M_{\kk}\phi^{3+}_{\bk\bk'}\big)\ , \quad
\phi^-_{\kk\kk'} =
\frac{1}{|M_{\kk\kk'}|}\big(M_{\kk'}\phi^{2-}_{\kk\kk'} - M_{\kk}\phi^{3-}_{\kk\kk'}\big)\ , \\
\chi^+_{\bk\bk'} &=
\frac{1}{|M_{\kk\kk'}|}\big(\bM_{\kk}\phi^{2+}_{\bk\bk'} + \bM_{\kk'}\phi^{3+}_{\bk\bk'}\big)\ , \quad
\chi^-_{\kk\kk'} =
\frac{1}{|M_{\kk\kk'}|}\big(\bM_{\kk}\phi^{2-}_{\kk\kk'} + \bM_{\kk'}\phi^{3-}_{\kk\kk'}\big)\ ,
\end{split}
\end{equation}
where $|M_{\kk\kk'}| = (|M_\kk|^2 + |M_{\kk'}|^2)^{1/2}$, and shifting the
vector superfield,
\begin{equation}
V^+_{\kk\kk'} \rightarrow V^+_{\kk\kk'} -
\frac{1}{\sqrt{2}|M_{\kk\kk'}|}\big(\chi^+_{\kk\kk'} -
\ochi^-_{\kk\kk'}\big)\ ,
\end{equation}
one obtains
\begin{align}
\mathcal{L}_{10} \supset &\int d^2 \theta \Big(\frac{1}{4} W_1W_1  +
\frac{1}{4} W_2W_2  \nonumber\\
&\hspace{1.2cm}+\sum_{\kk\kk'}\Big(\frac{1}{2} W^+_{\bk\bk'}W^-_{\kk\kk'}  + 
|M_{\kk\kk'}|\big(\phi^{1-}_{\kk\kk'}\phi^+_{\bk\bk'} 
- \phi^{1+}_{\bk\bk'}\phi^-_{\kk\kk'}\big)\Big)\Big) + \text{h.c.} \nonumber \\
 &+ \int d^4 \theta \sum_{\kk\kk'}\Big( |\phi^{1+}_{\kk\kk'}|^2 + |\phi^{1-}_{\kk\kk'}|^2 + |\phi^{+}_{\kk\kk'}|^2
 + |\phi^{-}_{\kk\kk'}|^2 +2|M_{\kk\kk'}|^2 V^-_{\bk\bk'} V^+_{\kk\kk'}\Big)\ .
\end{align}
The Goldstone chiral multiplets $\chi^\pm_{\kk\kk'}$ have been removed
from the Lagrangian, and a complex vector multiplet and four chiral
multiplets all have the same mass $M_{\kk\kk'}$, corresponding to
$\mathcal{N}=4$ supersymmetry.

The magnetic flux in the first torus mixes different Landau levels of
$\phi^{1\pm}_{n,\kk\kk'}$ and $\chi^\pm_{n,\kk\kk'}$. Now it is convenient to introduce the
superfields
\begin{equation}
\begin{split}
\Xi^+_{n,\bk\bk'} &= \frac{1}{\mu_{n,\kk\kk'}} \big(|M_{\kk\kk'}|\ \chi^+_{n,\bk\bk'}
-\sqrt{4gfn}\ \phi^{1+}_{n-1,\bk\bk'}\big)\ ,\\
\Xi^-_{n,\kk\kk'} &= \frac{1}{\mu_{n+1,\kk\kk'}} \big(|M_{\kk\kk'}|\ \chi^-_{n,\kk\kk'}
-\sqrt{4gf(n+1)}\ \phi^{1-}_{n+1,\kk\kk'}\big)\ ,\\
\Phi^+_{n,\bk\bk'} &= \frac{1}{\mu_{n+1,\kk\kk'}}
\big(\sqrt{4gf(n+1)}\ \chi^+_{n+1,\bk\bk'}
+|M_{\kk\kk'}|\ \phi^{1+}_{n,\kk\kk'}\big)\ ,\\
\Phi^-_{n,\kk\kk'} &= \frac{1}{\mu_{n,\kk\kk'}} \big(\sqrt{4gfn}\ \chi^-_{n-1,\kk\kk'}
+|M_{\kk\kk'}|\ \phi^{1-}_{n,\kk\kk'}\big)\ ,
\end{split}
\end{equation}
with
\begin{equation}\label{MnKK}
\mu_{n,\kk\kk'} = (4gfn + |M_{\kk\kk'}|^2)^{1/2}\ .
\end{equation}
Using Eqs.~\eqref{anncrea} and \eqref{fluxkk}, a straightforward
calculation yields for the 4d Lagrangian,
\begin{align}\label{L4dbc'}
\mathcal{L}_{10} \supset \int d^2 \theta 
 &\Big( \frac{1}{4} W_1W_1 + \frac{1}{4}W_2 W_2
+\sum_{n,\kk\kk'}\Big(\frac{1}{2} W^+_{n,\bk\bk'}W^-_{n,\kk\kk'}\nonumber\\
&+\mu_{n,\kk\kk'}\ \Phi^-_{n,\kk\kk'}\phi^+_{n,\bk\bk'} 
- \mu_{n+1,\kk\kk'}\ \Phi^+_{n,\bk\bk'}\phi^-_{n,\kk\kk'} \Big)\Big)
+ \text{h.c.}  \nonumber \\
 + \int d^4 \theta &\Big( 4f\big(V_1 +V_2\big)
+\sum_{n,\kk\kk'}\Big(|\phi^+_{n,\kk\kk'}|^2 + |\phi^-_{n,\kk\kk'}|^2 
+ |\Phi^+_{n,\kk\kk'}|^2 + |\Xi^+_{n,\kk\kk'}|^2 \nonumber\\
&+ |\Phi^-_{n,\kk\kk'}|^2 + |\Xi^-_{n,\kk\kk'}|^2 
+ g\big(|\phi^+_{n,\kk\kk'}|^2 + |\Phi^+_{n,\kk\kk'}|^2 +
|\Xi^+_{n,\kk\kk'}|^2\nonumber\\
&\hspace{1cm}- |\phi^-_{n,\kk\kk'}|^2 - |\Phi^-_{n,\kk\kk'}|^2 - |\Xi^-_{n,\kk\kk'}|^2\big)
\big(V_1 +V_2\big)\nonumber\\
&+\sqrt{2}\big(\big(\mu_{n,\kk\kk'}\ \Xi^+_{n,\bk\bk'}
- \mu_{n+1,\kk\kk'}\ \overline{\Xi}^-_{n,\bk\bk'}\big)V^-_{n\kk\kk'}
+ \text{h.c.}\big)\nonumber\\
&+2 M^2_{n,\kk\kk'}V^{+}_{\bk\bk'} V^{-}_{\kk\kk'}\Big)\Big)\ ,
\end{align}
where
\begin{align}\label{Amass}
M_{n,\kk\kk'} = (2gf(2n+1) + |M_{\kk\kk'}|^2)^{1/2}\ .
\end{align}
Like in the previous section the Goldstone bosons giving mass to the
vector bosons $A^{+\mu}_{n,\kk\kk'}$ are identified as\footnote{Note that
  we use the same notation for a chiral superfield and its scalar component.}
\begin{equation}
\Pi^+_{n,\bk\bk'} = \frac{1}{\sqrt{2}M_{n,\kk\kk'}}
\big(\mu_{n,\kk\kk'}\ \Xi^+_{n,\bk\bk'} + \mu_{n+1,\kk\kk'}\
\overline{\Xi}^-_{n,\bk\bk'}\big)\ ,
\end{equation}
with the orthogonal complex scalars
\begin{equation}
\Sigma^+_{n,\bk\bk'} = \frac{1}{\sqrt{2}M_{n,\kk\kk'}}
\big(\mu_{n,\kk\kk'}\ \Xi^+_{n,\bk\bk'} - \mu_{n+1,\kk\kk'}\
\overline{\Xi}^-_{n,\bk\bk'}\big)\ ,
\end{equation}
where we have used $\mu_{n+1,\kk\kk'}^2 + \mu_{n,\kk\kk'}^2 =
2M^2_{n,\kk\kk'}$.
The kinetic terms of the tower of Goldstone bosons are removed by
shifting the vector bosons,
\begin{equation}
A^{+\mu}_{n,\kk\kk'} \rightarrow A^{+\mu}_{n,\kk\kk'} 
+ \frac{i}{M_{n,\kk\kk'}}\ \pd^\mu \Pi^+_{n,\kk\kk'}\ .
\end{equation}
Eliminating all F-terms and the D-terms $D_1$, $D_2$, $D_\pm$ by their
equations of motion, one obtains for the bosonic mass terms
\begin{align}
\mathcal{L}^b_4 \supset -\sum_{n,\kk\kk'} \Big(&M^2_{n,\kk\kk'}
\big(A^+_{n,\kk\kk'\mu} A^{-\mu}_{n,\kk\kk'} + |\phi^+_{n,\kk\kk'}|^2 
+ |\phi^-_{n,\kk\kk'}|^2 + |\Sigma^-_{n,\kk\kk'}|^2\big) \nonumber\\
+ (&M^2_{n,\kk\kk'}-4gf)|\Phi^-_{n,\kk\kk'}|^2          
+ (M^2_{n,\kk\kk'}+4gf) |\Phi^+_{n,\kk\kk'}|^2\Big)\ . \label{bc2boson}
\end{align}
Note that the mass of $\Phi^-_{0,00}$,
\begin{equation}\label{tachyonbc'}
M^2[\Phi^-_{0,00}] = -2gf + g^2(|\xi_2|^2 + |\xi_3|^2)\ ,
\end{equation}
is tachyonic for $|\xi_2|^2 + |\xi_3|^2 < f/g$. The implications will be
studied in the subsequent section. The boson masses are consistent
with the string formula \eqref{master} for the internal helicities $(0,0,0)$,
$(\pm,0,0)$, $(0,\pm,0)$, $(0,0,\pm)$. 

Denoting the Weyl fermions contained in the superfields $\phi^\pm$, $\Phi^\pm$,
$\Xi^\pm$ and $V^+=V^{-\dagger}$ by $\psi^\pm$, ${\psi'}^{\pm}$,
$\omega^\pm$ and $\lambda^\pm$, respectively, one obtains for the
fermion mass terms (see Eq.~\eqref{MnKK}),
\begin{equation}\label{bc2fermion}
\begin{split}
\mathcal{L}^f_4 \supset - \Big(&\mu_{n,\kk\kk'}\big(\psi^-_{n,\bk\bk'}{\psi'}^{+}_{n,\bk\bk'}
 -i\omega^-_{n,\bk\bk'}{\lambda}^{+}_{n,\bk\bk'}\big) \\
&+ \mu_{n+1,\kk\kk'}\big(\psi^+_{n,\bk\bk'}{\psi'}^{-}_{n,\bk\bk'}
 -i\omega^+_{n,\bk\bk'}{\lambda}^{-}_{n,\bk\bk'}\big) \Big) + \text{h.c.} 
\end{split}
\end{equation}
For vanishing Wilson lines there are four vector-like zero modes, 
$\psi^-_{0,\bk\bk'}$, ${\psi'}^{+}_{0,\bk\bk'}$,
$\omega^-_{0,\bk\bk'}$ and $\lambda^{+}_{0,\bk\bk'}$. In the string
formula \eqref{master} the mass spectrum is obtained for the
helicities $(-1/2,\pm 1/2, \pm 1/2)$ and $(1/2,\pm 1/2, \pm 1/2)$.
There are two flux quanta in the first torus, hence the multiplicity
of all fields is two.
In the case $M_{n,\kk\kk'} = 0$, corresponding to a compactification
from six dimensions to four dimensions, the mass spectrum has
previously been obtained in \cite{Buchmuller:2016gib}.

\subsection{$bc$-sector}

The $bc$-sector is very similar to the $aa'$-sector. In both cases the
magnetic flux is non-zero only in the second and third torus. We
therefore do not treat this case in detail but only mention some key
features which are relevant for the discussion of tachyon condensation
in Section~\ref{sec:tachyon}. 

The sector contains the vector and chiral superfields 
$V^{+-}$, $\phi^{i+-}$, $V^{-+}$ and $\phi^{i-+}$, i.e., the charges with
respect to $H_1$ and $H_2$ are opposite. 
The commutators of the relevant $SO(32)$ matrices read (see Eq.~\eqref{combc}),
\begin{equation}
[H_1,T^{\pm\mp}] = \pm T^{\pm\mp}\ ,\quad
[H_2,T^{\pm\mp}] = \mp T^{\pm\mp}\ , \quad
[T^{+-},T^{-+}] = H_1 - H_2  \ .
\end{equation}
For zero Wilson lines, one obtains for the background fields given in
Eq.~\eqref{fq3} the total flux densities in the three tori
\begin{equation}
\begin{split}
&g(f^1_1-f^1_2) = 0\ ,\quad
g(f^2_1-f^2_2) = 2l\frac{\rho_2}{2\pi\alpha'}\bz_2 \equiv gf_2\bz_2\ , \\
&g(f^3_1-f^3_2) = -3\frac{\rho_3}{2\pi\alpha'}\bz_3 \equiv -gf_3\bz_3\ . 
\end{split}
\end{equation}
The crucial difference compared to the $aa'$-sector is the opposite
sign of the flux densities in the second and third torus. In the
derivation of the effective 4d action this exchanges annihilation and
creation operators in various steps of the calculation. Taking this
into account, all relevant $F$- and $D$-terms can be essentially read
off from Eq.~\eqref{Laa'basic}.

\section{Effective potential}
\label{sec:potential}

We are now ready to calculate the one-loop effective potential from the effective field theory. We start with the potential for
Wilson lines in the $bc'$-sector, then we discuss the potential in the $ab$-sector which is independent of Wilson lines and depends only on volume moduli. We shall perform the calculation for
the effective field theory discussed in the previous section, summing
over the full towers of Landau levels and KK modes, and we shall then
compare the result with a string calculation. 

\subsection{Field theory calculation}

The one-loop effective potential is given by the well known Coleman-Weinberg expression
\begin{equation}
V(\xi) = \frac{1}{2} \sum_I (-)^F \int \frac{d^4k}{(2\pi)^4} \ln{(k^2+
  M^2_I(\xi))} \ ,
\end{equation}
where the sum extends over all bosonic and fermionic states. The masses in the $bc'$-sector
are denoted by $M_I(\xi)$, $F$ denotes fermion number, $I$ accounts for Landau
levels and KK quantum numbers, and $\xi$ represents
real and imaginary parts of the Wilson lines in the second and third
torus, i.e.~$\xi = (\xi_1,\xi_2,\xi_3,\xi_4) = 
(\text{Re}\ \xi_2, \text{Im}\ \xi_2, \text{Re}\ \xi_3,
\text{Im}\ \xi_3)$. Using the Schwinger representation of propagators
one has
\begin{equation}
 \int \frac{d^4k}{(2\pi)^4} \ln{(k^2+M^2_I)} 
= -\frac{1}{16\pi^2}\int_0^\infty \frac{dt}{t^3}e^{-M_I^2 t}\ .
\end{equation}
According to Eqs.~\eqref{MnKK}, \eqref{bc2boson} and \eqref{bc2fermion} the mass
spectrum of the $bc'$-sector takes the form
\begin{equation}
M^2_I(\xi) = 2gfn + mgf + |M_{\kk\kk'}|^2\ ,
\end{equation}
where $n$ is the Landau
level, and $m$ takes the values $m=-1,0,1,2,3$; multiplicities $l_m$ and
fermion numbers $F_m$ for different values of $m$ are $(l_m) =
(1,4,6,4,1)$ and $(F_m) = (0,1,0,1,0)$.
The sum over Landau levels is easily carried out,
\begin{equation}
\sum_{n=0}^\infty e^{-(2gfn + gfm + |M_{\kk\kk'}|^2) t} = 
e^{- |M_{\kk\kk'}|^2 t-gf(m-1)t} \frac{1}{2\sinh(gft)}\ ,
\end{equation}
and the sum over all bosons and fermions yields
\begin{equation}
\begin{split}
\sum_m l_m (-)^{F_m} \sum_{n=0}^\infty e^{-(2gfn + gfm + |M_{\kk\kk'}|^2) t} = 
&\ e^{- |M_{\kk\kk'}|^2 t} \frac{1}{2\sinh(gft)} \\
&\ \times\left(e^{2gft} - 4 e^{gft} + 6 - 4 e^{-gft} + e^{-2gft} \right) \\
=&\ 16~e^{- |M_{\kk\kk'}|^2 t}~\frac{\sinh^4(gft/2)}{2\sinh(gft)} \ .
\end{split}
\end{equation}
There are two flux quanta in the first torus leading to a multiplicity
two for all states. Introducing radii for the second and third
torus as 
$(R_1,R_2,R_3,R_4) = (L_2,L_2',L_3,L_3')/2\pi$, the final result for
the potential takes the form
\begin{align}\label{potential}
V(\xi) = -\frac{1}{2\pi^2}\int_0^\infty \frac{dt}{t^3} 
\frac{\sinh^4\left(\frac{gft}{2}\right)}{\sinh\left(gft\right)}
\sum_{m_i}\exp{\Big(-t\Big(\frac{m_i}{R_i}+g\xi_i}\Big)^2\Big)\ .
\end{align}

The integral $V(\xi) = \int dt V(t,\xi)$ has an infrared as well as an
ultraviolet divergence. For large $t$ the contribution of the $m_i=0$
term to the integrand behaves as
\begin{equation}
V(t,\xi) \propto \frac{1}{t^3} e^{(gf - g^2\xi^2)t}\ .
\end{equation}
Hence, the integral diverges if $\xi^2 < f/g$. The same is true if
$\xi$ is closer than $\sqrt{f/g}$ to any lattice vector $m/R$. This 
infrared divergence is an effect of the tachyon in the spectrum. 
Moreover, there is an ultraviolet divergence. Although each term in
the sum is convergent, the sum over KK modes behaves as $R^4/t^2$
for small $t$ so that the integrand scales as
\begin{equation}
V(t,\xi) \propto \frac{(gf)^3R^4}{t^2} e^{(gf - g^2\xi^2)t}\ .
\end{equation}
Introducing an ultraviolet cutoff, $t>\delta \equiv 1/\Lambda^2_\text{UV}$,
the quadratic ultraviolet divergence becomes manifest,
$V \sim (gf)^3R^4\Lambda^2_\text{UV}$.

A convenient regularization of the potential can be obtained by
considering a Poisson resummation of the sum over KK modes,
\begin{equation}
\sum_{m_i}\exp{\Big(-t\Big(\frac{m_i}{R_i}+g\xi_i}\Big)^2\Big)
= \frac{\pi^2\prod_i R_i}{t^2}\sum_{l_i} 
\exp{\Big(i\sum_i l_iR_i\xi_i - \pi^2\sum_i(R_il_i)^2/t\Big)}\ .
\end{equation}
The ultraviolet divergence is now encoded in the term $l_i=0$. 
Adding to the sum a counter term
\begin{equation}
-c_1 e^{-\mu_1^2 t} - c_2 e^{-\mu_2^2 t} \ ,
\end{equation}
with
\begin{equation}
1 - c_1 - c_2 = 0\ , \quad c_1\mu_1^2 + c_2\mu_2^2 = 0 \ ,
\end{equation}
implies that 
\begin{equation}
\sum_{m_i}\exp{\Big(-t\Big(\frac{m_i}{R_i}+g\xi_i}\Big)^2\Big)
- \frac{\pi^2\prod_i R_i}{t^2}\Big(c_1 e^{-\mu_1^2 t} + c_2
e^{-\mu_2^2 t} \Big)
\end{equation}
is finite as $t\rightarrow 0$, yielding a finite integral $V(\xi) = \int dt V(t,\xi)$.
Note, that the terms
\begin{equation}
\int \frac{dt}{t^5} e^{-\mu_i^2 t} \propto 
\int \frac{dt}{t}\int\frac{d^8 k}{(2\pi)^8} e^{-(k^2 + \mu_i^2) t}
\propto \int\frac{d^8 k}{(2\pi)^8} \ln{(k^2 + \mu_i^2)} 
\end{equation}
correspond to Pauli-Villars regulators in 8d field theories.
\begin{figure}[t]
\begin{center} 
\includegraphics[width = 0.7\textwidth]{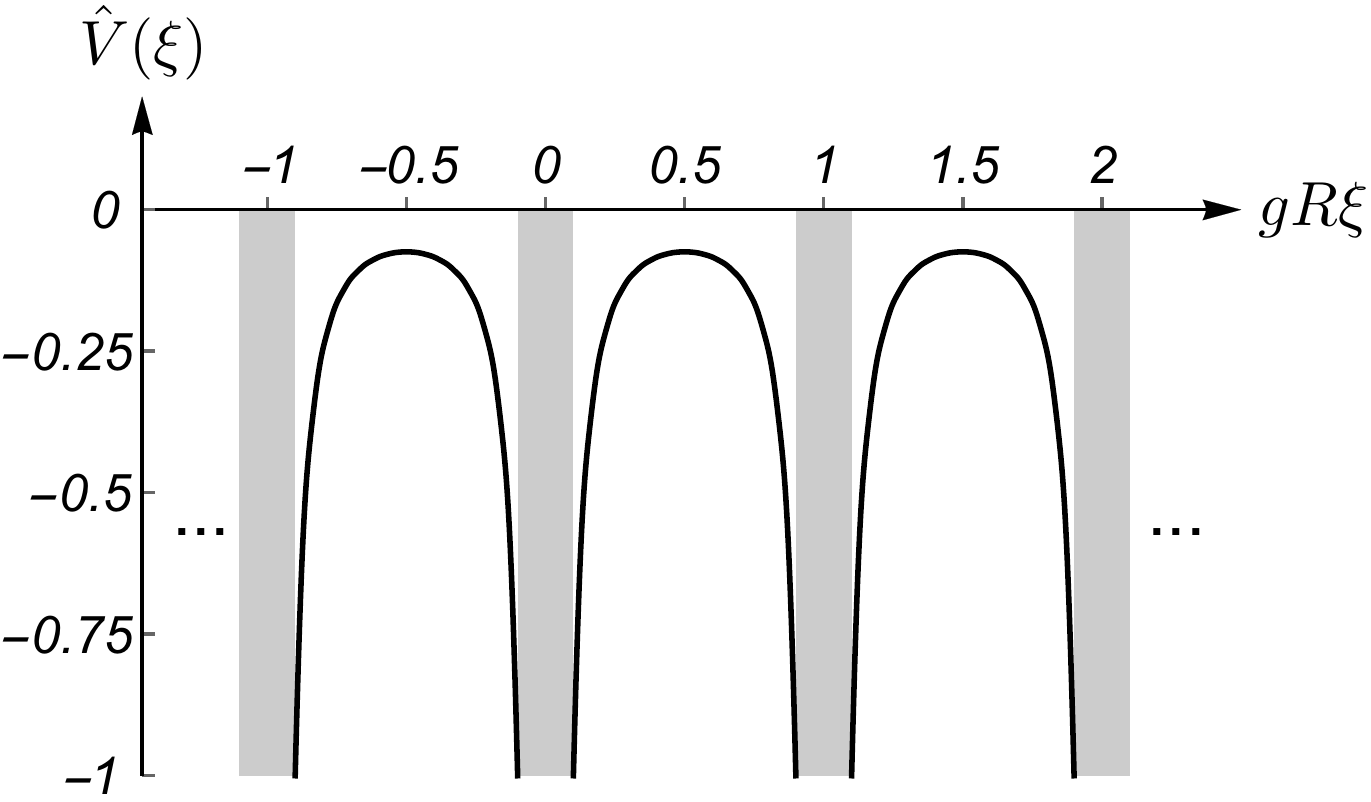}
\end{center}
\caption{Wilson-line potential, normalized to its value at the border
  to the tachyonic region, which is chosen to have the width
  $\Delta=2\sqrt{gfR^2}=0.2$ (see text).}
\label{fig:potential}
\end{figure}

Stationary points of the potential have to satisfy
\begin{equation}
\frac{\pd V(t,\xi)}{\pd \xi_i}  \propto \sum_{m_j}
\Big(\frac{m_i}{R_i} + g\xi_i\Big) \exp{\Big(-t\Big(\frac{m_j}{R_j}+g\xi_j}\Big)^2\Big) = 0\ .
\end{equation}
The solutions are given by
\begin{equation}
\hat{\xi_i} = \frac{n_i}{2gR_i}\ ,\quad n_i\in \mathbb{Z}\ ,
\end{equation}
since
\begin{equation}
\begin{split}
\sum_{m_j}
\Big(\frac{m_i}{R_i} + \frac{n_i}{2R_i}\Big)
&\exp{\Big(-t\Big(\frac{m_j}{R_j}+\frac{n_j}{2R_j}\Big)^2\Big)}\\
&= \sum_{m_j}\Big(\frac{m_i}{R_i} - \frac{n_i}{2R_i}\Big)
\exp{\Big(-t\Big(\frac{m_j}{R_j} - \frac{n_j}{2R_j}\Big)^2\Big)}\\
&=-\sum_{m_j}
\Big(\frac{m_i}{R_i} + \frac{n_i}{2R_i}\Big)
\exp{\Big(-t\Big(\frac{m_j}{R_j}+\frac{n_j}{2R_j}\Big)^2\Big)} = 0\ .
\end{split}
\end{equation}
In the vincinity of an extremum the potential can be approximated by
the contributions of a few neighboring lattice points. As an example, 
consider a one-dimensional case where $\xi$ points in one lattice
direction. For $gR\xi = 1/2$ the four nearest points yield
\begin{align}\label{approxpot}
\sum_{m}&\exp{\Big(-t\Big(\frac{m}{R}+\frac{1}{2R}\Big)^2\Big)} 
\simeq \ \exp{\Big(-t\Big(-\frac{2}{R} + \frac{1}{2R}\Big)^2\Big)} \\
&\quad + \exp{\Big(-t\Big(-\frac{1}{R}+\frac{1}{2R}\Big)^2\Big)} 
+ \exp{\Big(-t\Big(\frac{1}{2R}\Big)^2\Big)}
+ \exp{\Big(-t\Big(\frac{1}{R} + \frac{1}{2R}\Big)^2\Big)} + \ldots \nonumber
\end{align}
Using this approximation for the sum over KK modes the potential can
be evaluated numerically. As discussed above it is periodic with
period $gR\xi \sim gR\xi + 1$. Tachyonic regions $(gR\xi-n)^2 < gfR^2
\equiv \Delta^2/4$, where the potential is ill defined, have to be
excluded. 
The result for the
normalized potential $\hat{V}(\xi)=V(\xi)/V(\sqrt{f/g})$ is shown
in Figure~\eqref{fig:potential}. 
The approximation used in Eq.~\eqref{approxpot} is remarkably robust.
Changing the number of neighboring points from four to six, or even
two, does not lead to a visible change in Figure~\ref{fig:potential}.
At the boundary to the tachyonic region the potential looses its
meaning and one has to address the problem of tachyon condensation.

The computation in the $ab$-sector goes as follows. The two stacks $a$ and $b$ intersect in the three tori, therefore in the internal magnetic field framework charged states have Landau levels
in the three tori. Having checked that the mass formula (\ref{master})
is valid in the effective field theory (though the eigenvectors are linear combination of the states in the Fock space spanned by Landau levels), one can
compute the scalar potential after diagonalizing the mass matrix. The states contributing are charged gauge vectors $A_{\mu}$, three complex scalar fields $\Phi_i$ and four Weyl fermions $\lambda, \Psi_i$,
where $i=1,2,3$. As shown in detail in Section \ref{sec:ab}, not all scalars in $\Phi_i$ are physical, some of them being absorbed by the massive Landau levels of gauge fields $A_{\mu}$. It is however simpler
to consider separately the two degrees of freedom in $A_{\mu}$ and the absorbed scalars in the computation. Then the various contributions to the scalar potential are
\begin{align}
A_{\mu} \ &: \; 2 \sum_{n_i} e^{- t \sum_i (2 n_i + 1)  g f_i}  =
\frac{1}{4\prod_i \sinh (g f_i t)} \ ,\nonumber \\ 
\Phi_i  \ &: \; 1 \sum_{n_j \not=n_i} e^{- t \sum_{j \not=i} (2 n_j
   + 1)  g f_j}  \sum_{n_i} \left( e^{- t (2 n_i - 1)  g f_i} + e^{- t
     (2 n_i + 3)  g f_i} \right) 
= \frac{\cosh (2 g f_i t)}{4\prod_i \sinh (g f_i t)} \ ,\nonumber \\  
\lambda  \ &: \; 1 \sum_{n_i} \left( e^{- 2t \sum_i n_i  g f_i}  + e^{ - 2 t \sum_i (n_i + 1)  g f_i} \right) 
= \frac{\cosh (g (f_1+f_2+f_3) t)}{4\prod_i \sinh (g f_it)}\ ,  \label{ab1}\\
 \Psi_1 \  &: \;  1 \sum_{n_i} \left( e^{- 2t (\sum_i n_i  g f_i +
     gf_1)} + e^{ - 2 t (\sum_i (n_i + 1)  g f_i -gf_1)}  \right) 
= \frac{\cosh (g (-f_1+f_2+f_3) t)}{4\prod_i 
\sinh (g f_i t)}  \ , \nonumber
\end{align}   
with contributions of $\Psi_2,  \Psi_3$ similar to the one of  $\Psi_1$ with  appropriate obvious modifications.  Adding all the contributions, taking into account of the opposite sign contributions of bosons
versus fermions and multiplying also by the multiplicity $I_{ab}$ of zero modes and  Landau levels, one gets
\begin{align}\label{ab2}
V_{ab} = - \frac{I_{ab}}{16 \pi^2} &\int_0^{\infty} \frac{dt}{t^3 }
\frac{1}{\prod_i \sinh (g f_i t)} \Big(  1 + \sum_i \cosh (2 g f_i t) \nonumber\\
&-\cosh (g (f_1+f_2+f_3) t)  -\cosh (g (-f_1+f_2+f_3) t) \nonumber\\
&-  \cosh (g (f_1-f_2+f_3) t) -  \cosh (g (f_1+f_2-f_3) t) \Big) \ . 
\end{align}  
By using standard identities one can rewrite the result into the form
\begin{align}
V_{ab} =    \frac{I_{ab}}{2 \pi^2} &\int_0^{\infty} \frac{dt}{t^3 }   
\frac{1}{\prod_i \sinh (g f_i t)}   \sinh \Big(\frac{g (f_1+f_2+f_3)
  t}{2}\Big) \label{ab3}\\  
&\times \sinh \Big(\frac{g (-f_1+f_2+f_3) t}{2} \Big)
 \sinh \Big(\frac{g (f_1-f_2+f_3) t}{2}\Big) \sinh \Big(\frac{g (f_1+f_2-f_3) t}{2}\Big)
   \ . \nonumber
\end{align}

Notice that the scalar potential vanishes if
\begin{equation}
f_1 \pm f_2 \pm f_3 = 0   \ . \label{ab4}
\end{equation}
Whenever one of the four equations  (\ref{ab4}) is fulfilled, supersymmetry is restored in the corresponding ($ab$ in our case) sector, in agreement with the arguments given at the end of Section 3. 
More precisely, if $f_1 \pm f_2 = 0, f_3 = 0$, the four-dimensional
effective theory has ${\cal N}=2$ supersymmetry, whereas when all
$f_i$ are non-vanishing but one of  the equations (\ref{ab4}) is satisfied, the effective theory has ${\cal N}=1$ supersymmetry.
This is not always manifest in the effective actions written in the previous sections, except for the cases when $f_1 + f_2 + f_3 = 0$ . The reason is that for the other cases of supersymmetry restoration,
the preserved supercharge generates multiplets misaligned to our
superfield expansion. Indeed, in the superfield expansion we used pre-assigned superpartners in an universal
way, whereas the supersymmetries preserved by the internal magnetic fields generically generate supermultiplets in a different way. While this could seem surprising at first sight, it  is standard
in extended supersymmetric theories (see, for example, \cite{susyalign}).
One test of residual supersymmetries in the compactified theory is the
boson-fermion degeneracy at each mass level. However, this is realized non-trivially, since the eigenvectors of the mass matrix mix
different Landau levels, as shown explicitly in previous sections. Notice that this discussion matches known results on supersymmetry preservation in D-brane models at angles 
\cite{Berkooz:1996km} and  the T-dual version
of type I/type II  strings with internal magnetic fields. However, as
far as we know, this subtlety of the effective action has never been discussed in detail in the string literature.

\subsection{String calculation}

From the string theory perspective, the scalar potential coming from various sectors is given by (minus) the cylinder partition function found by usual string quantization with appropriate boundary conditions, in the
internal magnetic picture, or equivalently, the T-dual intersecting brane one. In particular,
\begin{equation}
 V_{bc'} = - A_{bc'} \ . \label{st1}
 \end{equation}
Let us start with the scalar potential in the $bc'$-sector. The corresponding brane stacks are parallel in the second and the third torus and intersect in the first torus.  Standard formulae \cite{Angelantonj:2002ct} lead to the partition function 
\begin{equation}
A_{bc'} = \frac{I_{bc'}^{1}}{2 (4 \pi^2 \alpha')^2} \int_0^{\infty} \frac{d \tau_2}{\tau_2^3} \frac{(V_8-S_8)(\epsilon_1 \tau | \tau)}{\eta^6} \frac{2 i \eta}{\theta_1 (\epsilon_1 \tau | \tau)} P_{{\bf m_2 + {\bf \xi_2}}} 
P_{{\bf m_3 + {\bf \xi_3}}}  \ , \label{st2}
\end{equation}
where 
\begin{equation}
 P_{{\bf m_2 + {\bf \xi_2}}}  = \sum_{\bf m_2}  e^{- \pi \tau_2 \alpha' \sum_i (\frac {m_i}{R_i} + \xi_i)^2} \  \label{st3}
 \end{equation}
is the Kaluza-Klein sum along the second torus, with a similar expression for $P_{{\bf m_3 + {\bf \xi_3}}}$. The parameter $\epsilon_1$ is related to the angle between the stacks in the first torus according
to $\theta_{bc'}^1 = \pi \epsilon_1$. 
Various modular functions are defined in Appendix~\ref{app:jacobi}. 
In Eq.~(\ref{st2}) we also used the character
\begin{equation}
  (V_8-S_8)(\epsilon_1 \tau | \tau) \equiv \frac{ \theta_3^3 \theta_3 (\epsilon_1 \tau | \tau)   -  \theta_4^3 \theta_4 (\epsilon_1 \tau | \tau) -  \theta_2^3 \theta_2 (\epsilon_1 \tau | \tau)  }{2 \eta^4} 
=  \frac{ \theta_1^4 (\frac{\epsilon_1 \tau}{2} | \tau)  }{\eta^4} 
   \ , \label{st4}
 \end{equation}
where the last equality is the  Jacobi identity \eqref{ji}, and
$\theta_i=\theta_i(0|\tau)$. The modular parameter of the doubly covering torus of the cylinder is defined as
\begin{equation}
 q = e^{2 \pi i \tau} \ , \ \tau = \frac{i \tau_2}{2} \ , \label{st5}
 \end{equation}
and the relation with the Schwinger proper time of the field theory
computation is $t = \pi \tau_2 \alpha'$.  

The connection with the field theory computation is done by decoupling the charged open
string oscillators in the formulae above, while keeping the Kaluza-Klein states and the Landau levels. This is achieved in the $\tau_2 \to \infty$ limit of the modular functions, for example,
 \begin{equation}
   \theta_1  ({\epsilon_1 \tau} | \tau) \to 2 i \sinh \Big(\frac{\pi \epsilon_1 \tau_2}{2}\Big) q^{1/8}   \ , \label{st6}
 \end{equation}
which is valid for $|\epsilon_1|< 1/2$. 
Notice that the Wilson-line dependence of the field theory expression
is accurate in the large volume limit, $v_i  \gg \alpha' $.  Indeed,
in this limit, Kaluza-Klein states and Landau levels are much lighter
than the charged open string oscillators. It is important that the UV divergence of the amplitude/scalar potential, which arises even after summing over all sectors due to the NSNS tadpole generated
by the magnetic fields, is independent of the Wilson lines. The scalar potential can therefore be regulated by the Pauli-Villars method discussed in the previous paragraph. 

 The analogous expression for the amplitude $A_{bc}$ is easily found to be
\begin{equation}
A_{bc} = \frac{I_{bc}^{2} I_{bc}^{3} }{2 (4 \pi^2 \alpha')^2} \int_0^{\infty} \frac{d \tau_2}{\tau_2^3} \frac{(V_8-S_8)(\epsilon_2 \tau ; \epsilon_3 \tau  | \tau)}{\eta^4} 
\frac{2 i \eta}{\theta_1 (\epsilon_2 \tau | \tau)} \frac{2 i \eta}{\theta_1 (\epsilon_3 \tau | \tau)}  P_{{\bf m_1 + {\bf \xi_1}}} 
  \ , \label{st7}
\end{equation}
where one can now use the Jacobi identity \eqref{ji}, 
 \begin{align}
(V_8-S_8) (\epsilon_2 \tau ; \epsilon_3 \tau  | \tau) &\equiv
\frac{1}{2 \eta^4} \big(\theta_3^2  \theta_3 (\epsilon_2 \tau | \tau)  \theta_3(\epsilon_3 \tau | \tau)  \nonumber\\
&\hspace{1cm}-  \theta_4^2 \theta_4 (\epsilon_2 \tau |\tau) \theta_4 (\epsilon_2 \tau | \tau)  
-  \theta_2^2 \theta_2 (\epsilon_2 \tau | \tau) \theta_2 (\epsilon_3 \tau | \tau)\big)  \nonumber \\
& =  \frac{1}{\eta^4}\theta_1^2 \Big(\frac{(\epsilon_2+ \epsilon_3) \tau}{2} \Big| \tau\Big)  \theta_1^2 \Big(\frac{(\epsilon_2- \epsilon_3) \tau}{2} \Big| \tau\Big) \ . \label{st8}
 \end{align}

\begin{figure}[t]
\begin{center} 
\includegraphics[width = 0.7\textwidth]{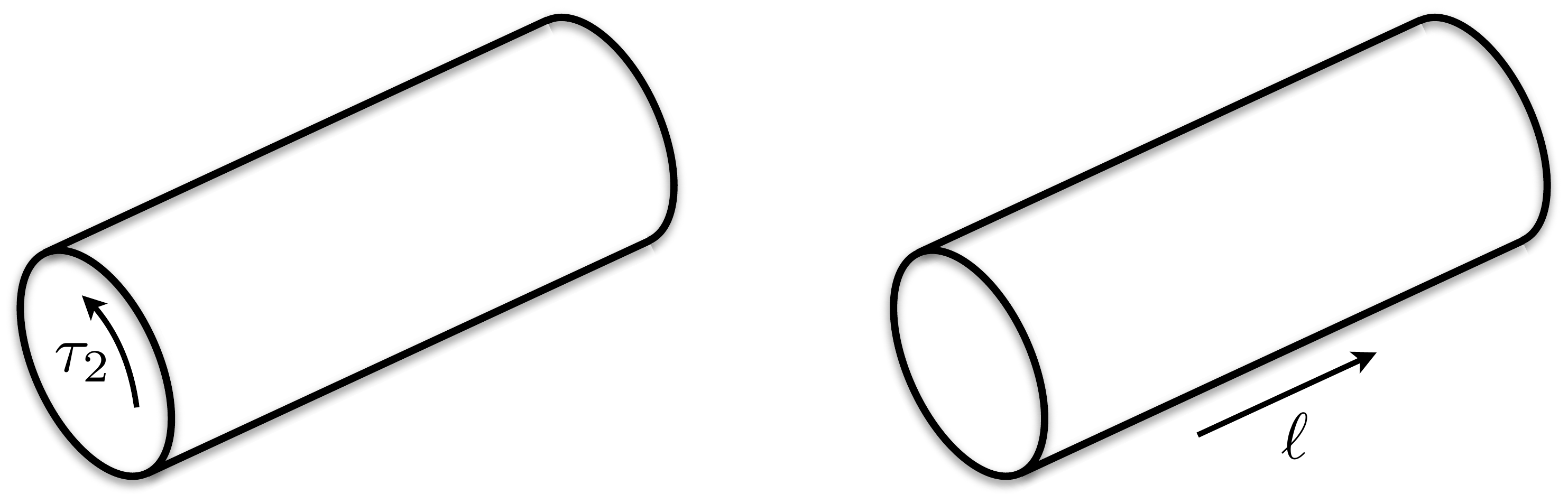}
\end{center}
\vspace{0.3cm}	
\caption{Left: loop-diagram for open string; right: tree-diagram for
  closed string.}
\label{cylinder}
\end{figure}

The stacks in the $ab$-sector intersect in all three tori. In this case, there are no standard Kaluza-Klein sums, but Landau levels in the three tori. The cylinder partition function reads
\begin{equation}
A_{ab} = \frac{I_{ab}^{1} I_{ab}^{2} I_{ab}^{3}}{2 (4 \pi^2 \alpha')^2} \int_0^{\infty} \frac{d \tau_2}{\tau_2^3} \frac{(V_8-S_8)(\epsilon_1 \tau ;  \epsilon_2 \tau ; \epsilon_3 \tau  | \tau)}{\eta^2} 
\prod_{i=1}^3 \frac{2 i \eta}{\theta_1 (\epsilon_i \tau | \tau)} 
  \ , \label{st9}
\end{equation}
 which can again be simplified with the help of the Jacobi identity \eqref{ji}, 
  \begin{align}
(V_8-S_8) (\epsilon_1 \tau ; \epsilon_2 \tau ; \epsilon_3
\tau|\tau) &\equiv \frac{\theta_3  \prod_{i=1}^3 \theta_3 (\epsilon_i \tau | \tau)   -  \theta_4 \prod_{i=1}^3  \theta_4 (\epsilon_i \tau | \tau)  -
  \theta_2 \prod_{i=1}^3   \theta_2 (\epsilon_i \tau | \tau)}{2\eta^4}\nonumber \\
&= -\frac{1 }{\eta^4}\theta_1\Big(\frac{(\epsilon_1 + \epsilon_2+
  \epsilon_3) \tau}{2} \Big| \tau\Big)   \theta_1 \Big(\frac{(-\epsilon_1 + \epsilon_2+ \epsilon_3) \tau}{2} \Big| \tau\Big) \nonumber\\
&\hspace{1cm} \times \theta_1 \Big(\frac{(\epsilon_1 - \epsilon_2+ \epsilon_3) \tau}{2} \Big| \tau\Big)    \theta_1 \Big(\frac{(\epsilon_1 + \epsilon_2- \epsilon_3) \tau}{2} \Big| \tau\Big)  
 \ . \label{st10}
 \end{align}
 Notice that the potential vanishes whenever 
 \begin{equation}
 \epsilon_1 \pm \epsilon_2 \pm \epsilon_3 = 0 \ , \label{st09}
  \end{equation}
 which encode the standard condition for supersymmetry restoration
 (see Eq.~\eqref{susyangles}),
\mbox{$\theta^1 \pm \theta^2 \pm \theta^3 = 0$}, as explained in  \cite{Berkooz:1996km}. 
   
After taking the field theory limit and by introducing Pauli-Villars
regulators for the UV part of the potential and using the field theory Schwinger proper time $t$, one finds the scalar
 potential 
\begin{align}\label{Vab}
V_{ab} =  \frac{I_{ab}^{1} I_{ab}^{2} I_{ab}^{3}}{2 \pi^2}
&\int_0^{\infty} \frac{d t}{t^3} \left(1 -c_1 e^{- \mu_1^2 t} -c_2 e^{- \mu_2^2 t} \right)\\
&\times\frac{  \sinh\big( \frac{ (\epsilon_1 + \epsilon_2+ \epsilon_3) t }{4}\big)   
  \sinh \big(\frac{ (-\epsilon_1 + \epsilon_2+ \epsilon_3) t }{4} \big) \sinh \big(\frac{ (\epsilon_1 - \epsilon_2+ \epsilon_3) t }{4}\big)  \sinh\big( \frac{ (\epsilon_1 + \epsilon_2- \epsilon_3) t }{4}\big)}
 {\sinh \big(\frac{ \epsilon_1  t }{2}\big)  \sinh \big(\frac{ \epsilon_2  t }{2}\big)   \sinh \big(\frac{ \epsilon_3  t }{2}\big)  }   \ , \nonumber
 \end{align}
 where $c_1 + c_2 = 1, c_1 \mu_1^2+ c_2  \mu_2^2= 0$.  The
 non-regularized potential 
matches, by using the field theory limit $\epsilon_i \to
2 g f_i$, the field theory result  (\ref{ab3}).
 As is well-known, the one-loop cylinder string partition functions can be also written, after a modular transformation, as a tree-level propagation
 of closed strings between two stacks of branes (see Figure \ref{cylinder}). This open-closed string duality is crucial for the consistency of the string theory partition functions. However, after taking the field theory limit and decoupling
 the open string massive oscillators, the field theory scalar potentials do not feature this duality. As a consequence, we choose for brevity to not write the scalar potentials in this dual formulation, which would
 otherwise be crucial for the full fledged string theory formulation.   
 
\begin{figure}
\begin{center} 
\includegraphics[width = 0.6\textwidth]{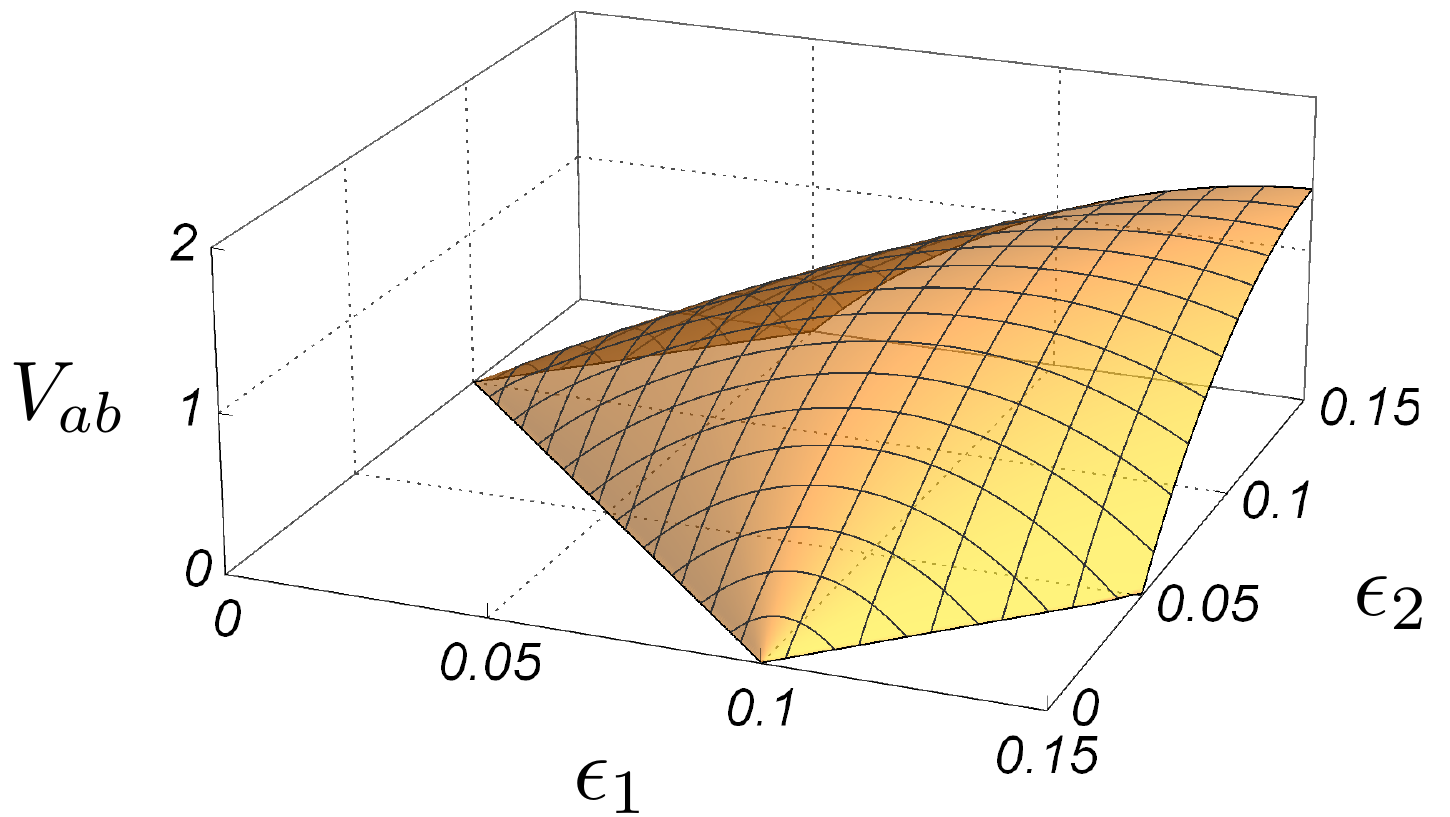}\\
\vspace{0.3cm}
\includegraphics[width = 0.6\textwidth]{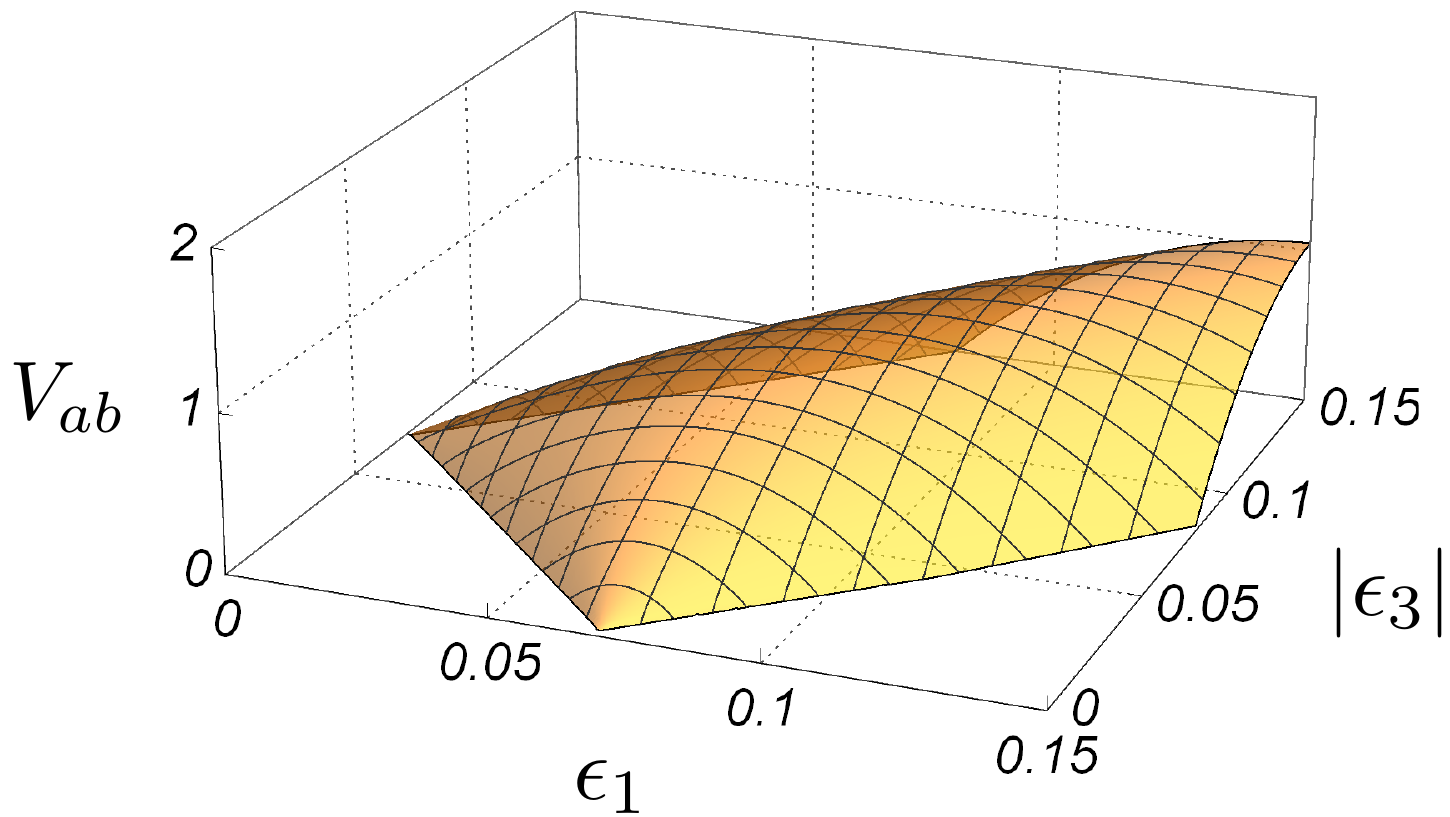}\\
\vspace{0.3cm}
\includegraphics[width = 0.6\textwidth]{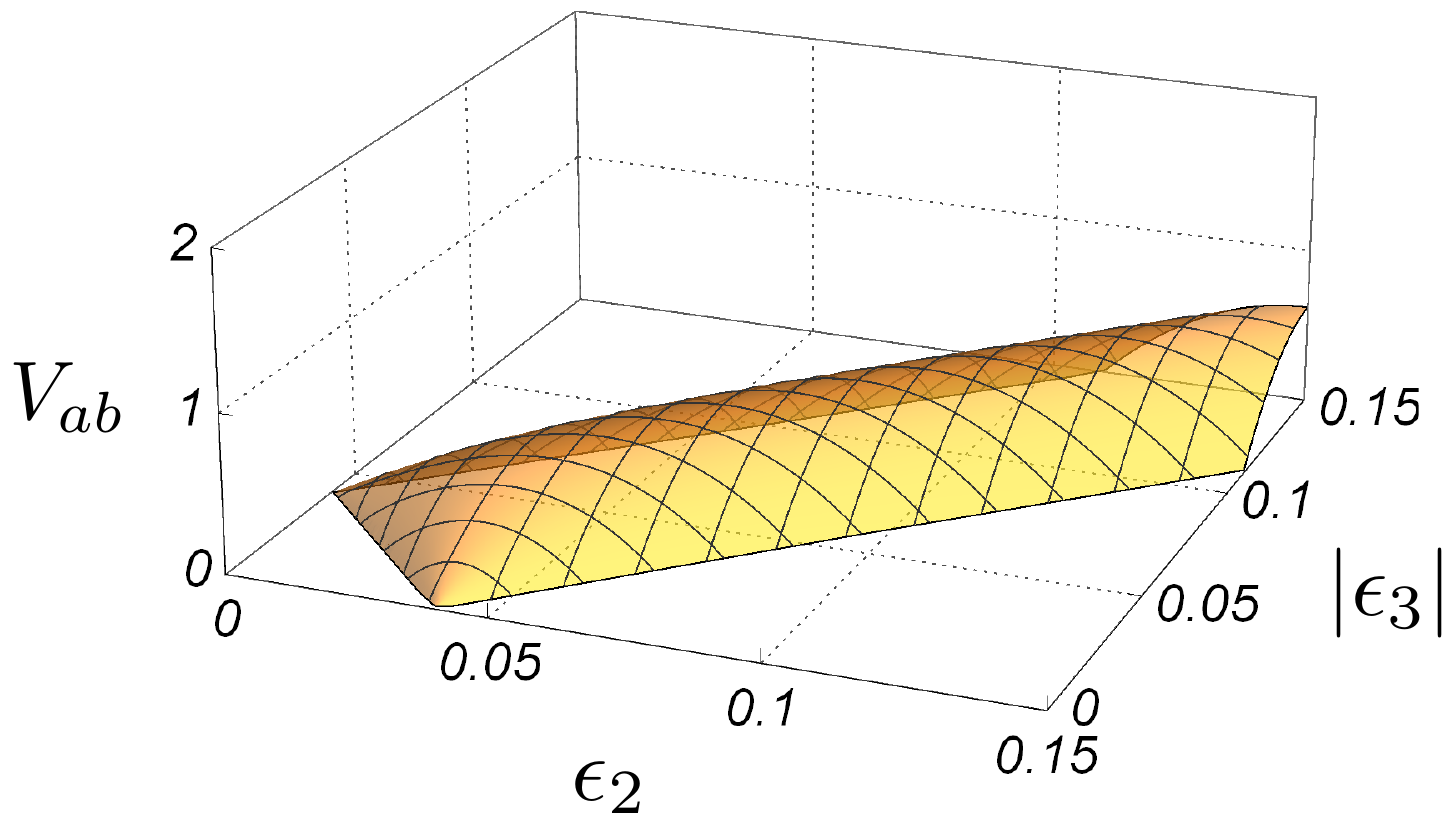}
\end{center}
\vspace{0.3cm}	
\caption{One-loop potential $V_{ab}$ for three slices in the
  three-dimensional space of the volume moduli $v_i$ of the three
2-tori $T^2_i$, with $\tan{\pi\e_i} \propto 1/v_i$. The slices are
defined by $\e_3 = \hat{\e}_3$ (top), $\e_2 = \hat{\e}_2$ (middle), and 
$\e_1 = \hat{\e}_1$ (bottom), where 
$(\hat{\e}_1, \hat{\e}_2, \hat{\e}_3) = (0.04,0.07,-0.1)$ is a point in
the tachyon-free region of moduli space. The potential
(arbitrary units) is evaluated numerically for an ultraviolet cutoff
$\delta^{-1} = 10^3$ and Pauli-Villars regulator masses $\mu_1^2 = 75$,
$\mu^2_2 = 25$.}
\label{fig:moduli}
\end{figure}

\subsection{Volume-moduli potential}

The effective potential \eqref{Vab} depends on the parameters of
$\e_i$. In the D-brane model they represent the brane
intersection angles, $\epsilon_i = \theta^i/\pi$ and in the T-dual
magnetic compactification they correspond to the torus volumes $v_i$,
with $\tan{\pi\epsilon_i} = m^i \rho_i = 4\pi^2\alpha' m^i/v_i$.

Consider first the case with vanishing flux in the first torus, which
is the case in the sectors $aa'$ and $bc$. The effective potential can
be obtained from Eq.~\eqref{Vab} by setting $\e_1=0$, which yields
\begin{equation}
\begin{split}
V_{aa'} &\propto -  \int_0^{\infty} \frac{d t}{t^3}  
\left(  1 - c_1 e^{- \mu_1^2 t} -  c_2 e^{- \mu_2^2 t} \right)
\frac{  \sinh^2 \big(\frac{ (\epsilon_2+ \epsilon_3) t }{4} \big)  
  \sinh^2 \big(\frac{ (\epsilon_2-\epsilon_3) t }{4}\big)  }
 {\sinh \big(\frac{ \epsilon_2  t }{2}\big)   \sinh \big(\frac{ \epsilon_3  t }{2}\big)  }   
  \ .
\end{split}
\end{equation}
On the line $\epsilon_2=\epsilon_3$ in moduli space (see
Figure~\ref{fig:modulispace}) the potential $V_{aa'}$
vanishes. However, as one easily verifies, for $\epsilon_2 \neq
\epsilon_3$ the  potential has an infrared divergence and approaches
$-\infty$. Hence, due to the existence of a tachyon for $\epsilon_2 \neq
\epsilon_3$, the line $\epsilon_2=\epsilon_3$ is unstable.

We can also evaluate the integral $V_{ab}$ for non-zero fluxes in all
three tori, and therefore no Wilson lines. In string theory, the result is UV divergent due to NSNS tadpoles which require a vacuum redefinition that is very challenging to perform 
explicitly \cite{Dudas:2004nd}.

In our field theory approach, the potential can be regulated
a l\`a Pauli-Villars, but now the result will depend on the regulator masses. We have checked numerically that for $\e_i \ll
\mu^2_{1,2} \ll 1/\delta$, where $1/\delta$ is the ultraviolet cutoff,
variation of $\mu^2_{1,2}$ essentially changes the normalization of
the potential and not the shape. Figure~\ref{fig:moduli} shows the
potential $V_{ab}$ for three slices of moduli space defined by 
$\e_3 = \hat{\e}_3$, $\epsilon_2 = \hat{\epsilon}_2$ and 
$\epsilon_1 = \hat{\epsilon}_1$, where 
$(\hat{\epsilon}_1, \hat{\epsilon}_2, \hat{\epsilon}_3) = (0.04,0.07,-0.1)$
is one allowed point in moduli space (see Figure~\ref{fig:moduli}).
At the boundary of the tachyon-free region the potential vanishes. The
figure clearly illustrates that the system is always driven to the
tachyonic region in moduli space. The same conclusion has previously
been reached in a related discussion in \cite{Blumenhagen:2001te} from
the viewpoint of the disc level scalar potential. This suggests that a
stabilization mechanism for the volume moduli is needed at or above
the compactification scale.

\section{Tachyon condensation}
\label{sec:tachyon}

Most sectors of the considered model have potentially tachyonic
charged scalars. A frequent assumption is that such tachyonic
instabilities can be avoided by means of Wilson lines. However, as we
demonstrated in the previous section for the $bc'$-sector, the
one-loop Wilson-line potential has no stable extrema and the system is
therefore driven to the tachyonic regime. For zero Wilson lines
tachyon condensation takes place. This is interpreted as brane-brane recombination
and it is expected that tachyon condensation restores supersymmetry,
at least partially (see, for example, \cite{Sen:2004nf,Hashimoto:2003xz,Epple:2003xt}).
In the following, we shall address for the first time tachyon
condensation in a compact space.

\subsection{$bc'$-sector}

The situation is particularly simple in the $bc'$-sector. According to
Eq.~\eqref{tachyonbc'} the field $\Phi^-_{0,00} = \phi^{1-}_{0,00}$
has a negative mass squared. The interesting question is whether its
condensation can restore supersymmetry. Inspection of \eqref{L4dbc'}
shows that the relevant $F$- and $D$-terms are given by (for
simplicity we restrict ourselves to $\kk=\kk'=0$),
\begin{align}
-\bar{F}^+_{n} &= |M_{n,00}|\ \Phi^-_{n}\ ,\\
-2D_{1,2} &=4f+g\sum_{n}\big(|\phi^+_{n}|^2 + |\Phi^+_{n}|^2 +
|\Xi^+_{n}|^2 - |\phi^-_{n}|^2 -
|\Phi^-_{n}|^2 - |\Xi^-_{n}|^2\big)\ ,\\
-\sqrt{2} D^+_{n} &=|M_{n}|\ \Xi^+_{n}
- |M_{n+1}|\ \overline{\Xi}^-_{n}\ .
\end{align}
The equation $D^+_{n} = 0$ is easily satisfied by
$\Xi^+_{n} = \Xi^-_{n} = 0$. The crucial point is that because of
$|M_{0,00}|=0$, the field $\Phi^-_{0}=\phi^{1-}_0$ decouples from the
superpotential, and therefore  $\bar{F}^+_{0} = 0$. Setting 
$\phi^+_{n} = \Phi^+_{n} = \Xi^+_{n} = \phi^-_{n} =
\Phi^-_{n+1} = \Xi^-_{n} = 0$, $D_{1,2} = 0$ can be satisfied by
$\phi^{1-}_0 = \sqrt{2f/g}$, and supersymmetry is restored. The D-term scalar
potential
\begin{align}
V_D = \frac{g}{4} \big(4f - |\phi^{1-}_0|^2\big)^2\ ,
\end{align}
yielding the tachyonic mass squared $-2gf$, in agreement with Eq.~\eqref{tachyonbc'}.

According to Eqs.~\eqref{yuk1}, \eqref{adj}, \eqref{expfi} and
\eqref{Wyuk1}, a vev of $\phi^{1-}_0$ leads to masses for all chiral fermions,
\begin{align}
\mathcal{L}_{mass} \propto y |\phi^{1-}_0| \Big(\sum_{j=1}^{3(l-2)} 
\bm{\bar{N}}^j_{1,0} \bm{N}^j_{0,1}  +
 \sum_{j=1}^{l+2} \bm{\bar{N}}^j_{0,1} \bm{N}^j_{1,0} \Big)\ ,
\end{align}
where $j$ labels the ground state wave functions.
Hence, after tachyon condensation, all fermions have masses of order $\sqrt{gf}$.

\subsection{$bc$-sector}

This sector is very similar to the $aa'$-sector, since the flux
vanishes in the first torus. However, an important difference is the
sign of the flux densities. In the $aa'$-sector one has positive flux
densities in the second and the third torus.
On the contrary, in the $bc$-sector the two flux densities
have opposite sign.
Taking this into account,  the relevant $F$- and $D$-terms can be
essentially read off from Eq.~\eqref{Laa'basic}. One finds, before forming
linear combinations for mass eigenstates,
\begin{align}
\bar{F}^{1+-}_{n,n'} &=
\sqrt{2gf_2n}\ \phi^{3-+}_{n-1,n'}-\sqrt{2gf_3(n'+1)}\ \phi^{2-+}_{n,n'+1}\
,\\
\bar{F}^{1-+}_{n,n'} &=
-\sqrt{2gf_2(n+1)}\ \phi^{3+-}_{n+1,n'}+\sqrt{2gf_3n'}\ \phi^{2+-}_{n,n'-1}\
,\\
-(D_1-D_2) &= f_2-f_3 +g\sum_{n,n'}\big(|\phi^{i+-}_{n,n'}|^2
-|\phi^{i-+}_{n,n'}|^2\big)\ ,\\
D^{-+}_{n,n'} &= \sqrt{gf_2}\big(\sqrt{n}\ \ophi^{2+-}_{n-1,n'} 
-\sqrt{n+1}\ \phi^{2-+}_{n+1,n'}\big) \nonumber\\
&\quad-\sqrt{gf_3}\big(\sqrt{n'+1}\ \ophi^{3+-}_{n,n'+1}
-\sqrt{n'}\ \phi^{3-+}_{n,n'-1}\big)\ .
\end{align}
Similar to the $bc'$-sector, now the fields $\phi^{2-+}_{0,0}$ and
$\phi^{3+-}_{0,0}$ decouple from the superpotential. Setting all other
fields to zero, $F^{1+-}_{n,n'}$, $F^{1-+}_{n,n'}$ and $D^{-+}_{n,n'}$
vanish and one is left with
\begin{align}
-(D_1-D_2) &= f_2-f_3 +g\big(|\phi^{3+-}_{0,0}|^2
-|\phi^{2-+}_{0,0}|^2\big)\ .
\end{align}
Depending on the sign of $f_2-f_3$, $D_1-D_2=0$ is achieved for a vev
of $\phi^{3+-}_{0,0}$ or $\phi^{2-+}_{0,0}$. Hence, as in the
$bc'$-sector, tachyon condensation restores supersymmetry. However,
according to Eq.~\eqref{yuk1}, these vev's do not generate mass terms
for chiral fermions. In the special case $f_2=f_3$, there are two
massless scalars and no tachyon condensation takes place.

Tachyon condensation in the $aa'$-sector is more complicated since the
$SU(N)$ D-terms and the superpotential couple the antisymmetric tensor
to chiral fields in the adjoint representation of $SU(N)$. Also Wilson
lines of the $U(1)_a$ gauge group have to be taken into account. This
allows for more complicated solutions of the $F$- and $D$-term
equations. Tachyon condensation involves fields of order $\sqrt{f/g}$.
Hence, the couplings between the various sectors by $D$- and $F$-terms
have to be taken into account in a complete analysis of the vacuum structure.

\section{Conclusions and open questions}
\label{sec:conclusion}

We have studied the effective field theory for an intersecting D-brane
model and its T-dual magnetic compactification, which has all features
wanted for extensions of the Standard Model with high-scale
supersymmetry breaking: the model has a `matter sector' with chiral
fermions, broken supersymmetry and massive scalars, and a `Higgs
sector' with vector-like fermions. For certain choices of fluxes,
in some sectors scalars are massless and supersymmetry is partially
preserved. Expectation values of Higgs scalars can give mass to the
chiral fermions. In general it is assumed that tachyons in the Higgs
sector can be avoided by means of Wilson lines. All these features are
well known from phenomenological
applications in the literature (see, for example, \cite{Ibanez:2001nd,
Antoniadis:2006eb}). 

The considered model is also
representative at the technical level. The different sectors are examples
of the three possibilities for background gauge fields, with flux in
one torus and Wilson lines in the other two, flux in two tori and
Wilson lines in one torus, and flux in three tori. The magnetic flux
mixes the towers of Landau levels, yielding also massless Goldstone
bosons that give mass to vector fields via the St\"uckelberg mechanism.
Physical 4d fields are linear combinations of fields from different Landau
levels. For each mass level the counting of bosonic and fermionic
states is consistent with the string mass formula.

The scalar masses depend on moduli, i.e., Wilson lines and the volume
moduli of the three tori. One of the main results of this paper is the
computation of
the one-loop effective potential for Wilson lines in the `Higgs
sector' based on the effective 4d field theory. 
Summing over the tower of Landau levels leads to a result which is
consistent with the 
string cylinder amplitude in the field theory limit. 
It turns out that the computation of the string amplitude is very convenient
to obtain the one-loop potential, and in this way we have therefore
evaluated the contributions of all sectors of the model to the
effective potential.

Notice, that in string theory, whenever the magnetic fluxes break supersymmetry, there are NSNS tadpoles that generate divergences. 
These divergences, that are UV from the loop viewpoint, are actually IR from the viewpoint of the tree-level gravitational exchange. Their existence implies that the computation is not performed in the
right vacuum, that has to be redefined (see, for example,
\cite{Dudas:2004nd}), which is technically very challenging (for
recent progress, see,  for example,  \cite{deLacroix:2017lif}).
This does not affect  the Wilson-line potential, since the divergence is independent of the Wilson lines. Our field theory approach with Pauli-Villars regulators allowed us to analyze also
the dependence of the potential on the volume moduli. We find the
expected instability of the perturbative vacuum. However, a more
detailed study is needed to obtain a definite result on the potential vacuum instability.

The one-loop Wilson-line potential in the Higgs sector is
concave. There are no stable extrema and the system is therefore
driven to the tachyonic regime. We showed that for vanishing Wilson
lines tachyon condensation indeed takes place, and the corresponding
vacuum expectation value gives masses to all chiral fermions of the
order of the compactification scale. It is quite
possible, however, that in other models some chirality remains after tachyon
condensation.

As we have seen, tachyon condensation in the Higgs sector 
restores supersymmetry. It is important to extend the first analysis in
this paper to all sectors of the model, since the restoration of
supersymmetry is closely related to the vacuum energy density and the
stability, or possibly metastability, of the model. Given the
phenomenological virtues of magnetic compactifications and
intersecting D-brane models, it appears mandatory to further pursue
these questions.

\section*{Acknowledgments}
We thank Ralph Blumenhagen, Luis Ib\'a\~nez, C.~S.~Lim, Dieter L\"ust, Hans-Peter Nilles,
Augusto Sagnotti and especially Markus Dierigl for valuable
discussions. E.D. was supported in part by the ``Agence Nationale de
la Recherche" (ANR). Y.T. is supported in part by Grants-in-Aid for JSPS Overseas Research Fellow (No.~18J60383) from the Ministry of Education, Culture, Sports, Science and Technology in Japan.

\begin{appendix}

\section{Embedding $U(N)$ into $SO(2N)$}
\label{app:N2N}

In Section~2 and Section~3 we discussed an intersection D-brane model
with gauge group $U(14)\times U(1)\times U(1)$  and a T-dual type I
string compactification on a magnetized torus, respectively. The
connection becomes particularly transparent if one uses step
generators for the $U(16)$ subgroup of $SO(32)$. In this appendix we
collect some formulae which extend the step generators of a $U(N)$
algebra to an $SO(2N)$ algebra by adding  generators that transform
as the antisymmetric complex representation of $U(N)$.

The $N^2$ generators of $U(N)$ are given by matrices $\hat{T}_{\A\B}$
that transform as $N\otimes\oN$, 
\begin{align}\label{step}
\left(\hat{T}_{\A\B}\right)_{\A'\Bb'} = \D_{\A'\A} \D_{\B\Bb'} \ .
\end{align}
Note that the $\hat{T}_{\A\B}$ are not hermitian but satisfy the relation
\begin{align}
{\hat{T}_{\A\B}}^T = \hat{T}_{\B\A}\ .
\end{align}
The step generators are related to $N(N+1)/2$ symmetric hermitean
generators $\hat{T}^1_{\A\B}$
and $N(N-1)/2$ antisymmetric hermitian generators $\hat{T}^2_{\A\B}$ by
\begin{align}\label{T12}
\hat{T}^1_{\A\B} = \hat{T}_{\A\B} + \hat{T}_{\B\A} \ , \quad \hat{T}^2_{\A\B} =
i\left(\hat{T}_{\A\B} - \hat{T}_{\B\A}\right)\ .
\end{align}
Infinitesimal $U(N)$ transformations of the  fundamental
representation $\psi\sim N$ read
\begin{align}\label{trafoT1}
\delta \psi = 
i\left(\E_{\A\B} \hat{T}_{\A\B} + 
  \E^*_{\A\B}\hat{T}_{\B\A}\right)\psi =
i\left(\E^1_{\A\B} \hat{T}^1_{\A\B} + 
  \E^2_{\A\B}\hat{T}^2_{\A\B}\right)\psi\ ,
\end{align}
where $\E_{\A\B} = \E^1_{\A\B} + i\E^2_{\A\B}$. Note that
$\E^1_{\A\B} \hat{T}^1_{\A\B}$ and $\E^2_{\A\B}\hat{T}^2_{\A\B}$ are symmetric and
antisymmetric $N\times N$ matrices, respectively.
An infinitesimal transformation of the complex conjugate
representation $\overline{\psi} \sim \oN$ reads
\begin{align}\label{trafoT2}
\delta \opsi = 
-i\left(\E^*_{\A\B} \hat{T}_{\A\B} + 
  \E_{\A\B}\hat{T}_{\B\A}\right)\opsi\ =
-i\left(\E^1_{\A\B} {\hat{T}^1}_{\A\B} + 
  \E^2_{\A\B}{\hat{T}^2}_{\A\B}\right)\opsi\ .
\end{align}
The step generators satisfy the commutator relations
\begin{align}
[\hat{T}_{\A\B},\hat{T}_{\G\D}] = \D_{\B\G} \hat{T}_{\A\D} - \D_{\D\A}
\hat{T}_{\G\B}\ ,
\end{align}
and are normalized as
\begin{align}
\text{tr}\left(\hat{T}_{\A\B}\right) = \D_{\A\B}\ , \quad
\text{tr}\left({\hat{T}_{\A\B}}^T \hat{T}_{\G\D}\right) = \D_{\A\G}
\D_{\B\D}\ .
\end{align}

The $N\times N$ matrices $\hat{T}_{\A\B}$ and $-\hat{T}_{\B\A}$
can be combined into
$2N\times 2N$ matrices
\begin{equation}\label{nohatT}
T_{\A\B} = 
\begin{pmatrix}
\hat{T}_{\A\B} & 0 \\ 0 & -\hat{T}_{\B\A}
\end{pmatrix}
= 
\begin{pmatrix}
\hat{T}_{\A\B} & 0 \\ 0 & -\hat{T}^{\ T}_{\A\B} 
\end{pmatrix} \ ,
\end{equation}
which act on the $2N$-component vector
\begin{align}
\Psi = \begin{pmatrix} \psi \\ \opsi \end{pmatrix} \ .
\end{align}
Note that
\begin{align}
\text{tr}\left(T_{\A\B}\right) = 0\ , \quad
\text{tr}\left({T^{}_{\A\B}}^\dagger T_{\G\D}\right) = 2\D_{\A\G}
\D_{\B\D}\ .
\end{align}
The generators $T_{\A\B}$ satisfy the same algebra as the generators
$\hat{T}_{\A\B}$, 
\begin{align}\label{K1}
[T_{\A\B},T_{\G\D}] = \D_{\B\G} T_{\A\D} - \D_{\D\A}
T_{\G\B}\ ,
\end{align}
and the corresponding $SO(2N)$ transformations read
\begin{align}\label{trafoT3}
\delta \Psi = i\left(\E^1_{\A\B} T^1_{\A\B} + 
  \E^2_{\A\B}T^2_{\A\B}\right)\Psi\ .
\end{align}

The generators of $SO(2N)/U(N)$ form a complex antisymmetric tensor of
$U(N)$. They can be chosen as
\begin{align}
X^+_{\G\D} =
\begin{pmatrix}
0 & \hat{X}_{\G\D} \\ 0 & 0
\end{pmatrix}\ , \quad
X^-_{\G\D} = 
\begin{pmatrix}
0 & 0 \\ -\hat{X}_{\G\D} & 0 
\end{pmatrix}\ , 
\end{align}
where
\begin{align}\label{anti}
(\hat{X}_{\G\D})_{\G'\D'} = \D_{\G\G'}\D_{\D\D'}-\D_{\G\D'}\D_{\D\G'}\ ,
\end{align}
with
\begin{align}
\hat{X}_{\G\D} = -{\hat{X}_{\G\D}}^T =  - {\hat{X}_{\G\D}}^\dagger = -  \hat{X}_{\D\G}  \ .
\end{align}
The generators $X^{\pm}_{\G\D}$ satisfy the relations
\begin{align}
{X^+_{\G\D}}^\dagger X^-_{\e\R} = 
{X^-_{\G\D}}^\dagger X^+_{\e\R} = 0 \ ,
\end{align}
and are normalized as
\begin{align}
\text{tr}\left({X^{\pm}_{\G\D}}^\dagger X^{\pm}_{\e\R}\right)
= 2(\D_{\G\e}\D_{\D\R} - \D_{\G\R}\D_{\D\e}) \ .
\end{align}
Together with $T_{\A\B}$ 
they form a closed algebra,
\begin{equation}\label{algebra}
\begin{split}
[T_{\A\B},X^+_{\G\D}] &= \D_{\B\G} X^+_{\A\D}
+ \D_{\B\D} X^+_{\G\A}\ ,\\
[T_{\A\B},X^-_{\G\D}] &= -\D_{\A\G} X^-_{\B\D}
- \D_{\A\D} X^-_{\G\B}\ , \\
[X^+_{\G\D}, X^+_{\e\R}] &= [X^-_{\G\D},
X^-_{\e\R}] = 0\ ,\\
[X^+_{\G\D}, X^-_{\e\R}] &=
\D_{\G\e} T_{\D\R} - \D_{\D\e} T_{\G\R} +
\D_{\D\R} T_{\G\e} - \D_{\G\R} T_{\D\e}\ .
\end{split}
\end{equation}
The corresponding $SO(2N)$ transformations read
\begin{align}\label{trafoX}
\delta \Psi = i\left(\tilde{\E}_{\G\D} X^+_{\G\D}
+\tilde{\E}^*_{\G\D} X^-_{\G\D}\right)\Psi =
i\left(\tilde{\E}^1_{\G\D} X^1_{\G\D}
+\tilde{\E}^2_{\G\D} X^2_{\G\D}\right)\Psi \ ,
\end{align}
where $\tilde{\E}_{\G\D} = \tilde{\E}^1_{\G\D} + i\tilde{\E}^2_{\G\D}$ and
\begin{equation}
\begin{split}
X^1_{\G\D} &= X^+_{\G\D} + X^-_{\G\D} =
\begin{pmatrix}
0 & \hat{X}_{\G\D} \\ -\hat{X}_{\G\D} & 0
\end{pmatrix}\ , \\
X^2_{\G\D} &= i \left(X^+_{\G\D} - X^-_{\G\D} \right)
= i 
\begin{pmatrix}
0 & \hat{X}_{\G\D} \\  \hat{X}_{\G\D} & 0
\end{pmatrix}\ . 
\end{split}
\end{equation}

From Eqs.~\eqref{trafoT3} and
\eqref{trafoX} one concludes that a general $SO(2N)$
transformation is given by the $2N\times 2N$ matrix
\begin{align}\label{X}
X = 
\begin{pmatrix}
S + iA_3 & A_1 + iA_2 \\ -A_1 +iA_2 & -S+iA_3
\end{pmatrix}\ .
\end{align}
Here $S = \E^1_{\A\B} \hat{T}^1_{\A\B}$ is a real symmetric $N\times N$ matrix,
and $A_3=-i\E^2_{\A\B}\hat{T}^2_{\A\B}$, $A_1=\tilde{\E}^1_{\G\D} \hat{X}_{\G\D}$ and $A_2
= \tilde{\E}^2_{\G\D} \hat{X}_{\G\D}$ are real antisymmetric $N\times N$ matrices.
This can be compared to the standard form of $SO(2N)$ generators \cite{Georgi:1982jb}
\begin{align}
\lambda = -i
\begin{pmatrix}
\eta_1 & \rho \\ -\rho^T & \eta_2
\end{pmatrix}\ 
= - \lambda^T \ ,
\end{align}
where $\eta_1$ and $\eta_2$ are antisymmetric real $N\times N$
matrices and $\rho$ is an arbitrary real $N\times N$ matrix. After a unitary
transformation,
\begin{align}\label{u1}
U = \frac{1}{\sqrt{2}}
\begin{pmatrix}
I & -i I \\  I & i I 
\end{pmatrix}\ ,
\end{align}
one obtains
\begin{align}\label{lambda}
\lambda' = U \lambda U^\dagger = \frac{1}{2}
\begin{pmatrix}
\rho + \rho^T - i (\eta_1+\eta_2) & -(\rho - \rho^T) -i (\eta_1 - \eta_2) \\ 
  (\rho - \rho^T) - i (\eta_1 - \eta_2) &  -(\rho + \rho^T) - i (\eta_1+\eta_2) 
\end{pmatrix} \ .
\end{align}
This expression for $\lambda'$ agrees with the one for $X$ in
Eq.~\eqref{X} with $S=(\rho+\rho^T)/2$, $A_3 = -(\eta_1+\eta_2)/2$,
$A_1 = -(\rho-\rho^T)/2$ and $A_2 = -(\eta_1-\eta_2)/2$. 

Notice that the transformation \eqref{u1} is also diagonalizing the magnetic flux. Indeed, in the $SO(2N)$ basis, the magnetic flux is of the type
\begin{equation}
\langle F \rangle =  \begin{pmatrix}
0 &  I \\  -I & 0 
\end{pmatrix} \ .  \label{u2}
\end{equation}
After the unitary transformation, the flux becomes
\begin{equation}
U \langle F \rangle U^\dagger = \begin{pmatrix}
i I & 0 \\  0 & - i I 
\end{pmatrix} \ . \label{u3}
\end{equation}

\section{Commutators}
\label{app:commutators}
In Sections~2--5 we have considered the groups $G = SO(2(N+2))
\supset U(N) \times U(1) \times U(1) = H$, and in Eqs.~\eqref{expV},
\eqref{expfi} and \eqref{expfib} we have expanded vector, chiral and
anti-chiral superfields in terms of $SO(2(N+2))$ generators, with the
identifications (cf.~\eqref{adj}),
\begin{equation}\label{u1s}
H_0 = \frac{1}{\sqrt{N}}  T_{\alpha\alpha} , \;
H_1 = T_{N+1,N+1} , 
H_2 = T_{N+2,N+2} ,\; T_{\A\B} = \tilde{T}_{\A\B} +
\frac{1}{\sqrt{N}}\D_{\A\B}H_0\  
\end{equation}
for generators of $H$ and
\begin{align}\label{commutators}
T^{-0}_\A &= T_{\A,N+1} , \; T^{0-}_\A = T_{\A,N+2} , \;
 T^{+0}_\A = T_{N+1,\A} , \; T^{0+}_\A = T_{N+2,\A} , \;
T^{+-} = T_{N+1,N+2} , \nonumber\\
T^{-+} &= T_{N+2,N+1} , \; X^{+0}_\A = X^+_{\A,N+1} = -X^+_{N+1,\A} ,\; 
X^{0+}_\A = X^+_{\A,N+2} = -X^+_{N+2,\A},\nonumber\\
X^{-0}_\A &= X^-_{N+1,\A} = -X^-_{\A,N+1},\; X^{0-}_\A = X^-_{N+2,\A} = -X^-_{\A,N+2},\;\nonumber\\
X^{++} &= X^+_{N+1,N+2} = - X^+_{N+2,N+1} ,\; 
X^{--} = X^-_{N+2,N+1} = - X^-_{N+1,N+2} \ ,
\end{align}
for generators of $G/H$.

Non-vanishing commutators needed in Sections~3~-~5 include
\begin{align}
[H_0,T^{\mp 0}_\A] &= \pm \frac{1}{\sqrt{N}} T^{\mp 0}_\A ,\; 
[H_1,T^{\mp 0}_\A] = \mp T^{\mp 0}_\A ,\;
[T^{- 0}_\A, T^{+ 0}_\B] = T_{\A\B} - \D_{\A\B} H_1 , \label{comab} \\
[H_0,T^{0 \mp}_\A] &= \pm \frac{1}{\sqrt{N}} T^{0 \mp}_\A ,\; 
[H_2,T^{0 \mp}_\A] = \mp T^{0 \mp}_\A ,\;
[T^{0 -}_\A, T^{0 +}_\B] = T_{\A\B} - \D_{\A\B} H_2  , \label{comac} \\
[H_0,X^{\pm 0}_\A] &= \pm \frac{1}{\sqrt{N}} X^{\pm 0}_\A ,\; 
[H_1,X^{\pm 0}_\A] = \pm X^{\pm 0}_\A ,\;
[X^{+ 0}_\A, X^{- 0}_\B] = -T_{\A\B} - \D_{\A\B} H_1  , \label{comab'} \\
[H_0,X^{0 \pm}_\A] &= \pm \frac{1}{\sqrt{N}} X^{0 \pm}_\A ,\;
[H_2,X^{0 \pm}_\A] = \pm X^{0 \pm}_\A ,\;
[X^{0 +}_\A, X^{0 -}_\B] = -T_{\A\B} - \D_{\A\B} H_2  , \label{comac'} \\
[H_0,X^{\pm}_{\A\B}] &= \pm \frac{2}{\sqrt{N}} X^{\pm}_{\A\B} ,\;
[X^{+}_{\A\B}, X^{-}_{\G\D}] = \D_{\A\G} T_{\B\D} - \D_{\B\G} T_{\A\D} 
+ \D_{\B\D} T_{\A\G} - \D_{\A\D} T_{\B\G} ,\label{comaa'} \\
[H_1,T^{\pm\mp}] &= \pm T^{\pm\mp} ,\;
[H_2,T^{\pm\mp}] = \mp T^{\pm\mp} ,\;
[T^{+-},T^{-+}] = H_1 - H_2 , \label{combc}\\
[H_1,X^{\pm\pm}] &= \pm X^{\pm\pm} ,\;
[H_2,X^{\pm\pm}] = \pm X^{\pm\pm} ,\;
[X^{++},X^{--}] = H_1+H_2 , \label{combc'}\\
[T^{-0}_\A,T^{0+}_\B] &= - \D_{\A\B} T^{-+} ,\;
[T^{-0}_\A,T^{+-}] = T^{0-}_{\A}  ,\;
[T^{-0}_\A,X^-_{\B\G}] = -\D_{\A\B} X^{-0}_\G +\D_{\A\G} X^{-0}_\B ,\;\\
[T^{-0}_\A,X^{+0}_\B] &= -X^+_{\A\B}  ,\;
[T^{-0}_\A,X^{0-}_\B] = -\D_{\A\B} X^{--}  ,\;
[T^{-0}_\A,X^{++}] = X^{0+}_\A  ,\;\\
[T^{0-}_\A,T^{+0}_\B] &= -\D_{\A\B} T^{+-} ,\;
[T^{0-}_\A,T^{-+}] = T^{-0}_\A\ ,
[T^{0-}_\A,X^-_{\B\G}] = -\D_{\A\B} X^{0-}_\G  + \D_{\A\G} X^{0-}_\B  ,\;\\
[T^{0-}_\A,X^{0+}_\B] &= -X^+_{\A\B}  ,\; 
[T^{0-}_\A,X^{-0}_\B] = \D_{\A\B} X^{--}  ,\;
[T^{0-}_\A,X^{++}] = -X^{+0}_{\A}  ,\; \\
[T^{+0}_\A,T^{-+}] &= -T^{0+}_\A ,\;
[T^{+0}_\A,X^+_{\B\G}] = \D_{\A\G} X^{+0}_\B - \D_{\A\B} X^{+0}_\G ,\;
[T^{+0}_\A,X^{0+}_\B] = \D_{\A\B} X^{++}  ,\; \\
[T^{+0}_\A,X^{-0}_\B] &= -X^-_{\A\B}  ,\;
[T^{+0}_\A,X^{--}] = -X^{0-}_{\A} \ ,\; \\
[T^{0+}_\A,T^{+-}] &= -T^{+0}_\A ,\;
[T^{0+}_\A,X^+_{\B\G}] = \D_{\A\G} X^{0+}_\B  - \D_{\A\B} X^{0+}_\G  ,\;
[T^{0+}_\A,X^{+0}_\B] = -\D_{\A\B} X^{++}  ,\; \\
[T^{0+}_\A,X^{0-}_\B] &= -X^-_{\A\B}  ,\;
[T^{0+}_\A,X^{--}] = X^{-0}_{\A} \ ,\; \\
[T^{+-},X^{0+}_\A] &= X^{+0}_\A ,\;
[T^{+-},X^{-0}_\A] = -X^{0-}_\A ,\;\\
[T^{-+},X^{+0}_\A] &= X^{0+}_\A ,\;
[T^{-+},X^{0-}_\A] = -X^{-0}_\A ,\;\\
[X^{+0}_\A,X^-_{\B\G}] &= \D_{\A\B} T^{+0}_\G - \D_{\A\G} T^{+0}_\B ,\;
[X^{+0}_\A,X^{0-}_\B] = -\D_{\A\B} T^{+-} ,\;
[X^{+0}_\A,X^{--}] = T^{0-}_\A ,\;\\
[X^{0+}_\A,X^-_{\B\G}] &= \D_{\A\B} T^{0+}_\G  - \D_{\A\G} T^{0+}_\B  ,\;
[X^{0+}_\A,X^{-0}_\B] = -\D_{\A\B} T^{-+} ,\;
[X^{0+}_\A,X^{--}] = -T^{-0}_\A ,\;\\
[X^{-0}_\A,X^+_{\B\G}] &= \D_{\A\B} T^{-0}_\G - \D_{\A\G} T^{-0}_\B ,\;
[X^{-0}_\A,X^{++}] = -T^{0+}_\A , \;\\
[X^{0-}_\A,X^+_{\B\G}] &= \D_{\A\B} T^{0-}_\G  - \D_{\A\G} T^{0-}_\B  ,\;
[X^{0-}_\A,X^{++}] = T^{+0}_\A .
\end{align}

\section{Superfield components}
\label{app:susy}

For $\mathcal{N}=1$ superfields we use the conventions of Wess and
Bagger \cite{Wess:1992cp}. In the following we list a couple of
formulae for charged superfields\footnote{Note, that we use the
  notation $\olambda^+ = \overline{\lambda^+}$, etc.} that are frequently needed in the derivation of the 
4d effective Lagrangian:
\begin{align}
V^{\pm} &= -\nt \sigma^{\mu}\ot A^{\pm}_{\mu} 
+ i\nt\nt\ot\olambda^{\mp} 
-i\ot\ot\nt\lambda^{\pm} 
+ \frac{1}{2}\nt\nt\ot\ot D^{\pm} \ , \\
\phi^\pm &= \phi^\pm + \sqrt{2}\nt\psi^\pm + i \nt\sigma^\mu \ot
\pd_\mu \phi^\pm + \ldots \, \\
V^+ V^- &= -\frac{1}{2} \nt\nt\ot\ot A^+_\mu A^{-\mu} + \ldots \ ,\\
W^{+}W^-&= \nt\nt D^+ D^- +\ldots ,\\
V^+ \phi^- &= \frac{i}{2} \nt\nt\ot\ot A^+_\mu \pd^\mu \phi^- 
+\frac{i}{\sqrt{2}} \nt\nt\ot\ot \lambda^+ \psi^- 
+ \frac{1}{2} \nt\nt\ot\ot D^+ \phi^- +  \ldots \ ,\\
V^+ \ophi^+ &= -\frac{i}{2} \nt\nt\ot\ot A^+_\mu \pd^\mu \ophi^+
-\frac{i}{\sqrt{2}} \nt\nt\ot\ot \olambda^- \opsi^+ 
+ \frac{1}{2} \nt\nt\ot\ot D^+ \ophi^+ +  \ldots \ .
\end{align}

\section{Jacobi functions}
\label{app:jacobi}

For the reader's convenience we collect in this Appendix the
definitions, transformation properties and some identities among the
modular functions that are used in the text. The Dedekind function is
defined by the usual product formula (with $q=e^{2\pi i\tau}$)
\begin{equation}
\eta(\tau) = q^{1\over 24} \prod_{n=1}^\infty (1-q^n)\ , \label{a1}
\end{equation}
whereas the Jacobi $\theta$-functions with general characteristic and
arguments  are
\begin{equation}
\theta \left[\alpha \atop \beta \right] (z | \tau) = \sum_{n\in \mathbb{Z}}
e^{i\pi\tau(n-\alpha)^2} e^{2\pi i(z- \beta)(n-\alpha)}\ . \label{a2}
\end{equation}
We give also the product formulae for the four special $\theta$-functions
\begin{eqnarray}
&& \theta_1(z | \tau) \equiv \theta \left[{{1\over 2} \atop {1\over 2} }\right]
  (z|\tau) = 2q^{1/8}{\rm sin}\pi z\prod_{n=1}^\infty
  (1-q^n)(1-q^ne^{2\pi i z})(1-q^ne^{-2\pi i z}) \ , \nonumber \\
&& \theta_2(z | \tau) \equiv \theta \left[{{1\over 2} \atop 0 }\right]
  (z|\tau) = 2q^{1/8}{\rm cos}\pi z\prod_{n=1}^\infty
  (1-q^n)(1+q^ne^{2\pi i z})(1+q^ne^{-2\pi i z}) \ , \nonumber \\
&& \theta_3(z | \tau) \equiv \theta \left[{0 \atop 0 }\right]
  (z|\tau) = \prod_{n=1}^\infty
  (1-q^n)(1+q^{n-1/2}e^{2\pi i z})(1+q^{n-1/2}e^{-2\pi i z}) \ , \nonumber \\
&& \theta_4(z | \tau) \equiv \theta \left[{0 \atop {1\over 2} }\right]
  (z|\tau) = \prod_{n=1}^\infty
  (1-q^n)(1-q^{n-1/2}e^{2\pi i z})(1-q^{n-1/2}e^{-2\pi i z}) \ . \label{a3}
\end{eqnarray} 
The modular properties of these functions are described by
\begin{equation}
\eta(\tau+1) = e^{i\pi/12}\eta(\tau)\ \ , \ \
\theta \left[{\alpha \atop {\beta}}\right] \left({z} |
  {\tau+1}\right)=
e^{-i\pi\alpha(\alpha-1)}\theta 
\left[{\alpha \atop {\alpha+\beta-{1\over 2}}}\right] \left({z}  |
  {\tau}\right) \nonumber 
\end{equation}
\begin{equation}
\eta(-1/\tau) = \sqrt{-i\tau}\; \eta(\tau)\ \ , \ \ 
\theta \left[{\alpha \atop {\beta}}\right] \Big({z \over \tau} \Big| {-1 \over \tau}\Big)=
\sqrt{-i \tau} \ e^{2 i \pi \alpha \beta +{i \pi z^2 / \tau}} \ 
\theta \left[{{\beta} \atop - \alpha}\right] (z | \tau ) \ . \label{a4}
\end{equation}
A useful identity for theta functions is the Jacobi identity 
\begin{align}\label{ji}
 \sum_{\alpha,\beta=0,1/2} (-1)^{2 \alpha + 2 \beta + 4 \alpha \beta} &\prod_{i=1}^4 \theta \left[\alpha \atop \beta \right] (z_i | \tau) =  \nonumber  \\
 - 2 \theta_1&\Big(\frac{-z_1+z_2+z_3+z_4}{2} \Big| \tau\Big)
\theta_1\Big(\frac{z_1-z_2+z_3+z_4}{2} \Big| \tau\Big)  \nonumber\\
\times&\theta_1\Big(\frac{z_1+z_2-z_3+z_4}{2} \Big| \tau\Big)  \theta_1\Big(\frac{z_1+z_2+z_3-z_4}{2} \Big| \tau\Big) \ .   
\end{align}
In computing partition functions, it is useful to define $SO(2n)$ characters.  Of particular relevance for us are
\begin{align}
V_8 (z_i \tau | \tau)  &= \frac{\prod_{i=1}^4 \theta_3 (z_i \tau | \tau) - \prod_{i=1}^4 \theta_4 (z_i \tau | \tau)}{2 \eta^4} \ , \nonumber \\ 
S_8 (z_i \tau | \tau)  &= \frac{\prod_{i=1}^4 \theta_2 (z_i \tau | \tau) + \prod_{i=1}^4 \theta_1 (z_i \tau | \tau)}{2 \eta^4}\ . \label{a4}
\end{align}
\end{appendix}

\newpage

\providecommand{\href}[2]{#2}\begingroup\raggedright\endgroup

\end{document}